\documentclass{aastex63}

\begin{document}                                                                     

\newcommand{\Gaia}{{\it Gaia}}
\newcommand{\HST}{{\it HST}}
\newcommand{\IUE}{{\it IUE}}

\title{{\it Hubble Space Telescope\/} Snapshot Survey for Resolved Companions of
Galactic Cepheids: Final Results\footnote{Based on 
observations with the NASA/ESA {\it Hubble Space Telescope,} obtained
at the Space Telescope Science Institute, which is operated by the Association
of Universities for Research in Astronomy, Inc., under NASA contract
NASA5-26555.}
\footnote{Based on observations made with the {\it Chandra X-ray Observatory}}
 }


\author{Nancy Remage Evans}
\affil{Smithsonian Astrophysical Observatory,
MS 4, 60 Garden St., Cambridge, MA 02138; nevans@cfa.harvard.edu}

\author{H. Moritz G\"unther}
\affil{Massachusetts Institute of Technology, Kavli Institute for Astrophysics and
Space Research, 77 Massachusetts Ave, NE83-569, Cambridge MA 02139, USA}

\author{Howard E. Bond}
\affil{Department of Astronomy and Astrophysics, Pennsylvania State University, 
University Park, PA, 16802}
\affil{Space Telescope Science Institute, 3700 San Martin Drive, Baltimore, MD
21218}

\author{Gail H. Schaefer}
 \affil{The CHARA Array, Georgia State University, P.O. Box 3965, Atlanta GA 30302-3965}

\author{Brian D. Mason}
\affil{US Naval Observatory, 3450 Massachusetts Ave., NW, 
 Washington, D.C. 20392-5420 }

\author{Margarita Karovska}     
 \affil{Smithsonian Astrophysical Observatory,
MS 4, 60 Garden St., Cambridge, MA 02138; nevans@cfa.harvard.edu}

\author{Evan Tingle}  
 \affil{Smithsonian Astrophysical Observatory,
MS 4, 60 Garden St., Cambridge, MA 02138; nevans@cfa.harvard.edu}

\author{Scott Wolk}
 \affil{Smithsonian Astrophysical Observatory,
MS 4, 60 Garden St., Cambridge, MA 02138; nevans@cfa.harvard.edu}

\author{Scott  Engle}
\affil{Department of Astronomy and Astrophysics, Villanova University, 800 Lancaster Ave., 
Villanova, PA, 19085, USA}

\author{Edward Guinan}
\affil{Department of Astronomy and Astrophysics, Villanova University, 800 Lancaster Ave., 
Villanova, PA, 19085, USA}

\author{Ignazio Pillitteri}
 \affil{INAF-Osservatorio di Palermo, Piazza del Parlamento 1,I-90134 Palermo, Italy}

\author{Charles Proffitt}
\affil{Space Telescope Science Institute, 3700 San Martin Drive, Baltimore, MD 21218}

\author{Pierre Kervella}
\affil{LESIA, Observatoire de Paris, Universit\'e PSL, CNRS, Sorbonne Universit\'e, Universit\'e de Paris, 5 Place Jules Janssen, 92195 Meudon, France}

\author{Alexandre Gallenne}
\affil{Nicolaus Copernicus Astronomical Centre, Polish Academy of Sciences, Bartycka 18, 00-716 Warszawa, Poland}
\affil{Departamento de Astronomia, Universidad de Concepcion, Casilla160-C, Concepcion, Chile}
\affil{Unidad Mixta Internacional Franco-Chilena de Astronomia (CNRS UMI 3386), Departamento de Astronomia, Universidad de Chile, Camino El Observatorio 1515, Las Condes, Santiago, Chile}

\author{Richard I. Anderson} 
 \affil{European Southern Observatory, Karl-Schwarzschild-Str 2, 85748 Garching, Germany} 

\author{Maxwell Moe}
 \affil{University of Arizona, Steward Observatory, 933 N. Cherry Ave., Tucson, AZ 85721, USA} 


\begin{abstract}

Cepheids in multiple systems provide information on the 
outcome of the formation of massive stars.
They can also lead to  
 exotic end-stage objects.  This study  concludes our 
survey of 70 galactic Cepheids using the {\it Hubble Space Telescope\/} (\HST)
Wide Field Camera~3 (WFC3) with images at two wavelengths to identify
 companions closer than $5\arcsec$.
In the entire WFC3 survey we identify 16 probable companions for 13 Cepheids.   
The seven Cepheids having resolved candidate 
companions within $2''$ all have the surprising property of themselves being 
spectroscopic binaries (as compared with a 29\% incidence of spectroscopic 
binaries in the general Cepheid population).
That is a strong suggestion that 
an inner binary is linked to the scenario of a
third   companion within a few hundred~AU\null.  This characteristic is continued 
for more widely separated companions.
Under a model where the outer
companion is formed first, it is unlikely that it can anticipate a
subsequent inner binary.  Rather it is more likely that a triple system
has undergone dynamical interaction, resulting in one star moving outward
to its current location. 
{\it Chandra\/} and {\it Gaia\/} data as well as 
radial velocities and  \HST/STIS and {\it IUE\/} spectra   are used
to derive properties 
of the components of the Cepheid systems.   
 The colors of the companion candidates show a change in distribution
at approximately 2000~AU separations, from a range including both hot and cool
colors for closer companions, to only  low-mass companions for 
wider separations.

\end{abstract}


\keywords{stars: binaries; stars: intermediate mass; stars: Cepheid variable; star formation: multiple 
star systems}



\section{Introduction}


Binary and multiple systems are very common products of 
star formation.  This is increasingly so for more massive 
stars.  Properties such as separation, 
mass ratio, and eccentricity are the ``footprints'' of star 
formation, providing observational insight into the process. 
 Separations between components of the systems
determine which observational approaches can be used to
study them, 
as summarized by Figure~1 in Sana (2017). 
These include 
radial velocities for close binaries and many new techniques for 
resolving more distant companions.  
Duch\^ene \& Krauss 
(2013) provide a recent summary of observations of multiple systems. 
Evolution of systems can occur at the pre-main sequence 
phase, or post-main-sequence phase through Roche-lobe 
overflow and dynamical interaction.  
Formation and evolution of massive stars in multiple 
systems is particularly complex (de Mink et al.\ 2014),
and  a high fraction of O~stars undergo interaction and
mergers.  
The formation scenario in multiple systems 
probably involves core fragmentation, 
leading to relatively wide separations followed by 
disk fragmentation for closer components and subsequent 
migration for very close components.  Complex multiple 
systems, however, can be produced by more than one 
scenario (Tokovinin 2018a).

The formation of wide binaries presents a puzzle. The 
most likely scenarios are the unfolding of triple systems
(Reipurth \& Mikkola 2012), the 
dissolution of clusters (Kroupa 1995), and formation from 
adjacent cores (Tokovinin 2017).

As we are increasingly realizing, O and B  stars 
often occur in systems with more than two stars, 
inevitably making  the discussion of individual 
systems somewhat complicated.
 Cepheids 
(former B~stars with typical masses of $5\,M_\odot$) fall in the 
lower part of the mass  range of O and B stars.  
However there are 
several features of Cepheid systems 
which mean they make an important contribution to 
an accurate picture of the distribution of masses and separations 
in multiple systems. 
They are a well-defined sample of evolved helium-burning stars.  
 They are less rare than O stars, providing
more examples which can be resolved by interferometric techniques.
The combination of a cool evolved star and a hot companion
leads to uncontaminated 
spectra of both the brightest and hottest stars in the system     
at different wavelengths. Finally, Cepheid parameters are 
easier to relate to their main-sequence B-star counterparts 
than more massive stars, since mass loss on the main sequence
is minimal.  
However,   formation and the main-sequence 
stages for Cepheids are not the end 
of the story.  From B-star binary statistics we know that approximately  
 a third of Cepheids are the products of mergers (Moe \& Di Stefano 2017; 
Sana et al.\ 2012).   This is a good example, however,
of a complication which we have information to understand, since it 
will have only affected short-period main-sequence binaries.   
In addition to understanding how multiple systems are formed
(and what they tell us about the process), multiple systems, 
especially those containing massive stars, are important 
in many exotic end stage objects.  
They are  implicated in creating many 
objects such as X-ray binaries, millisecond pulsars, supernovae, 
Algols, and perhaps gamma-ray bursts and gravitational radiation sources. Moe \& Di~Stefano (2017) review binary 
properties involved in such outcomes.  

 The current study  
is part of a series aimed at determining the binary/multiple properties of 
Cepheids. The first paper (Paper~I; Evans et al.\ 2013) 
is a discussion of the binary properties of a group of Cepheids
with mass ratios greater than 0.4, compiled from a list observed by the 
{\it International Ultraviolet Explorer\/} (\IUE) satellite. 
This list is equally 
sensitive to companions at any separation, making the distribution of separations 
unbiased. The second paper  discusses  a sample of 70 Cepheids observed 
with the {\it Hubble Space Telescope\/} ({\it HST}) Wide Field Camera~3 (WFC3)) in two 
filters (Paper II; Evans et al.\ 2016a).  This paper discusses candidate companions more 
distant than $5\arcsec$, where photometry of possible companions is not contaminated 
by the much brighter Cepheid.  The current paper (Paper~III) follows with a discussion 
of possible
companions closer to the Cepheid.   These studies are supplemented  with
a discussion of {\it XMM-Newton\/} X-ray 
observations of systems with possible companions, in order to determine whether 
the companion is young and active enough to be a physical companion of the 
Cepheid (Paper IV; Evans et al.\ 2016b).  The study of resolved companions is complemented by 
a discussion of radial velocities for northern Cepheids (Evans et al.\ 2015).



The sections below discuss the observations and data reduction, both for Cepheids with companions separated 
between $2\arcsec$ and $5\arcsec$, and also companions within $2\arcsec$.
 {\it Chandra\/} X-ray observations are then used to verify companions.
The discussion section summarizes candidate companions at all separations.
Special attention is given to companions  within $2\arcsec$, 
and the components of these complex systems are discussed in detail. 
This includes the color-magnitude diagram (CMD),  the distribution of 
colors as a function of separation, and inner binaries of  this group.  
Astrometric data from {\it Gaia\/} are used to identify wider companions, both 
gravitationally bound companions and comoving but unbound companions.

\section{Observations and Data reduction}

Observations were made in our \HST\/ Wide Field 
Camera 3 (WFC3) snapshot program, which imaged 70 of the brightest 
Cepheids.  The F621M and F845M filters were used; these can be 
transformed into the $V$ and $I$ bandpasses of ground-based photometry.  (Throughout this paper all magnitudes discussed are on the Johnson-Kron-Cousins system.)
Observations were made in a dithered sequence.   The survey 
is described in  Paper~II, including the target list.  
The aim was to identify possible companions of the Cepheids, which are typically
many magnitudes fainter.  Because the point-spread function (PSF), dominated by 
the bright Cepheid, is complex, our data analysis has been divided into three 
sections, according to the angular distance from the Cepheid.  For the area 
$\geq$5$\arcsec$ from the Cepheid, the background is modest and only changes
slowly, and hence photometry is straightforward to derive using standard tasks in
IRAF.\footnote{IRAF was distributed by the National 
Optical Astronomy Observatories,
operated by the Association of Universities for Research
 in Astronomy, Inc., under cooperative agreement with the National
 Science Foundation.}  The results for this separation range are discussed
in Paper~II.

In the current paper, we only treat the region within $5\arcsec$ of the
central Cepheid.
We subdivide this inner region again into two
zones. Between $2''$ and $5''$, we again perform simple photometry with standard IRAF tasks
(Section~\ref{sect:IRAF}). The innermost regions are dominated by the bright
Cepheid, where its extended PSF makes it challenging to perform  source detection and
measure photometry. In Section~\ref{sect:2as} we describe this in detail.

\section{Companions between $2''$ and $5''$}
\label{sect:IRAF}

Source detection and photometry between 2\arcsec\ and 5\arcsec\ separation  from the 
Cepheid was done in the same way as for the more distant companions; details are provided in Paper~II.  

Once the list of sources  between 2\arcsec\/ and 5\arcsec\/ had been
generated, we tested them for the appropriate combination of magnitude
and color for main-sequence stars at the distance of the Cepheid,
as in Paper~II\null.  For the main-sequence relation, we used the Dartmouth Stellar Evolution 
Database\footnote{\url{http://stellar.dartmouth.edu/models/index.html}}
to generate an isochrone for a 1~Gyr age, for the pair of WFC3 filters we used, i.e., F621M absolute magnitude versus F621M$-$F845M color.  Although this age---the minimum contained in the Dartmouth database (at the time of Paper II)---is older than that of the Cepheids
(typically 50~Myr), even most deviant stars in this study 
(the late-K  spectral-type stars) are only
$\sim$0.1 mag above this representation of the zero-age main 
sequence (ZAMS).  To compare this ZAMS with the observed WFC3 photometry, we have 
``reddened'' the color and ``absorbed'' the magnitude and ``moved''
it to the distance of the Cepheid.  
Full details (including the distances of the Cepheids) are provided in 
Paper~II. 




\begin{deluxetable}{ccccccc}
\tablewidth{0 pt}
\footnotesize
\tablecaption{Candidate Cepheid Companions at Separations Between $2\arcsec$ and $5\arcsec$  \label{poss}}
\tablehead{
\colhead{F621M} & \colhead{F621M$-$F845M}  
  & \colhead{$V$} & \colhead{$V-I$} &
\colhead{Sep.}  & \colhead{P.A.} & \colhead{Sep.} \\
\colhead{[mag]} & \colhead{[mag]}  
  & \colhead{[mag]} & \colhead{[mag]} &
\colhead{[arcsec]}  & \colhead{[$^\circ$]} & \colhead{[AU]} 
}
\startdata
\noalign{\medskip}
\multispan7{\hfil Y Car \hfil} \\
  16.45 $\pm$   0.02   &     0.60 $\pm$   0.02   &    16.91   &     0.93   &      2.6   &     55  & 3,820 \\
  16.79 $\pm$   0.03   &     0.86 $\pm$   0.03   &    17.40   &     1.29   &      3.2   &     112  & 4,700 \\
\noalign{\medskip}
\multispan7{\hfil V496 Aql \hfil} \\
  18.39 $\pm$   0.09   &     1.56 $\pm$   0.09   &    19.05   &     1.99   &      4.3   &    78  & 4,250  \\
\noalign{\medskip}
\multispan7{\hfil TT Aql \hfil} \\
  17.16 $\pm$   0.05   &     1.40 $\pm$   0.05   &    17.82   &     1.85   &      3.8   &     67  & 3,520  \\
\noalign{\medskip}
\multispan7{\hfil V350 Sgr \hfil} \\
  17.24 $\pm$   0.04   &     1.31 $\pm$   0.04   &    17.91   &     1.77   &      3.1   &     128 & 2,780  \\
\noalign{\medskip}
\multispan7{\hfil BB Sgr \hfil} \\
   17.41 $\pm$	0.06	&     1.31 $\pm$   0.07  &    18.08 &      1.77   &      3.3   &     158 & 2,740   \\
\noalign{\medskip}
\multispan7{\hfil RV Sco \hfil} \\
  15.37 $\pm$   0.01   &     0.90 $\pm$   0.01   &    16.00   &     1.34   &      3.6   &     173  & 2,710 \\
\enddata
\end{deluxetable}




To select candidate companions of the Cepheids, we required them to lie within 2$\sigma$ of a band lying between the ZAMS and a parallel line  0.75~mag above it (to allow for binaries). These candidates are listed
in Table~\ref{poss}.  Appendix~A contains   Figures~\ref{ttaql} 
to  \ref{rvsco}, which illustrate the WFC3 images for the candidates in 
Table~\ref{poss}, and plot the CMDs for each field.  The present paper focuses on possible
companions within 5\arcsec; consequently there are some 
additional stars within the ZAMS band which lie at larger separations from the Cepheid; these were discussed in the previous paper (Paper~II).

Photometry of 
the candidate 
companions is listed in Table~\ref{poss}. 
 The instrumental magnitudes have similar errors to 
those of more distant companions in the previous study  derived from
the statistics.  However, we expect that there is some additional 
uncertainty due to a more variable background.  Table~\ref{poss} lists
the instrumental F621M 
magnitude (F621M) and color (F621M$-$F854M)
(columns~1 and 2).  For convenience, these magnitudes 
have been transformed to ground-based 
$V$ and $V-I$, using the relations in Paper~II (Vegamags), and are given in columns~3 and~4.
The separation in arcsec, the position angle, and the projected separation in 
AU (using the distances to the Cepheids from Paper~II) are listed in columns~5, 6, and 7. (For V496~Aql, the position angle has been corrected in Table~\ref{poss}, from
than that in Table 5 in Paper~II.)


\section{Companions Within 2\arcsec\ \label{sect:2as}}

\subsection{Data Reduction}

For these close companions lying within the PSF wings of the bright Cepheid primaries, we proceeded as follows. The data processing started with the geometrically
distorted, flat-fielded individual frames (FLT images) from the \HST\/ archive. We then applied
a custom data-reduction procedure to deal with
saturation and charge bleed.

\subsubsection{Masking the Charge Bleed}
\label{sect:bleed}

The Cepheids are severely overexposed in our frames. Not only are the central pixels 
saturated, but the additional charge collected beyond saturation bleeds out to
adjacent pixels. This happens predominantly along the columns of the CCD\null. In the
vicinity of the Cepheid, pixels ``bled into'' also reach the saturation charge limit, and
are flagged by the pipeline appropriately. However, we find that charge
continues to bleed upwards and downwards along the columns, leading to severely
enhanced charge levels up to two pixels beyond the fully saturated region. In
addition, charge bleeds to the right (increasing $x$ value in image coordinates)
into the neighboring column. We thus add custom filtering to any pixel that is
found to the right or one or two pixels above or below a pixel that is already
marked as saturated by the standard pipeline. Gilliland et al.\ (2010) 
performed a similar analysis, and suggested increasing the region masked due to
saturation by one pixel in any direction. In our data,  on the one hand, that recipe 
leads to spurious enhanced emission above and below the bleed region where
charge bleeds more than one pixel beyond the fully saturated region. On the
other hand it masks out area to the left of the saturated regions that seems
usable.

\subsubsection{Combining Exposures}

We combined individual exposures for each filter and each object using \texttt{astrodrizzle}
from DrizzlePac (Hack et al.\ 2013). This algorithm projects the values
of all contributing exposures onto the sky, and then, for every sky position,
averages over all valid (non-masked) values. Since the individual exposures
have slightly different boresight coordinates, pixels fall onto the
sky differently. Thus, this procedure results 
in a smaller area masked due to bleed than in the
individual images. On the other hand, close to the Cepheid, only one of three
exposures might contribute valid data, while all three images are averaged at
larger distances. Since the signal close to the Cepheid is high, the
noise in the combined image is still acceptable. We chose $0\farcs0396$ as
the pixel size in the resampled images.

\subsubsection{Cutting Out Sub-images}

In this section, we concentrate on close companions within 2\arcsec. Because
the Cepheids are all well saturated, we cannot determine the position of peak
flux. Instead, we fit straight lines to the diffraction spikes (diagonals in
our images) and take the intersection of the diffraction spikes as the position
of the Cepheid. We then extract sub-images 120 pixels on each side, centered on
the Cepheid. Based on the uncertainties of the fit to the diffraction spikes,
we estimate that this procedure is accurate to about one pixel.

\subsection{PSF Subtraction and Fitting}

 Sources  more than
2\arcsec\ from the Cepheid are discussed in Paper II and Section~3 above.
In this
section, we study the region within 2\arcsec, where the image is dominated by the
PSF of the Cepheid. 

\subsubsection{Locally optimized combination of images (LOCI)}
\label{sect:loci}

\begin{figure}
\begin{center}
\includegraphics[width=0.49\textwidth]{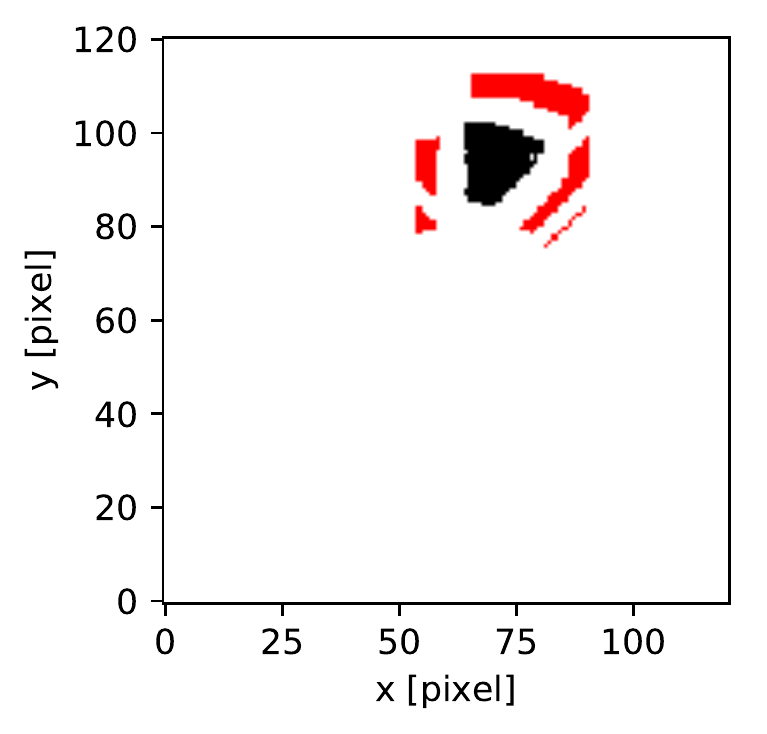}
\caption{Example of an optimization region in LOCI\null. To find the best possible
  combination of the PSF templates for the black region, we actually fit in the
  red region. This way, a moderately bright point source in the black region
  does not impact the results of the fit. Note how the red region does not
  completely envelop the black region, but has gaps avoiding regions that are
  contaminated by bleed or diffraction spikes or where one of the reference images 
cannot be used because it has a candidate companion itself.
\label{fig:fit_region}}
\end{center}
\end{figure}

In order to detect and perform photometry on potential
companions, we need to subtract the PSF of the Cepheid itself. This is
challenging since the {\it HST\/} PSF changes with time. Several
observational and computational approaches to detect faint sources close to a
much brighter primary have been developed in the past years, especially with an
application to exoplanets. One of the most common observational techniques
is angular differential imaging, where the same source is observed more than
once with a different orientation of the instrument, such that instrumental
features rotate on the sky, but true companions are fixed in sky
coordinates. The scheduling of a snapshot program did not allow for this
approach, so we are limited to computational methods to reduce the impact
of the Cepheid PSF\null. Lafreni{\`e}re et al.\ (2007)
introduced a method called
``locally optimized combination of images'' (LOCI); see their paper for a
mathematical description of the algorithm. In short, this algorithm describes
an image as a linear combination of companion-free reference images. The
coefficients that lead to the smallest residual between target image and the
combination of reference images can be found essentially by a matrix
inversion, which is computationally more efficient than numerical function
minimizers. Fitting the whole image at once would introduce two major problems.
First, the temporal change in the PSF might depend on the position, e.g., some
parts of the PSF change more than others due to a temperature change of the
telescope. Describing the full image at once thus requires a large number of
reference images. Second, if a companion exists and is included in the fit, the
fit will try to subtract it as much as possible, compromising the
photometry. Thus, for LOCI the image is divided into several smaller
regions. To find the best linear combination of templates for a specific
region, we perform the optimization on an area that is \emph{outside\/} the
region of interest (see Figure~\ref{fig:fit_region}).

\subsubsection{Modification of standard LOCI procedures}

Several enhancements of the LOCI method developed by Lafreni{\`e}re et al.\
have been suggested in the literature, including
``template LOCI'' (Marois et al.\ 2014),
which considers
spectral information, or ``matched LOCI'' (Wahhaj et al.\ 2015),
which inserts artificial point sources with known properties to inform the
fitting process. Another approach is to decompose the set of PSF templates into
eigenimages, thus using information from all images to isolate the true PSF
components and separate them from observational noise that varies from image to
image (Soummer et al.\ 2012).
After reviewing these ideas in the
literature, we modified the basic LOCI algorithm for the specific
characteristics of our dataset. 

\begin{figure}
\begin{center}
\includegraphics[width=0.49\textwidth]{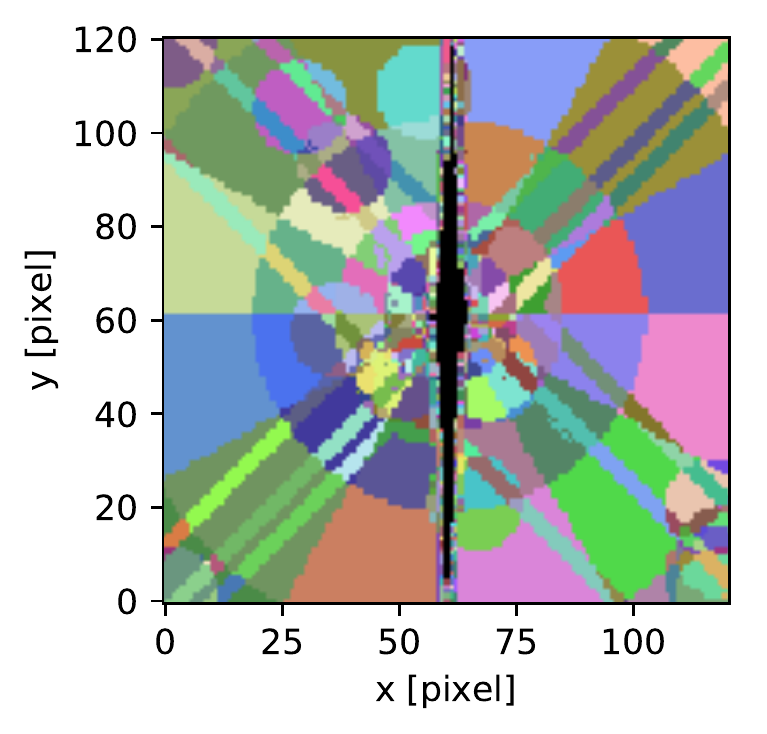}
\caption{An image is typically divided into more than 500 regions for
  fitting. Each color shows one region. Apparent circle shapes (e.g., the circle centered on
  (55, 110)) are due to potential companions in one of the images, forcing a
  smaller set of templates in that region. Similarly, diagonal lines are due to
  diffraction spikes from bright companions that we masked.
\label{fig:fit_regions}}
\end{center}
\end{figure}

For any one Cepheid observation, we use the set of the other 69 observations
taken in the same filter in our program as template images. This is justified because they are all taken
within a few months of each other, use an identical instrumental setup (except
for the exposure time), and all expose the central source well into saturation,
bringing out faint PSF features visible beyond $2''$, which are
generally hidden in the noise for other archival data taken in the same
filter. The drawback of this method is that some of the template images might
themselves contain companions, which would leave negative imprints in
the reduced images. We thus perform an initial source detection (see next
subsection) and mask regions potentially affected by a companion. For fits
involving this region, only the remaining 68 images are used as
templates. Similarly, the region affected by bleed differs from image to
image. To perform a fit close to the central source, we used only those images that
have valid data in this region as fitting templates. Following this
prescription, a typical image in our dataset is separated into more than 500
regions, each potentially with a different set of template images
(Figure~\ref{fig:fit_regions}).

\subsubsection{Initial source detection}
\label{sect:initialsources}

The very brightest companions (see Table~\ref{tab:comp.mag}) are themselves
saturated and are easily visible by eye. To identify other companions we need
to mask before using images as LOCI templates, we construct a median image. The
magnitude of the Cepheids is very different and this is not fully compensated
by choosing different exposure times. The flux level in the images differs by
about a factor of 1000 between the brightest and faintest image. We normalize
each image by dividing the image by its median value. As a second step, we form
a median image from all normalized images for each band. The median images show
the shape of the average PSF, not including any companions. We then divide each
normalized image by the median image. The result of this is an image that shows
how each dataset deviates from the median PSF\null. While not flux-conserving, this
procedure works well to highlight potential companions. We use the {\tt daofind}
algorithm from Stetson (1987)
as implemented in {\tt photutils}
to locate potential point sources and mask those before
performing LOCI fitting. At this stage we identify 17  potential companions and
mask their positions in both filters. Not all of those sources will turn out to
be real, but we can be conservative here because masking a potential source
just means that only one image cannot be used as PSF template in one region
(see Figure~\ref{fig:fit_regions} for an example), which has little impact on the
quality of the fit given that the set of potential templates consists of 69
images.

\subsubsection{Application of LOCI and PSF fitting}
\label{sect:psffitting}

\begin{figure}
\begin{center}
\includegraphics[width=1.0\textwidth]{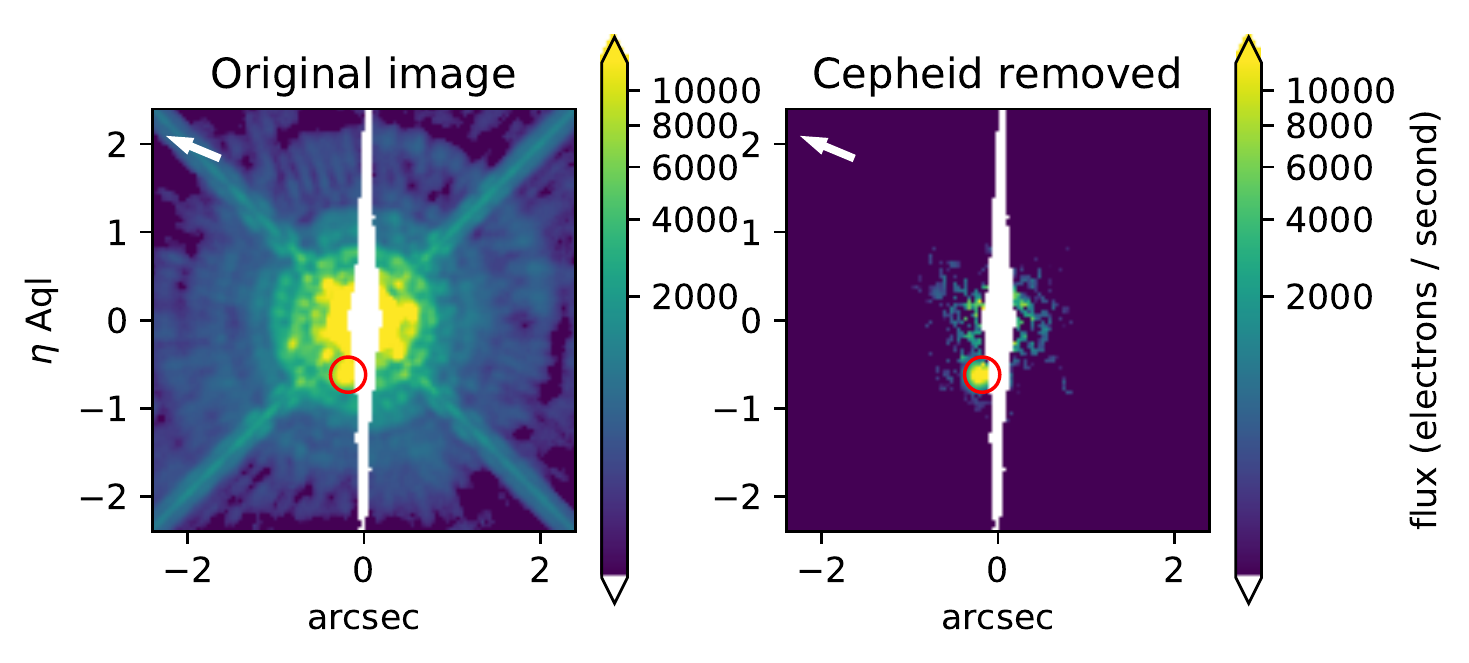}
\caption{ Image of $\eta$~Aql in the F845M band. Several bright
  PSF features are seen within 1\arcsec{} of the Cepheid position. The image is
  oriented along the detector coordinates. The white arrow points north. \emph{Right:}
  The companion is clearly visible after the LOCI subtraction.
\label{fig:EtaAql845}}
\end{center}
\end{figure}

We perform LOCI fitting to each of our images and subtract the result of the
best fit in each region from the original image. Figure~\ref{fig:EtaAql845} shows
an example. Most of the PSF features, including the diffraction spikes, are
significantly reduced, but within about 1\arcsec\ from the Cepheid, residuals
from the subtraction increase. We find that source-detection algorithms do not
perform well in this region. We thus inspect every image by eye and perform PSF
photometry on all source candidates from \S\ref{sect:initialsources}. In
principle, our procedure requires iteration, where any new sources identified
would now be masked and the LOCI fitting repeated. However, no new sources
beyond the previous candidates are identified. Table~\ref{tab:comp.mag} lists
the resulting source detections.

\subsubsection{Uncertainty estimates}

The uncertainty of the properties of the companion is dominated by systematics
introduced through the LOCI subtraction. Additionally, there is a risk of
oversubtraction in the LOCI procedure. We quantify those effects through
simulations where we insert fake sources with known fluxes into the images, run the same LOCI subtraction and measure the flux of the detected companions.
From our images, we select sources not visibly impacted by the PSF of the Cepheid. We checked the {\it Gaia\/} Data Release~2 (DR2) catalog
(Gaia Collaboration et al.\ 2016, 2018)  for these sources. 
 While this release does not explicitly mark extended or multiple sources, the errors on the coordinates are a good proxy. We reject any sources with {\it Gaia\/} coordinate errors above 0.1~milliarcsec as likely binaries. We fit a beta function as analytical PSF to the remaining sources and reject sources where fitted parameters are significantly different than the sample mean. We end up with a set of 18 PSF template stars. We fix the parameters of the shape and use the resulting analytical PSF for fits to both the real data (\S\ref{sect:psffitting}) and the LOCI subtracted images with fake companions inserted.

\begin{figure*}
\begin{center}
\includegraphics[width=1.0\textwidth]{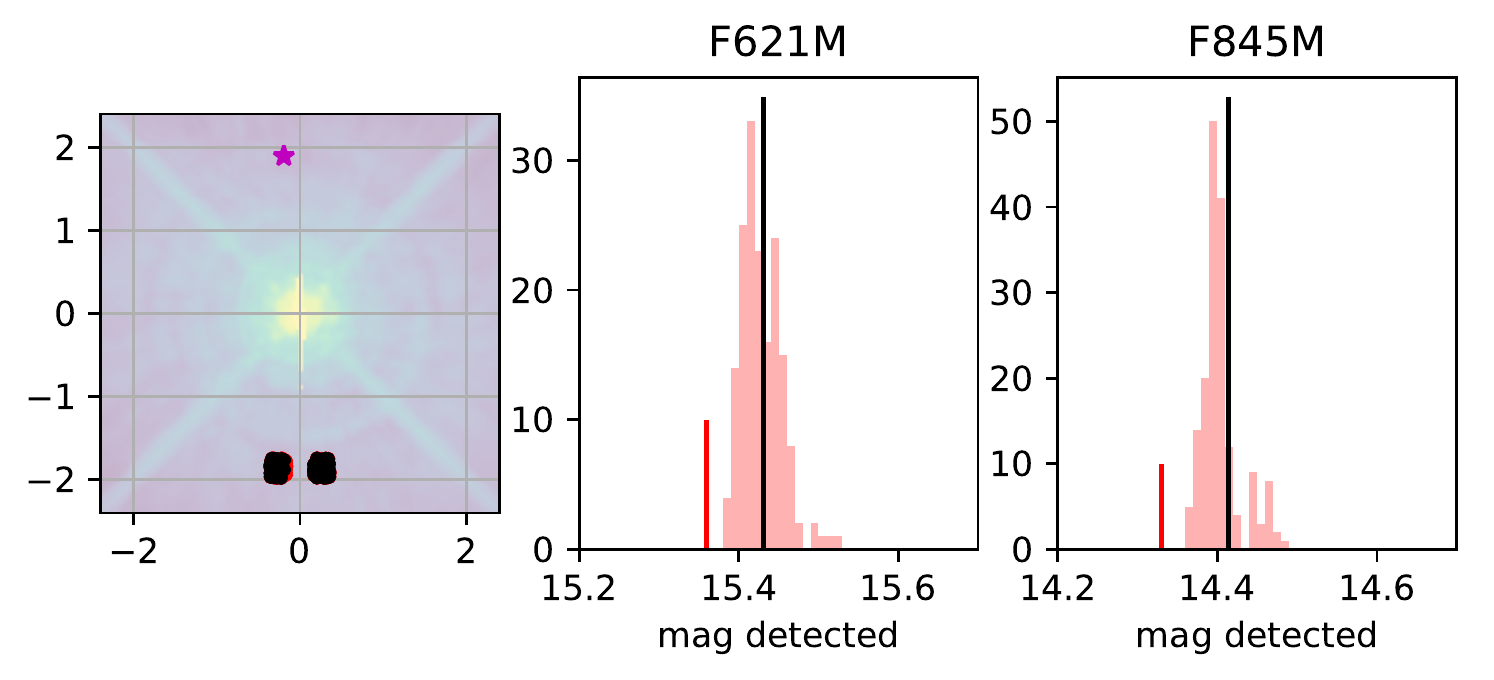}
\caption{\emph{left:} Image of R~Cru in the F621M band. The image is
  oriented along the detector coordinates. The purple star marks the location of the detected companion.  Fake companions are inserted at a location where the PSF of the Cepheid is comparable, here below the Cepheid at a similar distance to Cepheid and bleed column. Black dots are inserted companions that are recovered by the fit procedure, red dots are companions not recovered (e.g.\ because they are located too close to the bleed column). \emph{center and right:} Histogram of the recovered magnitudes of inserted companions for each filter. The red line marks the true magnitude of the inserted companions; the black line measured magnitude of the companion in the real image.
  \label{fig:RCruerror}}
\end{center}
\end{figure*}

For each template star, we extract squares 25
pixels wide, centered on those sources, from background-subtracted images.  For
each simulation, we select one of those sources, scale its \texttt{F621M} and
\texttt{F845M} flux as required and add it to one of the original Cepheid
images. We then perform LOCI subtraction on the image with the artificially
inserted companion and measure the source flux through PSF photometry. For each
source in Table~\ref{tab:comp.mag} we run 200 simulations. Depending on the
position of the companion in Table~\ref{tab:comp.mag}, we select positions
similarly effected by the Cepheid PSF\null.  For instance, for R~Cru, where the companion is
found almost 2\arcsec\ from the Cepheid, but only a few pixels from the bleed
column, we inject the artificial companions into the image on the opposite side
of the Cepheid with a similar distance to the bleed column
(Figure~\ref{fig:RCruerror}) and vary the insert location slightly to probe any background fluctuations statistically. Typically, some of the companion flux is subtracted in the LOCI procedure and thus the measured magnitudes (pink histograms in Figure~\ref{fig:RCruerror}) are fainter than the input magnitudes (red line). We thus conclude that the true flux of the companion in the real image must also be higher than the measured flux and we apply the offset determined from the simulations to correct the measured companion magnitude (black line). We take the standard deviation of the simulated magnitudes as the error estimate for the companion flux because the distribution of magnitudes is typically peaked, but the distribution has a significant tail towards fainter magnitudes for S~Nor, which means the true magnitudes could be up to 0.5~mag brighter than the number given in Table~\ref{tab:comp.mag}.

\subsubsection{Saturated companions}

This procedure cannot be applied for the very brightest companions that are themselves saturated, because we do not have PSF template stars of comparable brightness. If the companion's core is saturated, PSF photometry can in principle determine the flux by fitting the wings of the PSF\null. However, our template stars are not bright enough to determine the shape of the wings accurately enough. Similarly, the template stars would have to be scaled up significantly to be inserted into images for our simulations that determine the flux uncertainties, but this would significantly enhance the noise. However, for a measurement of the position alone, the exact shape of the PSF wings is not important and we thus provide positions for the saturated companions in Table~\ref{tab:comp.mag}.

\subsection{Results}

\subsubsection{Completeness}

\begin{figure}
\begin{center}
\includegraphics[width=0.49\textwidth]{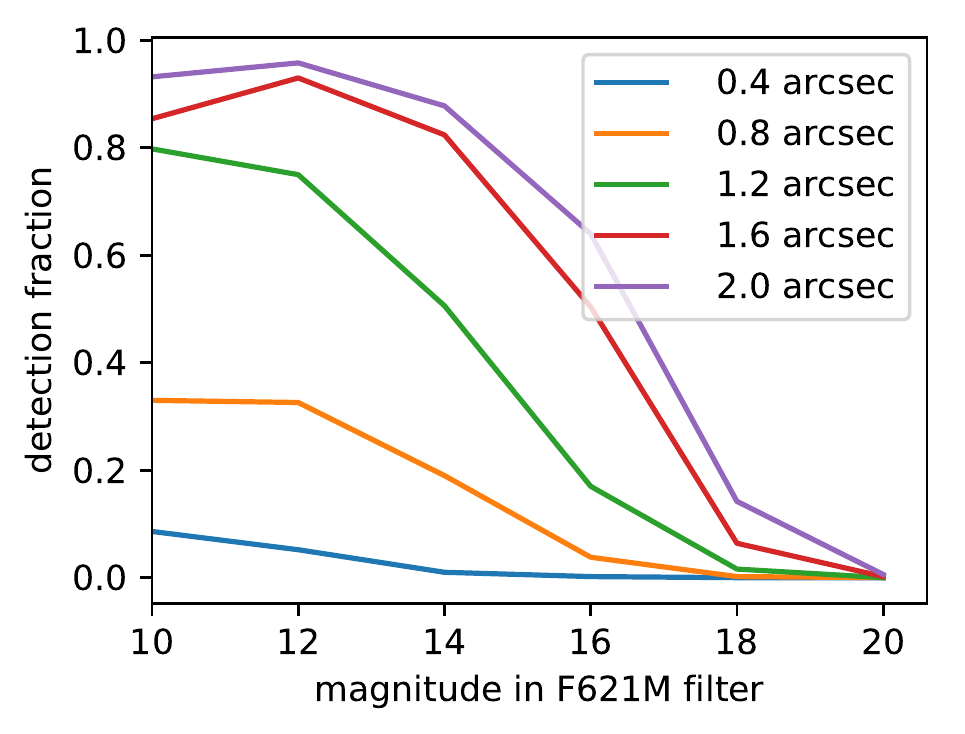}
\caption{Detection fraction for companions depending on distance from the Cepheid and flux from the companion.
\label{fig:detectlim}}
\end{center}
\end{figure}

To quantify our detection limits, we employ a procedure similar to our uncertainty estimate for detected companions. We insert companions into Cepheid images at a certain radius from the Cepheid, apply our source-detection algorithm, and measure the fraction of companions recovered in both bands. The inserted companions have a color $m_{621} - m_{845} = 1$, similar to the values observed in the detected companions (Table~\ref{tab:comp.mag}). We verified that the detection fraction changes little for reasonable choices of this color.
Figure~\ref{fig:detectlim} shows the  limits. We note that these limits describe the detectability in our whole sample. The flux level at a given distance from the Cepheid differs by three orders of magnitude between the brightest and faintest image due to different exposure times and Cepheid magnitudes and thus a companion that could be detected in one image, can be hidden in the background in another.
We note further (in the image segments we discuss outside of  $0\farcs5$)
that  between 1\arcsec\/ and $0\farcs5$ the bleed column obliterates a significant part of the image.

At $0\farcs4$, only a small fraction of companions is detected, even for the brightest sources. For bright sources, this number rises quickly with increasing distance and beyond 1\arcsec, most companions with $m < 12$ that do not overlap the diffraction spikes or the bleed columns are detected. Fainter companions require larger distances. Only for distances beyond $1\farcs$ are more than half of the $m_{621} = 16$ companions  detected.

\subsubsection{Detected companions}\label{detect.2}

The images in which companion candidates were detected are shown in Figure~\ref{fig:EtaAql845},
and Figures~\ref{fig:AXCir845} to \ref{fig:coaur845} in Appendix~\ref{appcom2}, both 
before and after PSF correction.

Table~\ref{tab:comp.mag} lists the detected companions. It gives both the raw, measured magnitude and the corrected magnitude based on the simulations. The uncertainty given is also taken from the simulations. No correction can be derived for the brightest three companions (around $\eta$~Aql, V659 Cen, and AX Cir), because those companions are themselves saturated. Fluxes are highly uncertain, but separation and position angle only rely on the assumption that the PSF is rotationally symmetric. The uncertainties in the position angle,  like the uncertainties in the fluxes, are dominated by residuals from the LOCI procedure. We analyze fake inserted sources and find that essentially all inserted sources are recovered within 5 pixels ($0\farcs2$) of the inserted position, 
while the large majority of sources are found within half that distance. These are upper limits to the positional uncertainty for true companions, because our insertion procedure can add sources only at integer-valued positions, introducing an additional shift up to half a pixel.

The photometry in Table~\ref{tab:comp.mag} has been converted to $V$ and $I$ in the same 
way as in Paper II, using the corrected photometry.
 The results are listed in Table~\ref{tab:prop.comp}.  Distance 
and reddening are also from Paper~II, with the distance based on the Benedict et al.\
(2007)  Leavitt Law.  The separation in AU is also listed, based on this distance.

The LOCI search also identified three additional sources in the postage stamps, 
 which are listed in the last three rows of Table~\ref{tab:comp.mag} and 
Table~\ref{tab:prop.comp}.  In the case of BG Cru, the source lies on a diffraction 
spike, and required LOCI processing to identify it. 
$V$ and $V-I$ for CO~Aur are based on uncorrected magnitudes in 
Table~\ref{tab:prop.comp}.  However, if we use the distance to the Cepheid, all three 
companions are too faint to be consistent with a ZAMS relation.  For BG Cru, 
for instance, the $(V-I)_0$ corresponds to an F8~V star (Drilling \& Landolt 2000), 
which would have $M_V = +4.0$.  The $V$, $E(B-V)$, and distance, however, correspond 
to $M_V = +6.7$. We conclude that 
they are all chance alignments, and will not discuss them further.

\begin{deluxetable*}{lcccccccc}
\tablecaption{Candidate Companions Detected after PSF Correction \label{tab:comp.mag}}
\tablehead{
\colhead{Cepheid} & \colhead{Sep.} & \colhead{P.A.} & \colhead{$m_{621}$ (raw)} & \colhead{$m_{621}$ (corr)} & \colhead{$\sigma_{621}$} & \colhead{$m_{845}$ (raw)} & \colhead{$m_{845}$ (corr)} & \colhead{$\sigma_{845}$} \\ 
\colhead{ } & \colhead{[arcsec]} & \colhead{[$^\circ$]} & \colhead{[mag]} & \colhead{[mag]} & \colhead{[mag]} & \colhead{[mag]} & \colhead{[mag]} & \colhead{[mag]}   
}
\startdata
R Cru & 1.9 & 344.5 & 15.43 & 15.36 & 0.03 & 14.41 & 14.33 & 0.02 \\
U Aql & 1.6 & 225.1 & 12.01 & 11.91 & 0.06 & 11.12 & 11.04 & 0.04 \\
U Vul & 1.5 & 319.7 & 16.26 & 16.18 & 0.04 & 14.92 & 14.85 & 0.03 \\
S Nor & 0.9 & 259.2 & 11.69 & 11.40 & 0.49 & 11.47 & 11.20 & 0.43 \\
V659 Cen & 0.6 & 234.9 & 10.54 & -- & -- & 10.12 & -- & --\\
$\eta$ Aql & 0.7 & 95.7 & 9.34 & -- & -- & 8.90 & -- & -- \\
AX Cir & 0.3 & 332.4 & 7.24 & -- & -- & 6.98 & -- & -- \\
 & & & & & & & & \\
AV Cir & 2.1 & 288.5 & 20.11 & 20.00 & 0.03 & 19.19 & 19.10 & 0.07 \\
BG Cru & 3.1 & 294.8 & 15.52 & 15.11 & 0.08 & 14.90 & 14.70 & 0.04 \\
CO Aur & 2.8 & 8.8 & 18.65 & -- & -- & 17.66 & -- & -- \\
\enddata
\end{deluxetable*}


\begin{deluxetable*}{lcccccc}
\tablecaption{Properties of Candidate Companions Detected after PSF 
Correction\label{tab:prop.comp}}
\tablehead{
\colhead{Cepheid} & \colhead{Sep.} & \colhead{$V$} & \colhead{$V-I$} & \colhead{$(V-I)_0$} & \colhead{$E(B-V)$} & \colhead{Dist.} \\ \colhead{ } & \colhead{[AU]} & \colhead{[mag]} & \colhead{[mag]} & \colhead{[mag]} & \colhead{[mag]} & \colhead{[pc]} 
}
\startdata
R Cru & 1580  & 16.02   &  1.49  &  1.27  &  0.19   & 829   \\
U Aql & 981 &  12.53 & 1.30   &  0.90  & 0.35  & 613   \\
U Vul & 822 & 16.85  & 1.79  &  1.04   & 0.65   & 548   \\
S Nor & 819  &  11.55  &  0.30  &  0.08    &  0.19  & 910   \\
V659 Cen & 452 & -- &  & -- &   0.21   & 753  \\
$\eta$ Aql & 191  & -- &  & -- &  0.12 & 273   \\
AX Cir & 158  & -- &  & -- & 0.25   &  527   \\
 & & & & & &   \\
AV Cir & 1472 & 20.63  &  1.34 & 0.88  & 0.40  & 701   \\
BG Cru & 1615  & 15.42  & 0.64   & 0.58  & 0.05   & 521    \\
CO Aur & 2229  & 19.30 uc &  (1.45) & (1.19)  & 0.23  & 796   
\enddata
\end{deluxetable*}

\subsection{Caveats}

As discussed in Paper II, the goal of the survey was to detect companions down to dwarf~K spectral 
types, which limits contamination by field stars. 

As discussed below (Table~\ref{com.mix}), in a number cases an absolute magnitude from other 
sources, such as {\it IUE\/} spectra, is more accurate than the \HST\/ photometry for very close companions.

\section{{\it Chandra} Observations}

Analysis of the properties of members of the multiple systems containing a
Cepheid frequently requires multiwavelength observations.  X-ray observations 
are often an important diagnostic. Specifically, {\it Chandra} observations 
of two stars R Cru and S Mus followup {\it XMM} observations in Paper IV 
with higher 
spatial resolution, providing additional information about system components. 
The details of the observations are provided in Appendix~\ref{chandra}.

\section{Discussion}
\label{comp.2}

\subsection{Summary of Candidate Companions}

Because of differences in the data-analysis procedures dictated by the angular separations,
 in the preceding work we divided the resolved companions found in our 
{\it HST\/} WFC3 imaging 
survey into three groups (separations $>$5\arcsec, 2--5\arcsec, and $<$2\arcsec).  In the 
following discussion, we will combine the first two groups, designating them the 
{\it ``Resolved Wide''} companions. The latter group will be called the 
{\it ``Resolved Close''} companions. In terms of physical projected separations, 
the Resolved Wide companions lie more than $\sim$2000~AU from the Cepheid, and the 
Resolved Close ones are closer to the primary star. We now examine  whether the 
properties of the stars in the two groups differ.  





Table~\ref{com.sum} presents the results for the combined
sample of 16 candidate companions.  
These data are assembled from Table~2 in Paper~II for the top group, with separations greater than $5''$. For the 
middle group, with separations of $2''$ to $5''$, the entries are from  Table~\ref{poss}.
The data for the close companions below $2''$ are more complicated, and these companions are discussed individually below. Our measures of separation and magnitude are listed in 
Tables~\ref{tab:comp.mag} and \ref{tab:prop.comp}. However, for very close companions, and 
systems with multiple companions, the most reliable companion parameters come
from several sources, which are discussed and assembled in Table~\ref{com.mix}.
 Table~\ref{com.sum} lists the projected separations in both arcseconds and AU, the $V$ and $V-I$ magnitudes and colors, the dereddened $(V-I)_0$ color, the $E(B-V)$ reddening, and the distance.  The distances  and reddenings are from Paper~II\null. As in Paper~II, the $V-I$ color excess is found from the relation $E(V-I) = 1.15\,E(B-V)$.

Among the possible companions separated by more than $5''$ from the primaries, only five systems remained after we eliminated those candidates undetected in X~rays, implying that they are not young enough to be Cepheid companions. The remaining five are R~Cru, FF~Aql, AP~Sgr, RV~Sco, and TT~Aql. 
  The discussion in Paper~II concluded that a Cepheid 
with a crowded line of sight (a field  with more than
two possible companions) is more likely to have chance alignments.  On that basis we 
remove TT Aql.  The candidate companion of AP~Sgr has the second-largest physical separation in the remaining sample, and it is a K~dwarf, which is the most common type of field-star 
chance alignment. Thus AP~Sgr is the most suspect of the remaining retained candidates in the list.   
  As discussed in Appendix~\ref{chandra},  
we are retaining the candidate wide companion for R~Cru.


We also checked the {\it Gaia\/} DR2 catalog for parallaxes and proper motions of our candidate 
companions.  Because the candidates are so close to the much-brighter Cepheids,
in most cases parallaxes are not available, particularly for the closer
companions. The candidate companions of V737~Cen, R~Mus, and Y~Sgr were found to be 
much more distant than the Cepheids, consistent with the negative results of the X-ray 
observations reported in Paper~IV; they are not included in Table~\ref{com.sum}.  For the possible companions with separations $>5''$ in Table~\ref{com.sum}, the DR2 results are as follows. The {\it Gaia\/} DR2 parallax for R~Cru itself is negative, but the parallax of the wider of its two candidate companions is consistent with the distance given in Table~\ref{com.sum} (and with the {\it Hipparcos\/} parallax of R~Cru), and its DR2 proper motion is similar to that of the Cepheid. The parallax of the FF Aql companion agrees within the errors with that of the Cepheid, 
but as noted by Kervella et al.\ (2019b), its proper motion is significantly different; this implies
 a relative velocity too large for it to be bound, unless it is itself a binary. 
The companion for AP Sgr is listed in DR2, but no parallax is given.
Finally, the DR2 parallax for the wider of the two RV~Sco companions agrees with that of the Cepheid within the errors, and its proper motion is similar. 


The middle group in  Table~\ref{com.sum}
lists the candidate companions with separations between $2''$ and 5$\arcsec$. The only companion listed in {\it Gaia\/} DR2 is that of RV~Sco, but no parallax or proper motion is given. 
In the case of V496~Aql,
the DR2 parallax is significantly smaller (4.3$\sigma$) than that of the Cepheid, and hence is likely to be 
a chance alignment.  It has been omitted from Table~\ref{com.sum}. 




\begin{deluxetable}{lccccccc}
\footnotesize
\tablecaption{Summary of Resolved Candidate Companions \label{com.sum}}
\tablehead{
\colhead{Cepheid}
  & \colhead{Sep. } &  \colhead{Sep. } &  \colhead{$V$} &   \colhead{$V-I$} & 
\colhead{$(V-I)_0$} &
 \colhead{$E(B-V)$}  &   \colhead{Dist.}      \\
\colhead{}
  & \colhead{[arcsec]} & \colhead{[AU]}
& \colhead{[mag]} & \colhead {[mag]} &
\colhead{[mag] }  & \colhead{ [mag]}    &   \colhead{[pc]} 
}
\startdata
\noalign{\medskip}
\multispan8{{\bf Resolved Wide} \hfil}\\
\multispan8{\hfil Separation $>5''$ \hfil} \\
 R Cru   &       7.6    & 6330   & 16.28  &  1.17   & 0.95    & 0.19 & 829 \\
  FF Aql   &      6.9    & 2520 &   11.22 &    0.85 & 0.60 & 0.22 & 365 \\
 AP Sgr   &     6.3   &  5320 & 17.85 &    1.72 &  1.50 & 0.19  & 845 \\
  RV Sco   &      6.0   &   4520 & 12.68 &    0.63 &  0.24 & 0.34 & 753   \\
\noalign{\medskip}
\multispan8{\hfil $2'' < \rm Separation < 5''$ \hfil} \\
  RV Sco   &      3.6   &   2710 & 16.00 &   1.34 & 0.95 & 0.34  & 753  \\ 
  BB Sgr   &       3.3   &   2740 &  18.08 &    1.77 & 1.45 & 0.28 & 831 \\ 
   Y Car   &      3.2   &    4700 & 17.40 &    1.29 &  1.20  & 0.08 & 1468 \\
   \hfil$''$\hfill   &      2.6   &    3820 & 16.91 &    0.93 &  0.84  & 0.08 & 1468 \\
V350 Sgr   &       3.1   &   2780 & 17.91 &    1.77 & 1.40  & 0.32 & 896 \\
\noalign{\medskip}
\multispan8{{\bf Resolved Close} \hfil}\\
\multispan8{\hfil Separation $< 2''$ \hfil} \\
   R Cru   &      1.9   &    1580 & 16.02  & 1.49 &  1.27  & 0.19 & 829 \\ 
   U Aql   &      1.6   &    981 & 12.53 & 1.30  &  0.90  &  0.35 & 613  \\
   U Vul   &      1.5   &    822 & 16.85 & 1.79  &  1.04  &  0.65 & 548 \\
   S Nor   &      0.9   &    819 & 11.55 & 0.30   &  0.08  & 0.19  & 910  \\
 $\eta$ Aql   &      0.7   &    191 & --    &  --  & -- & 0.12 & 273 \\
V659 Cen   &      0.6   &    452 & --  &  --    & -- & 0.21 & 753 \\  
  AX Cir   &      0.3   &     158 & -- & --  & --  &  0.25  & 527  \\
\enddata
\end{deluxetable}



\subsection{Resolved Close Companions: Companions Closer than $2''$}
 \label{sect:sys.clos}
 

A number of the systems with a companion closer than
2$\arcsec$ to the Cepheid have at least one additional 
component.  This means disentangling the mass and temperature 
of the companions often requires information from a number
of techniques.  Here we  discuss what is known about the 
components of these close systems in the third group in 
Table.~\ref{com.sum}.  This 
becomes important below, in first examining the CMD
of the companions, 
and subsequently the distribution of colors as
a function of separation, 
and the multiplicity properties of 
the sample. 


{

Because massive star systems often have multiple components,
in Appendix~\ref{sys.2}
we go into considerable detail about each system. 
Derivation of the parameters of each component (mass, 
temperature, and luminosity)  can be done using 
observations over a wide wavelength range, and also 
as a function of separation.  There are many approaches
and tools currently available to do this.  Radial-velocity 
studies extend over many decades.  The emphasis in this study 
is on separations on the 1$\arcsec$ scale, but a number 
of systems have been observed via interferometry (e.g., 
Gallenne et al.\ 2019), accessing closer separations. 
Both {\it HST\/} and {\it IUE\/} provide information about 
hot companions in the satellite ultraviolet. {\it Chandra\/}
and {\it XMM\/} X-ray observations test the ages of 
companion candidates.  Finally 
{\it Gaia\/} and {\it Hipparcos\/} data have contributed in two
ways (Kervella et al.\ 2019a, 2019b). First, the comparison of 
proper motion from the two satellites in some cases results
in ``proper-motion anomalies,'' indicative of orbital motion. Second, 
comoving companions can be identified and classified either as 
bound or not.  These data are discussed further in Section~\ref{gaia.comp}, 
 ``{\it Gaia\/} Components''. 
 

In  the discussion of some of the systems 
we have  estimated masses of the companions.
 The approach used is  discussed 
in Evans et al.\ (2018), based on data from 
detached eclipsing binary systems (Torres et al.\ 2010).  In particular, since the companion to a more massive
primary  should not have evolved significantly, the lower envelope to the 
temperature--mass data is used.

{

Magnitude measurements for sources within 2$\arcsec$ of the Cepheid have substantial 
uncertainties.  This is particularly important in determining the color or temperature
using the difference of two filters, resulting in an uncertainty which is a 
significant fraction of the range of color.  In many cases for the companions, there
is an alternate measurement which provides a more accurate color or temperature. 
Specifically,  many of the
stars have an {\it  IUE\/} 
spectrum, which provides a sensitive determination of the spectral type of the companion; the companion can completely dominate the flux at 1500~\AA\null.   Table~\ref{com.mix} summarizes 
the best  (``preferred'') combination of $V_0$ and $(V-I)_0$ available for the companions within 
2$\arcsec$ of the Cepheid. $V_0$ is taken from Table~\ref{tab:prop.comp} 
 for S Nor, R Cru, U Aql, and U Vul.
 From this,
and the distance in Paper~II, $M_V$ is computed (column~3).  Column~4 is the spectral
type from the instruments and literature references in columns~6 and~7. 
Additional details are 
given for each star in Appendix~\ref{sys.2}. The closest companions, 
V659 Cen B and AX Cir B, provide 
magnitudes which are too uncertain on the WFC3 images, so the value of $M_V$ based 
on the \IUE\/ spectral type has been substituted.   For $\eta$ Aql, the $M_V$ from the VLT/NACO
imaging is in Appendix~\ref{sys.2}.


This  information for complex systems is discussed in 
Appendix~\ref{sys.2}.  We summarize the
component parameters   in Table~\ref{mcomp} for 
 reference.  This includes the resolved companions (Table~\ref{com.mix}), 
as well as inner spectroscopic binaries and possible wider system  
members. While there is a lot of detail in
the linked inferences, the available observations provide
well constrained properties for most of the stars involved. 
The top entry for each Cepheid is for the resolved companion identified 
in this study.  
The second column is the binary/multiple star notation.
$V_0$, $(V-I)_0$, and $M_V$ are taken from 
Table~\ref{com.mix}. The final right column lists results for very wide 
companion candidates from {\it Gaia\/} DR2 (Kervella et al.\ 2019b), where WB is 
bound companions, WC is comoving companions. 
Systems in Tables~\ref{com.mix} and \ref{mcomp} are ordered by decreasing separation 
of the resolved companion.


\begin{deluxetable}{lcclclc}
\footnotesize
 \tablecaption{Preferred Properties of Resolved Close Companions \label{com.mix}}
\tablehead{
 \colhead{ Primary }  & \colhead{$V_0$}
  & \colhead{$M_V$} & \colhead{Spec.\ Type} & \colhead{$(V-I)_0$} &  Source & Ref.\tablenotemark{a}  \\
\colhead{ (Cepheid)}  & \colhead{[mag]} & \colhead{[mag]} &  \colhead{} &\colhead{[mag]}
  & \colhead{} & \colhead{} \\
\colhead{}  &  \colhead{   Comp} & \colhead{  Comp} &  \colhead{  Comp}
 & \colhead{  Comp} 
  & \colhead{} & \colhead{} 
}
\startdata
R Cru & 15.4  & +5.8 & & 1.27 & & (1) \\
U Aql & 11.3  & +2.4 & A5 & 0.16\tablenotemark{b} & {\it HST}/STIS & (2) \\
U Vul & 14.6 & +5.9 &  & 1.04 & & (1) \\
S Nor & 10.89 & +1.1  & B9.5 V  & $-0.04$\tablenotemark{b} &  \IUE &  (3) \\
V659 Cen & 8.72 & ($-0.3$) & B6.0 V & $-0.16$\tablenotemark{b} & \IUE & (3) \\
$\eta$ Aql & 10.0 & +3.2  & F3 & 0.44\tablenotemark{b}  & VLT/NACO & (4) \\
AX Cir & 6.91 & ($-0.3$) & B6.0 V & $-0.16$\tablenotemark{b} & \IUE &  (5) \\
\enddata
\tablenotetext{a}{References:
(1) Table ~\ref{com.sum};
(2) Appendix D
(3) Evans 1992b;
(4) Gallenne, et al.2014b
(5) Evans 1994
}
\tablenotetext{b}{From Drilling \& Landolt 2000, their Table 15.11 (which is quoted from Bessell 1979).}
\end{deluxetable}





\begin{deluxetable}{llccccclcc}
\footnotesize
\tablecaption{Multiple Components in Systems with Resolved Close 
Companions  \label{mcomp}}
\tablewidth{0pt}
\tablehead{
\colhead{Cepheid} & \colhead { Multiple}
  & \colhead{Sep.}  & \colhead {Sep.} &  
  \colhead{SB\tablenotemark{a}} &
\colhead{$V_0$}     & \colhead{$M_V$}   & \colhead {Spec.} & \colhead{Ref.\tablenotemark{b}}
& \colhead{Wide\tablenotemark{c}}  \\
\colhead{ }  & \colhead{ ID}
& \colhead{[arcsec]} & \colhead{[AU]} &  \colhead{} &
\colhead{[mag]} & \colhead{[mag]} & \colhead{Type} & \colhead{} & \colhead{} 
}
\startdata
   R Cru   &   AB &  1.9    & 1580 & & 15.4  &       +5.8 &    K & 1 & WB \\ 
   &   Aa,Ab & & 3: & SB &  &  & $<$A2 & 9 & \\
 &  AC & 7.6                    &  6300 & & 15.6 &       +6.0 & K2 V & 11 & \\
   U Aql   &   AB &   1.6   &    981 &  & 11.3  &   +2.4  & A5 V   & 6   &   \\
  &  Aa,Ab  & 0.107  & 66 &  & & +1.8  & A3-4 V &  6  &  \\
  &  Aa1,Aa2  & 0.0101  & 6.2 & SB &  & +1.2  & B9.8 V   &  6  &  \\
   U Vul  &  AB    &      1.5   &   822 &  & 14.6  &   +5.8 & K0    & 1, 9    &   \\  
      &  Aa,Ab       &   & 7.1 & SB & & &  $<$A1  &  3, 9  &   \\ 
   S Nor   &   AB &    0.9        & 819 &  & 10.9  &  +1.1  &  B9.5 V  & 1, 10 & WC     \\ 
   &  Aa,Ab  &  &  8.87 & SB: & & &  F0 V & 3 &  \\
  &  AC & 14.6 &  13300 &  & 13.29 & +3.5  & G3 V & 11 &  \\
V659 Cen   &  AB  &   0.6   &     452 &  & 8.7    & $-0.3$ &  B6.0  & 1, 9  & WB   \\
  & Aa,Ab  &  &    3: &  SB  &  &   &     & 5 &     \\
 $\eta$ Aql   &   AB   &    0.7   & 191 & & 10.0  &    +3.2 &  $\sim$F3 V    &    1, 2  &  \\
   &  Aa,Ab &  &  1.3: & SB & 8.39 & +1.2 &    B9.8 V  & 4, 8  & \\
  AX Cir   &  AB &     0.3   &   158 & & 6.9  &  $-0.3$ &  B6.0  & 1, 9   &   WB, WC   \\
  &  Aa,Ab   & 0.029 & 15.4 & SB & &  +0.2 & B9.0     &  7, 12  &  \\
  & Ab1,Ab2  &  &   &  & &   &  & 7, 12  &       \\
\enddata
\tablenotetext{a}{
Spectroscopic Binary Companion}
\tablenotetext{b}{References:
1: Table~\ref{com.mix}; 
2: Gallenne et al.\ 2014a; 
3: Kervella et al.\ 2019a; 
4: Appendix C; 
5: Anderson, R. I. 2020, in preparation; 
6: Appendix D; 
7: Gallenne et al.\ 2014b; 
8: Evans 1991; 
9: Evans 1992a; 
10: Evans 1992b;
11: Paper IV;
12: Gallenne et al.\ 2020 in preparation
}
\tablenotetext{c}{Wide companion types: WB gravitationally bound; WC comoving.}
\end{deluxetable}
}

The important characteristic that emerges from the 
discussion of the complex systems is 
 that all the systems with Resolved Close companions
(companions within $2''$) have 
an {\it interior spectroscopic binary\/} (or at least one is strongly suspected, in the case of S~Nor).  
This is discussed in more detail
in Section~\ref{Multiplicity}.

\subsection{Resolved Close Companion Properties: CMD}


We now return to a discussion of the close but resolved 
companions (Table~\ref{com.mix}), which are the first entry for 
each of the systems in  Table~\ref{mcomp} as discussed in Appendix~\ref{sys.2}.  
As a first step in assessing the properties of
the resolved close companions 
(within 2$\arcsec$ of the Cepheid [Table~\ref{com.mix}])
we consider the CMD  (Figure~\ref{cmd}),  which provides  
 a first examination of the properties of these companions. 
 The companions are fairly evenly distributed 
in color, $M_V$, and hence in mass.  Unlike the field Initial Mass Function (IMF), 
they are not concentrated to low-mass stars. For wider companions, comparisons with 
the ZAMS are the way 
 physical companion candidates are identified.  In  Figure~\ref{cmd}
this is only a test for the two coolest companions, R Cru and U Vul, since they are the
only ones which are based on WFC3 photometry.  While these points sit above the ZAMS
indicating some errors, they are consistent with the companions being low-mass stars.   

\begin{figure}
\centering
\includegraphics[width=5in]{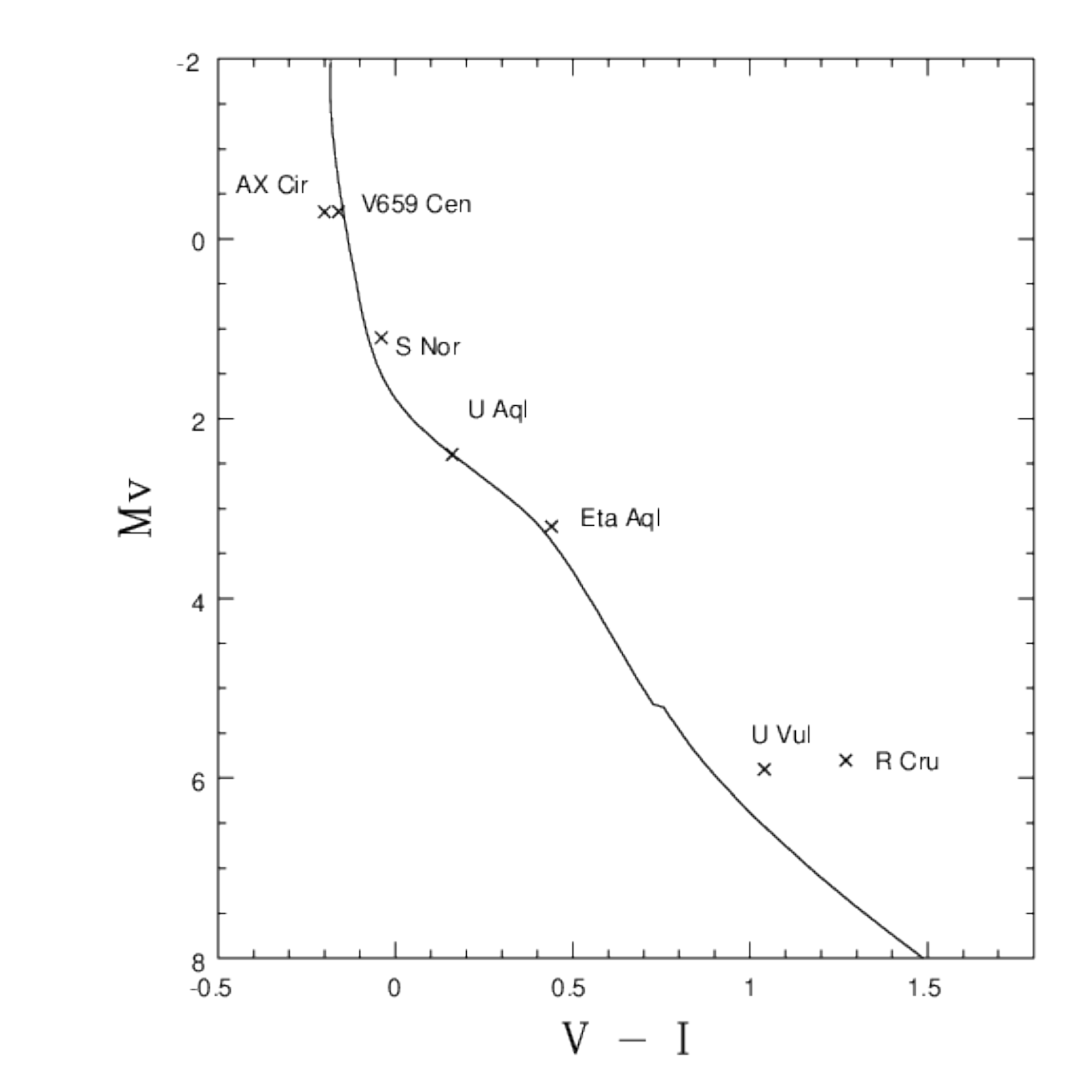}
\caption{The $M_V$ versus $(V-I)_0$ color-magnitude diagram
for the resolved close companions 
(Table~\ref{com.mix}). 
Colors and magnitudes are from Table~\ref{com.mix}, for 
which most are taken from WFC3 images. 
For $\eta$~Aql, 
the data are from Gallenne et al.\ (2014a), although it agrees
with the WFC3 data.  For AX Cir and V659 Cen, data are from 
{\it IUE\/} spectra. Identifications of the  stars 
are included  on the plot. 
The solid line is the ZAMS from Paper~II.
 \label{cmd}}
\end{figure}

\subsection{Systems with a Resolved Wide Companion (Wider than $2''$)}

Because the systems with a companion closer than  $2''$ have the 
distinctive property of an inner spectroscopic binary 
(Table~\ref{mcomp}),  we  
examine the multiplicity properties of the systems with a companion 
wider than $2''$ to see if they share this characteristic.
Components of these systems are summarized in Table~\ref{mcompw}.
Unless discussed below, the spectral types for the companions  
are derived from $(V-I)_0$ in Table~\ref{mcompw}, using the calibration 
of Drilling \& Landolt (2000; their Table 15.11).  The details of the 
components of these systems are discussed in Appendix~\ref{sys.5}.

{\bf 
\begin{deluxetable}{llcccccclcl}
\footnotesize
\tablecaption{Multiple Components in Systems with Resolved Wide
Companions   \label{mcompw}}
\tablewidth{0pt}
\tablehead{
\colhead{Cepheid} & \colhead { Multiple} 
  & \colhead{Sep.}  & \colhead {Sep.} 
  & \colhead{SB\tablenotemark{a}} &
\colhead{$V_0$}   & \colhead{$(V-I)_0$}  & \colhead{$M_V$}   & \colhead {Spec.} 
& \colhead{Ref.\tablenotemark{b}}  & \colhead{Wide\tablenotemark{c}}  \\
\colhead{}  &   \colhead{ ID} &   \colhead{[arcsec]} & \colhead{[AU]} &
\colhead{} &
\colhead{[mag]} & \colhead{[mag]} & \colhead{[mag]} & \colhead{Type} & \colhead{}  &     \colhead{}   
}
\startdata
AP Sgr &  Aa,Ab &  &  12:  & SB: &  & & & $\leq$A5 & 5,2,4  &  \\
 &  AB  & 6.3  & 5320  & & 17.19 & 1.50 & +7.56 & K7 & 3  & \\
Y Car &  Aa,Ab   & 0.0025 &    3.6 & SB &  &    & &     B9.0 V & 6,7,4  &   WC  \\
  & AB   &   2.6 &  3820 & & 16.63 & 0.84 & +5.80  & K0 & 3 & \\   
   or  &  AB &  3.2 &  4700 &  & 17.12 & 1.20  & +6.29 & K5 & 3 & \\     
RV Sco  &   Aa,Ab & & 14:  & SB & & & & $<$A3  & 4,2    & \\
  &  AB & 6.0  & 4520   &  & 11.50 & 0.24 & +2.12  & A7 V & 3  & \\
  or  &  AB  & 3.6  & 2710  &  & 14.82 & 0.95 & +5.44  &  K2 V & 3  & \\
V350 Sgr &  Aa,Ab  & 0.0030  & 2.7 &  SB &   & &         & B9.0 V & 8,9 & WB, WB \\
      &  AB &  3.1 & 2780 &  & 16.80 & 1.40 & +7.04  & K6 & 3 &  \\  
BB Sgr &  AB  & 3.3  & 2740  &  & 17.11 & 1.50 & +7.51 & K7 & 3,4,5  & \\
FF Aql    &  Aa,Ab  & 0.0082  & 4.5  & SB &  & & &  K0   & 1,4  & \\
  &  AB    & 6.9  & 2520  &  & 10.46  & 0.60 & +2.65 &  G0 V & 3  & \\
\enddata
\tablenotetext{a}{Spectroscopic Binary Companion}
\tablenotetext{b}{References:
1. Gallenne et al.\ 2019; 2. Evans 1992a; 
3. Table~\ref{com.sum}; 4. Kervella et al.\ 2019a; 
5. Szabados 1989; 6. Petterson et al.\ 2004; 7. Evans et al.\
2005; 8. Evans et al.\ 2018; 9. Kervella et al.\ 2019b.}
\tablenotetext{c}{Wide companion types: WB gravitationally bound; WC comoving.}
\end{deluxetable}
}

 The possible companions with separations larger than $2''$ are more 
likely to contain a line-of-sight coincidence with an unrelated 
field star than the inner group.  However, of the six Cepheids 
for which a possible companion has been identified in this 
group, five have strong evidence of an inner spectroscopic 
binary, and only for one (BB Sgr) is an inner binary questionable
and it is a cluster member. That is, the resolved wide systems share
this characteristic of the resolved close companions. This is an additional
property consistent with physical system membership.

\subsection{Distribution of Colors}

Ultimately we want to examine the properties of possible companions
as a function of the (projected) separation in AU, which is shown in Figure~\ref{sep}.   
Companion candidate properties are taken from Table~\ref{com.mix}  when available 
or from Table~\ref{com.sum}.

Companion candidates closer than 2000 AU span the full range of $(V-I)_0$.  
The hottest companion candidates, however, do not occur at larger separations. 
The hottest of the more distant companions, FF Aql and RV Sco [with $(V-I)_0$ colors 
corresponding approximately to G0 and A7 dwarfs, respectively]
have  evidence 
that they are associated with the Cepheid (Appendix~\ref{sys.5}).
The other companion candidates with  separations larger than 2000 AU are all cooler
 (approximately K0 or later).  Thus there appears to be a trend in the color or 
temperature and mass of the companion candidates to less-massive stars at wider 
separations.    This raises the question whether wider companions 
actually are preferentially less massive (or have been dynamically ``widened''),
or whether they are chance alignments with field stars or possibly a mixture of both.

One of the goals of this study is to determine how far from the Cepheid 
 physical companions  occur. The diminishing number with separations greater
then 6000 AU is consistent with the results of X-ray studies to identify 
active stars young enough to be Cepheid companions.
  X-ray observations of 14 of the 39 
Cepheids with companion candidates with separations greater than 2$\arcsec$
in Paper II were discussed in Paper IV. The only two Cepheids with X-ray confirmed companions 
greater then 6000 AU were R Cru (above) and S Nor.  However, since S Nor 
is in a cluster the X-ray source may be a cluster member rather than a
gravitationally bound companion.  Possible wide but bound companions revealed by {\it Gaia}
are discussed in Section~\ref{gaia.comp}.  Seven Cepheids in the WFC3 survey 
($\delta$ Cep, AX Cir, AW Per, V350 Sgr, V659 Cen, 
BP Cir and V950 Sco) have possible companions (Tables~\ref{gcomp} and \ref{gcomp2})
as does the cluster Cepheid  U Sgr. This hints at  further extended low-density 
companions. 

\begin{figure}
\centering
 \includegraphics[width=5in]{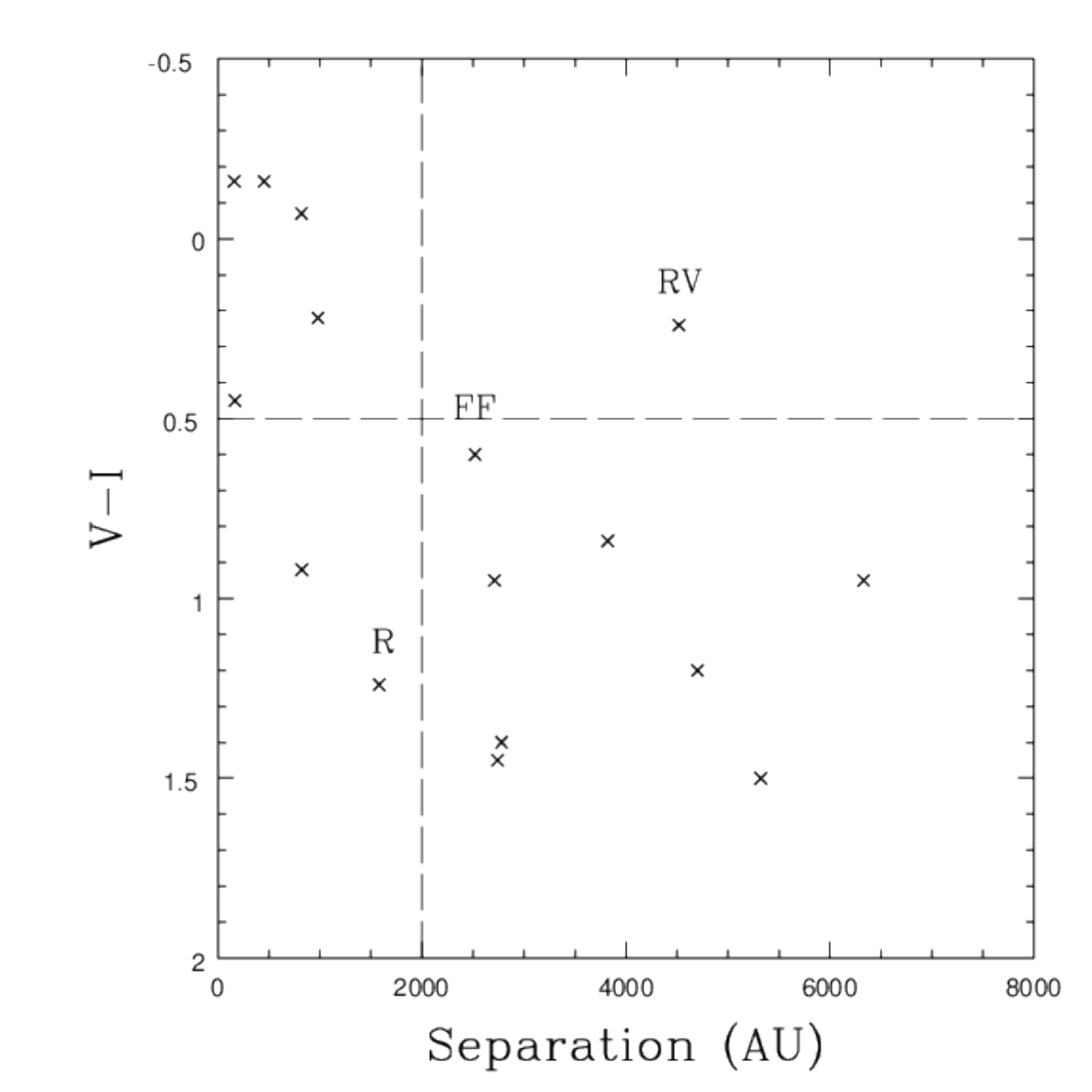}
\caption{Colors and  separations of the resolved close companions 
from the Cepheids.  
$(V-I)_0$ is in mag.  The horizontal and vertical dashed lines 
are arbitrary divisions of the companion properties, for 
discussion purposes.  A component of the R Cru, RV Sco, and FF Aql 
systems is indicated.
 \label{sep}}
\end{figure}

\subsection{Multiplicity}\label{Multiplicity}

The group of stars with resolved close companions 
 contains   companions which are highly likely to be  physically 
related to the Cepheid.
As discussed above, the systems are complex, and it requires 
a number of approaches in addition to the {\it HST} WFC3 observations 
to build up a picture of the masses and 
separations of the components.  Table~\ref{mcomp} summarizes the 
information on the members of each system.   
 In addition, as discussed in Section~\ref{gaia.comp}, Kervella 
et al.\ (2019b) used {\it Gaia} data to identify wide companions 
from distances and proper motions   and test whether they 
are likely to be gravitationally bound to the Cepheid.  The final right-hand
column in Table~\ref{mcomp}
indicates whether a wide bound (WB) or wide comoving (WC) 
companion candidate was found.  

Figure~\ref{comp.mult}
 further summarizes the components schematically.  We stress that it 
was developed from the list of Cepheids with resolved companions at
separations of typically a few hundred AU (Table~\ref{mcomp}).  
It is striking that in the Cepheid list
compiled in this way, {\bf all} have evidence of an inner binary, 
often with a known orbit.
As noted above, because S Nor is in a cluster it may have a 
different history of formation or interaction than the field stars
and also possible chance alignments with cluster stars. 
 Precise masses are not available for all the companions; however  
limits or estimates can often be made from spectral types (Table~\ref{mcomp}).  
Figure~\ref{comp.mult} shows 
the masses schematically either as comparable to the Cepheid 
itself  (a B  star: large blue squares) or a lower-mass  A, F or G star (small red squares).  
While this is a crude division, several 
things are apparent in Figure~\ref{comp.mult}.  There is a variety 
in the component masses.  For instance, R Cru and U Vul have only lower-mass 
companions, whereas other systems have a mixture of masses.  For AX Cir 
and V659 Cen, the 
most massive companion is not the closest companion (Section~\ref{sect:sys.clos}).
 The more massive 
companions are typically the closer ones, although AX Cir and V659 Cen are  exceptions.  
This might be 
due to the formation process or subsequent dynamical evolution. 


 It is surprising that all the systems with  resolved close 
companions in Table~\ref{mcomp} also have an inner binary system,
or at least a suspected one.
This is markedly different from the fraction of spectroscopic binaries 
among  Cepheids with 
periods less than 20 years, which is 29\% $\pm$ 8\% (Evans et al.\ 2015),
less than a third of a Cepheid sample.  
In a scenario where systems with wider separations are formed before
more closely separated components (in a process such as core fragmentation
and subsequent disk fragmentation) it is hard to explain how wide 
components ``know'' an inner binary will be formed later.  A more
probable explanation for the {\it requirement\/} that a wide component in 
Figure~\ref{comp.mult} has an inner binary is that the wide component of 
the triple (or higher) system was moved outwards through dynamical 
interactions.




\begin{figure}
\centering
 \includegraphics[width=5in]{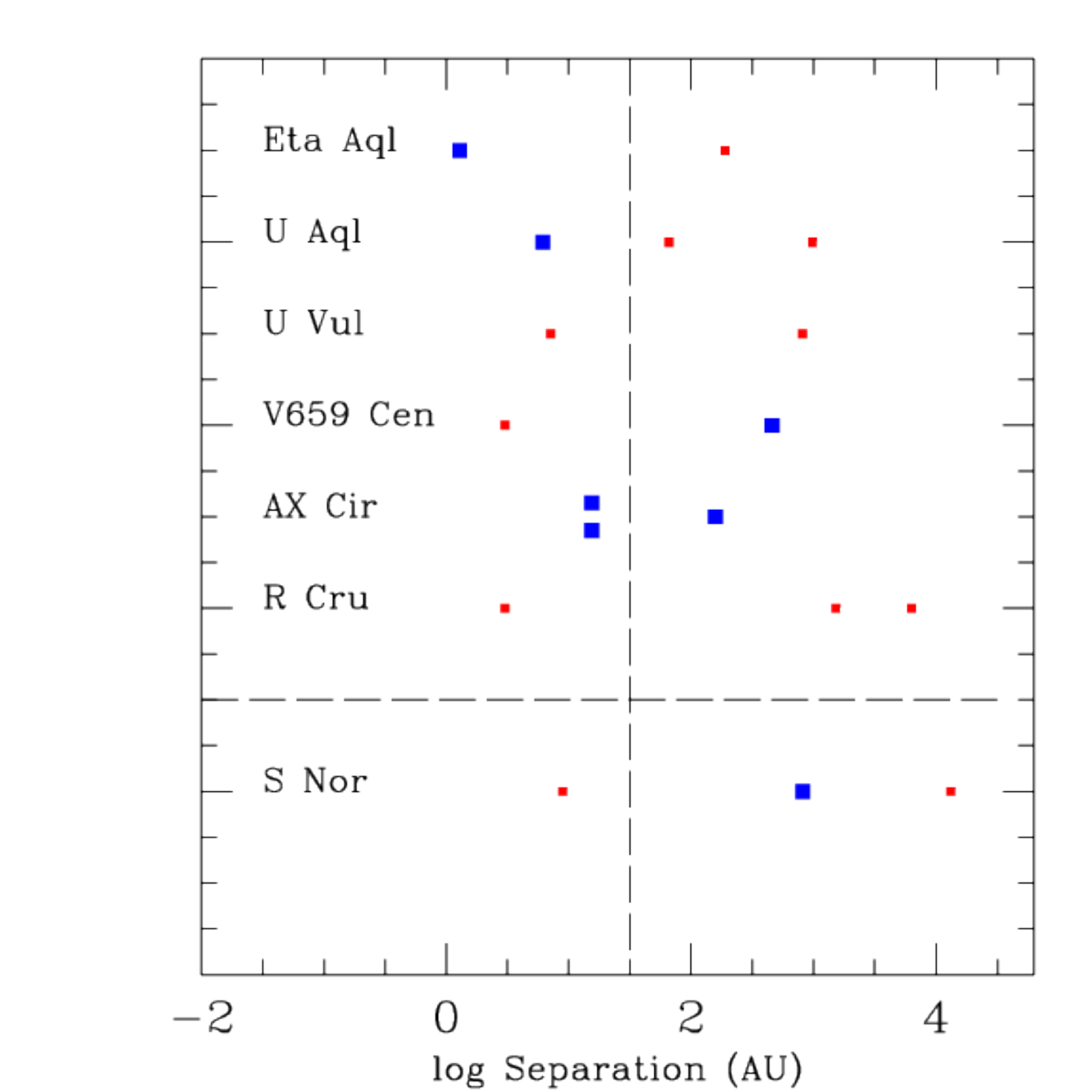} 
\caption{Schematic of the separations of companions
from the Cepheids for resolved close companions (within $2''$)
from Table~\ref{mcomp}. 
The vertical dashed line 
separates the spectroscopic companions (to the left of the line) 
from the resolved companions  (to the right). 
Blue (large) 
and red (small)
symbols represent companions earlier and later than A0 respectively.
S Nor (below the horizontal dashed line) is a cluster member, and more 
difficult to interpret.
 \label{comp.mult}}
\end{figure}

 The  resolved close companions  in Figure~\ref{comp.mult} all 
appear to be  distant third stars in triple systems.  Because they 
are the closest resolved companions in the survey, 
they are the most likely to be gravitationally bound 
to the Cepheids.  However, the striking result in Figure~\ref{comp.mult}
of inner spectroscopic binaries
encouraged us to check the slightly more distant 
resolved wide companions (from 
$2''$ to $7''$).  System parameters are summarized in Table~\ref{mcompw}, 
and represented schematically in Figure~\ref{comp.mult.w}.  The system
summaries in Figure~\ref{comp.mult.w} have been simplified in two
ways.  For two stars from Table~\ref{mcompw} (RV Sco and Y Car), there are 
two possible companions in this range.  Since it is dynamically 
unlikely that they are both system members because they are 
at very similar apparent separations,  Figure~\ref{comp.mult.w}
shows only the one considered to be the most probable. 
Second, wider companions discussed below are not shown.  From the 
discussion of the Cepheids with resolved wide companions 
above (Table~\ref{mcompw}) all except BB Sgr 
have good evidence of an inner spectroscopic binary system.
In addition, as with S Nor, BB Sgr is a cluster
member. This provides an alternate possibility for the resolved wide companion, 
a chance alignment with a cluster member, rather than a bound system member.   
It is also notable that the outer companion is significantly 
less massive than the Cepheid.  This again fits the picture that a resolved 
companion with a projected separation of several thousand AU 
was moved outward through dynamical interactions within the triple 
system.

\begin{figure}
\centering
 \includegraphics[width=5in]{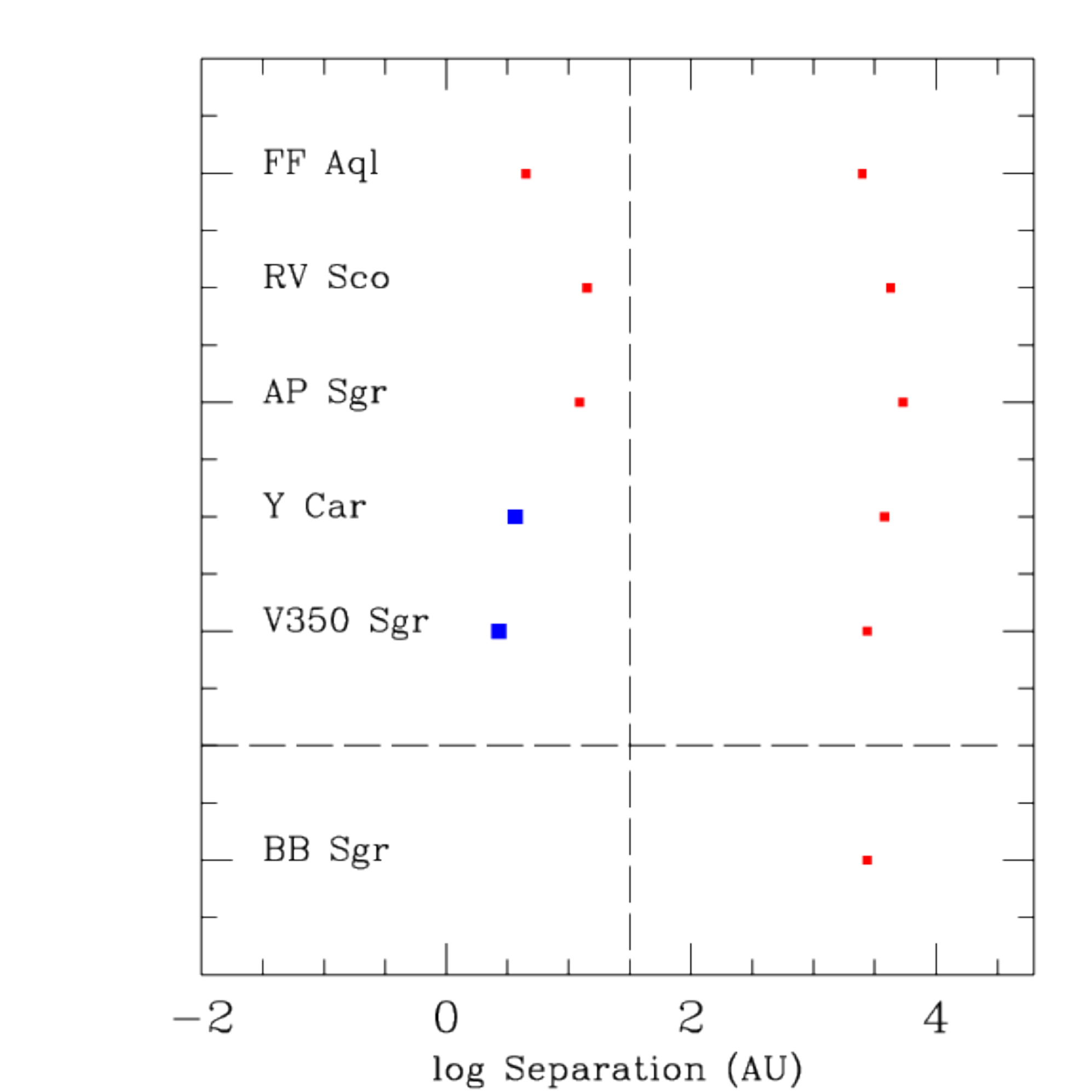} 
\caption{Schematic of the separations of companions
for resolved wide companions (wider than $2''$). Symbols are the same as for 
Figure~\ref{comp.mult}.  BB Sgr (below the horizontal dashed line)
is a cluster member.
 \label{comp.mult.w}}
\end{figure}

\subsection{Dynamical Stability}

The inner spectroscopic binaries for the 13 Cepheids in Tables~\ref{mcomp} and 
\ref{mcompw} certainly have long-term dynamical stability with respect to the 
resolved companions from the WFC3 survey because of the very different orbital  
periods. However, two stars in Table~\ref{mcompw}
(RV Sco and Y Car) have two companions which  have  separations from the
Cepheid which correspond to period ratios much less than 5, found by 
 Tokovinin (2018b) in stable triple systems. 
Anticipating the {\it Gaia} results for likely wide but bound companions, V350 Sgr 
may also have two similar companions.  While future studies may rule out one of these
possible companions, and hence remove the question of stability, the fact that  
half  the stars in Table~\ref{mcompw} 
are in this situation may reflect something about star 
formation.  One way to have a pair of companions at apparently similar separations but 
in a stable configuration is if they are not coplanar, but 
have very different inclinations.  In summary, RV Sco, Y Car, and V350 Sgr 
may provide a clue to additional complexity in the formation processes.   

 The population of triple systems can have important effects on the evolution of the systems, such
as in the Kozai-Lidov mechanism (Naoz 2016) as discussed in  
 Section~\ref{implic} 

\subsection{Detection Probability} 

The following example provides an indication of the completeness of the survey. 
For a typical distance in the WFC3 survey (700 pc) and a typical reddening [$E(B-V) = 0.20$; 
Paper~II], we  estimate  apparent magnitudes for a range of spectral 
types in Table~\ref{det.prob}; the $M_V$ values are  from Drilling \& Landolt (2000).
Using these magnitudes, 
 the detection fraction from Figure~\ref{fig:detectlim} shows that 
nearly all the B9 and A3 companions would 
be detected outside $1''$.   Later spectral types (K0) 
are severely missing inside $1\farcs6$. Thus, for the 
example of 700 pc, essentially all early-type companions 
would be detected at separations larger than 
700~AU\null.  For lower-mass companions, as represented by the 
K0 star, they would be largely detected only 
outside 1100~AU.   



\begin{deluxetable}{lcc}
\tablecaption{Typical Magnitude for  Probable Detection \label{det.prob}}
\tablewidth{0pt}
\tablehead{
   \colhead{Spectral}   & \colhead {$M_V$}   & \colhead {$V$ at 700 pc\tablenotemark{a}}  \\
   \colhead{Type}   & \colhead {[mag]}   & \colhead {[mag]}  
}
\startdata
B9 V &    +0.2 &     10.1 \\
A3 V &     +1.5  &    11.4 \\ 
K0 V &     +5.9  &    15.8  \\
\enddata
\tablenotetext{a}{Assuming $E(B-V)=0.2$}
\end{deluxetable}

\subsection{Companions Identified by 
{\it Gaia}} \label{gaia.comp}


The {\it Gaia\/} spacecraft has provided new data to investigate 
multiple systems, particularly to identify
wide components from 
the outer regions of a star-forming cloud. This 
is also relevant to the question of whether there are stars
formed at the same time which are neither
 gravitationally bound system members nor
 in a recognizable star cluster.  In particular, studies have 
been done of proper motions to study orbits (Kervella et al.\ 2019a) and 
to identify wide companions (Kervella et al.\ 2019b)

Kervella et al.\ (2019a)
have used the difference between proper motions from {\it Gaia\/}
and {\it Hipparcos\/} to look for ``proper-motion anomalies'' resulting
from orbital motion between the two epochs.  Of the 70 stars in our 
WFC3 sample, they list data for 63, or 90\%.  (Of these 16, or 25\%, 
had less-than-perfect {\it Gaia\/} data, often because the Cepheid is too
bright.) The  {\it Gaia\/} data showed proper-motion anomalies for 
28 stars (44\%) at the level of  2$\sigma$ or greater, many of which are 
known spectroscopic binaries. We stress that this is only a first 
exploration of the {\it Gaia\/} results, which are expected to improve
in the future.  Specifically, other stars have indications of 
proper-motion anomalies, but at a lower significance level.  As 
an example, S Mus is a well-known binary with an orbit, but does not 
meet the proper-motion-anomaly criterion.   


\subsubsection{Proper Motion Anomaly} \label{pma}

{\bf Proper Motion Anomaly plus Orbit}

 Comparison of stars in the WFC3 survey and the {\it Gaia\/} 
  proper-motion study is presented  in   
 Table~\ref{gcomp}, 
 based on Tables A1, A2, and A3 in Kervella et al.\ (2019a), which
 lists the Cepheids which have 
proper-motion anomalies and orbits. 
The columns in Table~\ref{gcomp}  contain:
2 and 3: 
the semimajor axis in mas and AU; 4: whether there is a resolved companion from
 Table~\ref{com.sum} (Y); 5: whether a gravitationally bound (B) 
companion or a wide (W) comoving companion has been identified (see below); 6 and 7: the 
separation for a gravitationally bound companion in AU and mas;
8: the spectral type of the companion;  9 and 10: the separation for a comoving companion
(W), and its spectral type; and 11: whether the star is a cluster member (C) from 
Anderson  (2013).  The semimajor axes are from Kervella et al.\ (2019a; Table 2). Because
of the additional information from {\it Gaia}, particularly the individual parallaxes 
(which will be updated in DR3), they differ somewhat from the values in 
Table~\ref{mcomp}.
Approximate spectral types have been inferred from 
the temperatures using the calibration of Pecaut \& Mamajek (2013).  (The symbol 
$<$ M means cooler than M spectral types.) Two stars appear in Table~\ref{com.sum}
but not in either Tables~\ref{gcomp} or \ref{gcomp2} ($\eta$ Aql
and BB Sgr).  For $\eta$ Aql  the {\it Gaia} quality flags are poor, probably because 
it is very bright, so the information on a proper-motion anomaly is also poor.
  BB Sgr has no proper-motion anomaly and no wide companions.   

When a spectroscopic orbit is known, the proper motions allow the  orbit 
to be fully determined, including the inclination.  By assuming a mass for the Cepheid, 
a mass for the secondary can be determined.  In 
 Table~\ref{gcomp} we list the semimajor
axes in AU and mas for stars with orbits 
(from Kervella et al.\ 2019a Table 2). 
Even when 
spectroscopic orbits are available, those shorter than 1000 days  
 are considered preliminary (marked with :).  

{\bf Proper Motion Anomaly without Orbit}

 In addition to the stars in Table~\ref{gcomp} a number of stars were found to have 
 a proper-motion anomaly which do not have an orbit: AV Cir, T Cru, X Cyg, $\beta$ Dor
R Mus, AP Pup, RV Sco, SZ Tau, LR TrA, AH Vel, and T Vul.  However, for a number of these
stars, the {\it Gaia} parameters are imperfect (X Cyg, $\beta$ Dor, AP Pup, and T Vul).
  
It is likely that many of these have longer-period orbits
than stars for which orbits are known. This means they are likely to sample the 
orbital period range of the WFC3  stars  with
resolved close  companions 
(Table~\ref{tab:prop.comp}), making this range accessible by a second technique.   
 Further information on these stars, 
for instance from subsequent {\it Gaia} releases, is expected to provide  
new results for several sources, partly because the errors on the separations 
are large.  For 
instance, X Cyg, SZ Tau, and T Vul (which have proper-motion anomalies) 
have accurate radial-velocity data 
which does not indicate orbital motion for periods less than 20 years (Evans et al.\
2015).  On the other hand, T Vul has a hot companion, which must be in a wider 
binary orbit (Evans et al.\ 2013).  T Cru and AH Vel are found by Bersier (2002) to be 
binaries.  However,  Gallenne et al.\ (2019) had
difficulty identifying an orbital period from new velocity data.  
LR TrA shows orbital motion (Szabados et al.\ 2013).
There is also some evidence  of orbital velocity variation for R Mus, AP Pup,
and RV Sco, as discussed by Szabados (1989).  All the stars in this section (except AV Cir 
and LR Tra) have been observed with {\it IUE\/} to search for hot companions.
Only T Vul was found to have such a companion.  The other stars had upper limits 
on spectral types in the early A range, corresponding to companions less 
massive than 2.9 to 2.0~$M_\odot$.
The whole  group warrants further velocity 
observation; however, orbital periods of more than 20 years are difficult to determine
 because they require a long time span of very accurate velocities.   

There is a striking difference between system multiplicity of the proper motion anomaly 
stars with and without orbits.  In Table~\ref{gcomp}, the stars with orbits frequently 
have resolved companions from the WFC3 survey.  In the group without orbits, only RV Sco 
has a WFC3 resolved companion.  Furthrmore, stars with orbits frequently have a {\it Gaia}
companion either bound or comoving.  None of the stars without orbits have these companions.
Only X Cyg is identified as a cluster member by Anderson (2013).    

{\bf No Proper Motion Anomaly}

In addition to the stars with a proper motion anomaly 
  (Table~\ref{gcomp} with orbits, and those without a known orbit), 
a number of stars in the WFC3 sample showed
no proper-motion anomaly in the {\it Gaia\/} analysis. 
In the case of $\eta$ Aql, RT Aur, {\it l} Car, V659 Cen, $\zeta$ Gem, T Mon, 
S Mus, W Sgr,  Y Sgr, and X Vul the {\it Gaia} astrometry in DR2 has large uncertainties, often 
because the stars are  bright.  The following stars have good astrometry solutions, but no 
proper-motion anomalies: TT Aql, V636 Cas, V Cen, V737 Cen, IR Cep, S Cru, BG Cru, DT Cyg, 
W Gem, BF Oph, V440 Per, RS Pup, MY Pup, BB Sgr, V482 Sco, S TrA, V Vel,  and
 SV Vul.


\subsubsection{ Wide Bound and Comoving Companions} \label{wbc}

In addition to the proper-motion anomalies, Kervella et al.\ (2019b) 
have used {\it Gaia\/} data to identify more widely separated companions.  
Candidates are selected because the companions and Cepheids have 
 similar parallaxes, small  differential tangential 
velocities, and small projected linear separation. (The list is drawn from
Table 1 in Kervella et al.\ 2019b for companions rated likely on the 
basis of visual inspection.)
They further divide 
the candidates into gravitationally bound companions and unbound but 
comoving companions  based on differential tangential velocities.  

 These companions are added to the entries
in  Table~\ref{gcomp} if the Cepheids also have
proper-motion anomalies, or are listed in Table~\ref{gcomp2}
if they have no proper-motion anomalies. 
The top three entries in  Table~\ref{gcomp2}
(V659 Cen, R Cru, and Y Sgr) have imperfect {\it Gaia\/} quality 
parameters, so their status might change in the DR3 release.
In Table~\ref{gcomp2}, V659 Cen and 
R Cru have spectroscopic-binary companions as previously discussed. 
BP Cir also shows orbital motion (Anderson et al.\ 2020, in preparation).
 As can be seen in Table~\ref{gcomp}, 
 few of the Cepheids with proper-motion anomalies have {\it wide 
 comoving} companions.  This is also true of the whole sample 
in Kervella et al.\ (2019b), where there are also distinctly fewer comoving 
companions than bound companions. (These wide companions would not, in general,
have been in the field of our WFC3 survey.)

  The comoving wide companions provide a first 
opportunity to consider companions formed at the same location as the intermediate-mass Cepheids, but independent of the binary-star formation process.  This is 
a part of star formation that we have been unable to explore previously because
of the difficulty of recognizing a single star as compared with a gravitationally
bound companion or a cluster.  It is possible that some of the     gravitationally
bound candidates may ultimately be reassigned  
to the comoving group since the present analysis does not 
include radial-velocity information.


The occurrence of wide but gravitationally bound companions in  Tables~\ref{gcomp}
and \ref{gcomp2} varies 
markedly between the groups.  In Table~\ref{gcomp}
 (proper-motion anomaly with 
an orbit), 4 out of 17 systems (23\%) have wide bound 
companions (B).  The companions
in this group have a range of spectral types from B through M, similar 
to the companion spectral types in Figure~\ref{cmd}.  Thus the companions 
do not appear to be dominated by field-star ``impostors.'' 
In Table~\ref{gcomp2}  (no proper-motion anomaly but a wide
companion or comoving star), 5 systems have a gravitationally bound 
companion.  This makes a total of 9 systems with a wide companion, or 
13\% of the whole WFC3 sample.  The fact that the fraction of wide gravitationally
bound companions is smaller in the whole WFC3 sample than in the subsample with 
proper-motion anomalies and orbits (Table~\ref{gcomp}) could be due to a 
different density in the star-forming environment.

For the comoving candidate companions, there is no particular reason to 
relate their occurrence to the details of the multiple systems.
The comoving candidates, however,
are heavily dominated by K and M dwarfs.  This is similar to the 
color distribution of companions at more than 2000 AU in Figure~\ref{sep}, 
indicating a larger fraction of ``impostor'' field stars, a 
change in properties of companions produced in star formation, 
or a dynamical evolution in multiple systems pushing smaller 
members outward.

We summarize the results for the Cepheids in the WFC3 study with 
resolved close and resolved wide companions 
in Table~\ref{res.sum}. The Tables from which the separations
are taken is indicated.  
Three stars (Y Car, RV Sco, and V350 Sgr) have two  possible companions 
with roughly the same separation from the Cepheids, making it dynamically unlikely that
both are physical companions. We have marked one of the companions as an alternate 
(alt). 
As noted in Table~\ref{mcomp} and 
Figure~\ref{comp.mult}  stars with a  close resolved companion
 also have an inner spectroscopic binary.  It is also notable 
that three of the seven (or six, omitting S Nor) have wider gravitationally bound candidates, 
or 43 or 50\%, much larger than the Cepheid sample as a whole (9 systems or 13\%).
Of the six resolved wide companions sample in Table~\ref{res.sum} only one has a 
{\it Gaia\/} bound companion.

\begin{deluxetable}{lrrrrrrrrrr}
\footnotesize
\tablecaption{System Information from {\it Gaia}: PM Anomaly  \qquad\qquad\qquad\qquad \label{gcomp}}
\tablewidth{0pt}
\tablehead{
\colhead{}
  & \colhead{Sep }   & \colhead {Sep } &  \colhead {Res} & \colhead {Wide} &
\colhead{B Sep }   & \colhead{B Sep }  & \colhead {SpTy}
& \colhead{W Sep  }   & \colhead {SpTy}  & \colhead {Clust}**   \\
   \colhead{} & \colhead{mas} &  \colhead{AU} & \colhead{Comp} & \colhead{Comp}
& \colhead{AU} & \colhead{mas} 
&     \colhead{}   &\colhead{AU}
& \colhead {} & \colhead {} 
}
\startdata
{\bf PM Anomaly: Orbits}  & &   &  & & & & & & & \\
U Aql     &          5.64 &   20.47  & Y & &   &   &   &   & & \\
FF Aql    &          4.47  &  8.22 & Y & & & & &  & & \\
V496 Aql  &          3.83  &  3.72  & & & & & & &  & \\
RX Cam    &          4.21  &  3.41 & & & & & & &  & \\
Y Car     &        3.73: & 2.57: & Y &  W  & & & & 32300 & K6 V & \\
SU Cas    &         1.65: & 3.54: & & & & & & & &  \\
$\delta$ Cep   &         5.85 & 19.86 &  & B &    10800 & 40740 &  B7-8 III-IV & & & \\
 & &   &  & & & & +F0 V & & &  \\
AX Cir    &       14.66  &  26.00  & Y &  B, W & 42500 &   81490  &   M3.5 V & 23000 & M3 V & \\
SU Cyg    &       2.73: &  3.27: &  & W   &       &      &   & 34400 & K6 V & C \\
V1334 Cyg &        6.18  &   7.29  & & W &    &     &    & 43700 &  $<$ M & \\
S Nor     &       8.87   &  9.67  & Y & W & & & & 41100 & K3 V & C \\
 & &   &  & W & & & & 38800 & -- & \\ 
S Sge     &        3.08: &  2.07: &  & & & & &  & & \\
X Sgr     &        2.40:  & 8.32: & & & & & &  & & \\
AW Per    &        27.22  &   29.15 & &  B  &  8400  &   10270  &   K3.5 V  & & &  \\          
V350 Sgr  &        5.16    &  5.24 & Y &  B   &  26200  &  29850  &   A2 V & & & \\
  & &   &  & B &                                  30600  & 34920   &   K4.5 V & & &  \\
V636 Sco  &         4.59   &   4.61 & & & & & & & &  \\
U Vul     &         7.15   &  7.74 & Y & & & & &  & &  \\
 \enddata 
B Sep: Separation for gravitationally bound companions

W Sep: Separation for Comoving Companions


** From Anderson  
\end{deluxetable}

\begin{deluxetable}{lrrrrrrrr}
\footnotesize
\tablecaption{System Information from {\it Gaia}: Bound and Comoving 
Companion Candidates  \qquad\qquad\qquad\qquad \label{gcomp2}}
\tablewidth{0pt}
\tablehead{
\colhead{}
   & \colhead {Res} &  \colhead {Wide} &
\colhead{B Sep }   & \colhead{B Sep }  & \colhead {SpTy}
& \colhead{W Sep  }   & \colhead {SpTy}   & \colhead {Clust**} \\
   \colhead{}  & \colhead{Comp} & \colhead{Comp}
& \colhead{AU} & \colhead{mas} 
&     \colhead{}   &\colhead{AU}
& \colhead {} & \colhead {} 
}
\startdata
{\bf No PM Anomaly}   &  &  & & & & & &  \\   
V659 Cen*   & Y & B &  48300 & 62220 &   M3 V  & & & \\ 
R Cru*      & Y & B &  6600 &  7700  &   G8 V   &   & &  \\    
Y Sgr*      & &   W & & &  & 33300 & $<$ M & C \\
           & &   W & & &  & 44300 & $--$ & \\
           & &    & & &  &  &  & \\
BP Cir     &  &  B &  39000 &  66300 &    M2 V  & 48700 & M9 V & \\
     &  & & & & &  33500 & M3 V & \\   
     &  & & & & & 29100 & M2 V & \\   
Y Oph      &  &  W &  &  &  & 44000  &  K4 V & \\
U Sgr      &  & B &  43100 & 71990  &  A0 IV-V  &  40000    &  M0 V & C \\
           &  & B &     46100 & 76960 &   A0 V  & 21200 &  K7 V & \\
V950 Sco   &  & B &  15000 & 16070 &  G1 V  &  15800 & G2 V & \\ 
   &  & & & & & 27300  & K3 V & \\ 
R TrA       & &  W &  & &  & 32600 & K1 V  & \\
\enddata 

B Sep: Separation for gravitationally bound companions

W Sep: Separation for Comoving Companions

* {\it Gaia} quality parameters imperfect

** From Anderson  
\end{deluxetable}

\begin{deluxetable}{lrrrrrrrr}
\footnotesize
\tablecaption{Summary: Separations for Systems with Resolved Companions  \qquad\qquad\qquad\qquad \label{res.sum}}
\tablewidth{0pt}
\tablehead{
\colhead{}
   & \colhead {Sep }    & \colhead {Sep } & \colhead {B Sep} 
   & \colhead {SpTy}
& \colhead{W Sep }   & \colhead {SpTy} & \colhead {Other}  & \colhead {Clust**}   \\
   \colhead{} & \colhead{SB} &  \colhead{Res} & \colhead{} 
& \colhead{} & \colhead{} 
& \colhead {} & \colhead {}  & \colhead {}  \\
   \colhead{} & \colhead{AU} &  \colhead{AU} & \colhead{AU} 
& \colhead{} & \colhead{AU} 
& \colhead {} & \colhead {AU}  & \colhead {}  
}
\startdata
%
Close  &   &   &  & & &  &  \\
    & Table~\ref{mcomp} & Table~\ref{mcomp} & Table~\ref{gcomp} / ~\ref{gcomp2}
&  & Table~\ref{gcomp} /~\ref{gcomp2}  &  & Table~\ref{mcomp} &  \\
R Cru    & 3: &   1580    &  6600 &     G8 V   &  & & & \\   
U Aql     &   6.2  & 981 &    &   &   &  & 66 & \\ 
U Vul     &      7.1 & 822  & &  &  & & & \\
S Nor     &   8.87  &  819 & & & 41100 & K3 V & 13300 &  C \\
 & &   &  & & 38800 & -- & & \\ 
V659 Cen  & 3: & 452     &  48300 &    M3 V  & & & & \\ 
$\eta$ Aql  & 3.9: &  191  & & & & & & \\ 
AX Cir    &   15.4 & 158  & 42500 &    M3.5 V & 23000 &  M3 V & & \\
  & 15.4  &   &  & & &  &  \\
 & &   &  & &  &  & \\ 
Wide  &   &   &  & & &  &  & \\
  & Table~\ref{mcompw} & Table~\ref{mcompw} & Table~\ref{gcomp} / ~\ref{gcomp2} 
&  & Table~\ref{gcomp} / ~\ref{gcomp2}   & {} & Table~\ref{mcompw}  & \\
AP Sgr    & 12:      & 5320  & &  &  & & & \\
Y Car    &  3.6  & 3820  & &  & 32300 & K6 V  & & \\
         &       & alt 4700   & &  &  & & & \\
RV Sco    & 14:    & 4520 &    &   &   &  & & \\ 
          &   & alt 2710   & &  &  & & & \\
V350 Sgr  & 2.7  & 2780   & 26200 & A2 V & & & & \\  
         &       &    & alt 30600  & K4.5 V &  & & & \\
BB Sgr  &       & 2740  &  &  &  & & & C \\
FF Aql     &  4.5   & 2520 &    &   &   &  & & \\
\enddata

B Sep: Separation for gravitationally bound companions

W Sep: Separation for Comoving Companions

alt: alternate: not both dynamically stable

** from Anderson
\end{deluxetable}

\subsection{Multiplicity: Implications}
\label{implic}

Triple stars can form via two modes. First is an ``outside-in'' process, 
whereby core fragmentation on large scales initially produces the wide 
companions with $a_{\rm out}> 100$~AU, followed by disk fragmentation on smaller 
scales that forms the close inner binaries with $a_{\rm in}< 100$~AU (Moe \& 
Di Stefano 2017; Tokovinin 2017). Second is an ``inside-out'' process, where 
two companions fragment within the same disk near 
$a_{\rm in} \simeq a_{\rm out}\simeq 10$~AU in 
an initially unstable configuration, and then subsequent dynamical interactions 
throw one of the components (typically the least-massive) to larger separations 
or eject it entirely (Reipurth \& Mikkola 2012;  Moe \& Kratter 2018). 
 The Kozai-Lidov mechanism (Naoz 2016) also results in dynamical evolution of the 
system.

As discussed in Section~\ref{Multiplicity}, all companions to Cepheids 
with separations from 
 100 to 2,000 AU, and nearly all companions with separations between
 2,000 and 7,000 AU, 
are outer tertiaries to inner spectroscopic-binary companions with separations from 1 to 20 AU\null. The strong link between outer tertiaries and inner 
binary companions to Cepheids suggests a causal link, e.g., the 
dynamical-unfolding scenario. Nearly all of the outer tertiary companions 
are less massive than the inner spectroscopic binary companions, consistent 
with the triple-star dynamical-unfolding scenario. In this dynamical-unfolding scenario, the outer tertiaries are expected to be highly 
eccentric, which may be tested in the future with \Gaia\/ astrometry. 

Nonetheless, correlation does not necessarily imply causation. The 
binary fraction of O-type (Sana et al.\ 2014) and early-B (Moe \&
Di Stefano 2017) stars is nearly 100\% within $a < 100$~AU, and therefore 
nearly all wide companions beyond $a > 100$~AU to massive stars are outer 
tertiaries in hierarchical triples. Similarly, Evans et al.\ (2005) showed 
that triples are quite common among Cepheids, and  again Evans et al.\ (2015) found 
that 30\% of Cepheids have companions for $a$ between 1 and 20 AU, probably 
because Cepheids began life as {\it late\/} B stars.  In addition B stars in 
short-period orbits will have merged before the Cepheid stage, removing them 
from the binary statistics.   
The preponderance of Cepheids in hierarchical 
triples with $a_{\rm in}$ between 1 and 20 AU and $a_{\rm out}$ between 100 and 7000 AU 
could simply be 
due to the efficient formation of companions to massive stars via both disk 
fragmentation and core fragmentation on small and large scales, respectively. 
On the other hand, 
 the estimates in Tables~\ref{mcomp} and \ref{mcompw} show that the 
spectroscopic companions are within 20 AU, the regime well studied in 
Evans et al.\ (2015).
The binary fraction from that study is 30\%. Thus, in Table~\ref{mcomp}, 
of the 7 Cepheids with
companions 2 would be expected to have spectroscopic-binary companions, rather than 
all 7.  Similarly, in Table~\ref{mcompw}, 2 of the 7 stars would be expected to 
have companions in this range rather than 6. In sum, 13 of these 14 
stars have spectroscopic
companions within 20 AU, where 4 are expected, strongly implying that the system 
structure (triples) is not produced by the standard ``outside-in'' process.

      Wide companions   that derive from core 
fragmentation have systematically lower masses than inner binary companions 
that come from disk fragmentation (Moe \& Di Stefano 2017).
That is the case for Tables~\ref{mcomp} and \ref{mcompw}. 
Our wide tertiaries are 
systematically less massive than the inner binaries, but are still top-heavy 
compared to random pairings from the IMF.

Although a significant majority of {\it very\/} wide companions to solar-type stars are 
tertiaries, Tokovinin (2017) argued that most derive from fragmentation of adjacent 
cores, not dynamical unfolding, and therefore there is no causal link between 
inner binaries and outer tertiaries. This is related to the possibility that some 
Cepheids are the remnants of dispersed clusters (Anderson \& Riess 2018).
The identification of 
wide binaries formed from separate but adjacent cores derives from the fact that beyond 
about 10,000~AU, the number of companions increases strongly, 
beyond those presumably formed
by core or disk fragmentation.
 Insight here comes from the {\it Gaia\/} wide companions. 
Wider companions from {\it Gaia\/} (Tables~\ref{gcomp} and \ref{gcomp2})
from adjacent cores 
are expected to show some degree of correlation in component masses compared 
to independent cores (Tokovinin 2017). The frequency of these very wide companions 
rises steeply beyond about 10,000~AU\null.
  Tables~\ref{gcomp} and \ref{gcomp2} show that 16 Cepheids have either bound
or unbound {\it Gaia\/} companions,  very similar to  the 13 Cepheids 
with resolved companions in the WFC3 study. The most likely explanation is that the 
{\it Gaia\/} survey only extended to 50,000~AU, where core/disk fragmentation stars are
mixed with adjacent core stars. This may be an indication of the density of the 
formation environment (Deacon \& Kraus 2020).

In reality, both an ``outside-in'' formation via the relatively independent 
processes of core fragmentation followed by disk fragmentation and an 
``inside-out'' mode via dynamical unfolding likely contribute to our Cepheid 
triple sample. The Cepheid sample discussed here contributes to the
 sample of massive triples needed
 to fully compare the companion mass distributions 
to the two different models and determine which scenario is dominant. As 
indicated above, astrometric orbital solutions (or at least a sense of relative 
orbital motion) of the outer tertiaries will also help constrain their formation
 mechanism.

\section{Summary from This Series of Papers}

This paper  completes the discussion  of the series reporting results of our  {\it HST}/WFC3  
two-color search for resolved companions of 70 Galactic  Cepheids.  In the present paper  
we identified candidates stars lying at separations between $0\farcs5$ and $5\farcs$0, and 
present a list of the most probable companions.  Our major results from the series are 
summarized here. 


1.  Detection of companions of bright Cepheids lying within $2''$
requires sophisticated image processing. 
In our sample, seven Cepheids have 
companions within this range.  

2.  Discussion of the identifications,
 separations, and spectral types of the companions has been aided by 
{\it Chandra\/} observations of R Cru and S Mus, and also {\it HST}/STIS spectra
of U Aql and AX Cir. 
We provide a discussion of components for these systems (R Cru, $\eta$ Aql, 
S Nor, U Aql, U Vul, V659 Cen, and AX Cir).

3.  The $(V-I)_0$--$M_V$ CMD shows that
the companions      are fairly evenly distributed in mass (as 
opposed to being dominated by low-mass companions). However, this may 
be partly because low-mass companions are more difficult to discover 
within $1''$ of the Cepheid. In
contrast, companion candidates 
with wider separations are less massive than $1.5\,M_\odot$.  Cepheids with
companions within  $2''$  with separations of a few hundred AU 
comprise 10\% of those in the survey.

4.   It is particularly 
striking that companion candidates in the $0\farcs5$ to $2\farcs0$ range,
corresponding to separations less than 2000~AU, all have an inner binary 
system.  This is in contrast with the typical fraction of Cepheids in 
spectroscopic binaries of 29 $\pm$ 8\%
found in a velocity sample.
It is also clear that the resolved companion may be either 
more or less massive than the spectroscopic-binary companions.  

5.  Companions wider than $2''$ also have an inner spectroscopic binary for all or most
systems (Table~\ref{mcompw}  and Figure~\ref{comp.mult.w}).  

6. As discussed in Section~\ref{implic},   in a standard 
scenario,  it seems more 
likely that the resolved companion is formed first in a core-collapse 
process, followed by the formation of the spectroscopic binary through disk
fragmentation.  In this scenario it is unlikely that the formation of 
the outer component can ``anticipate'' the formation of an inner binary.
The most likely explanation that the resolved companion seems
to ``require'' a spectroscopic binary is that that combination results from 
dynamical evolution, which can, of course, occur in triple systems.  However, 
processes of ``outside-in'' formation, adjacent cores, and the dissolution of 
clusters may be involved in some cases.

7.  In order to put the Cepheid multiple systems in context, we have also used 
{\it Gaia\/} results  to incorporate proper-motion anomalies into 
orbits (Kervella et al.\ 2019a) and to identify wider companion candidates
(gravitationally bound and comoving; Kervella et al.\ 2019b). 
 The occurrence of wide 
companions is about 14\%.

 This paper focuses on Cepheids with companions at separations from approximately 
100 AU to  several thousand AU\null.  
This group has 
several  characteristics relevant to the formation of multiple stellar systems.
All have an inner (spectroscopic) binary.  The distribution of companion
spectral types (temperatures) or masses is distinctly different 
for companions within about 2000 AU and those at wider separations   
(Fig.~\ref{sep}).  As discussed in Item 6 above, the properties of companions 
provide clues to the formation and evolution of the systems. 
Thus, although the properties of 
companions in multiple systems are difficult to obtain, this group 
warrants further observation.

\acknowledgments

This research is based on observations made with the NASA/ESA {\it Hubble Space
Telescope\/} obtained from the Space Telescope Science Institute, which is
operated by the Association of Universities for Research in Astronomy, Inc.,
under NASA contract NAS 5-26555. 
These observations are associated with program
12215, with support for this work was provided by NASA
through  grants from the Space Telescope Science Institute
(GO-12215.01-A and GO-13368.01-A).
 Support was provided to NRE by the {\it Chandra}  X-ray  Center  NASA  Contract  NAS8-03060,
and for NASA grants for proposal 17200363.
Support for this work was provided by the National Aeronautics and Space Administration 
through the Smithsonian Astrophysical Observatory contract SV3-73016 to MIT for Support 
of the Chandra X-Ray Center, which is operated by the Smithsonian Astrophysical Observatory 
for and on behalf of the National Aeronautics and Space Administration under contract NAS8-03060.
The authors acknowledge the support of the French Agence Nationale de la Recherche (ANR), 
under grant ANR-15-CE31-0012-01 (project UnlockCepheids). The research leading to these 
results  has received funding from the European Research Council (ERC) under the European 
Union's Horizon 2020 research and innovation program (grant agreement No. 695099).
This work has made use of data from the European Space Agency (ESA) mission
{\it Gaia} (\url{https://www.cosmos.esa.int/gaia}), processed by the {\it Gaia}
Data Processing and Analysis Consortium (DPAC,
\url{https://www.cosmos.esa.int/web/gaia/dpac/consortium}). Funding for the DPAC
has been provided by national institutions, in particular the institutions
participating in the {\it Gaia} Multilateral Agreement.
 The SIMBAD  database, and NASA's Astrophysics Data System Bibliographic Services 
were used in
the preparation of this paper.

\facilities{{\it HST} (WFC3) {\it Chandra}}

\software{DrizzlePac (Hack, et al. 2012), 
Astropy (The Astropy Collaboration 2013),  
 DAOPHOT (Stetson, P. 1987),  
  Sherpa (Freeman, et al. 2001; Doe, et al. 2007) 
 Photutils (Bradley, L. et al. 2019),
and IRAF (Tody, D. 1986, 1993)
}

\clearpage

\appendix

\section{Candidate Companions with Separations Between $2''$ and $5''$: Resolved Wide Companions}
\label{appcom5}

The images and CMDs for companion candidates with separations between $2''$ and $5''$,  discussed in 
Section~\ref{sect:IRAF}, are presented in this Appendix.  The CMDs include stars from the full WFC3 images, but the companion candidates indicated by arrows are only from the $2''$ to $5''$ region.


\begin{figure}[hb!]
\plottwo{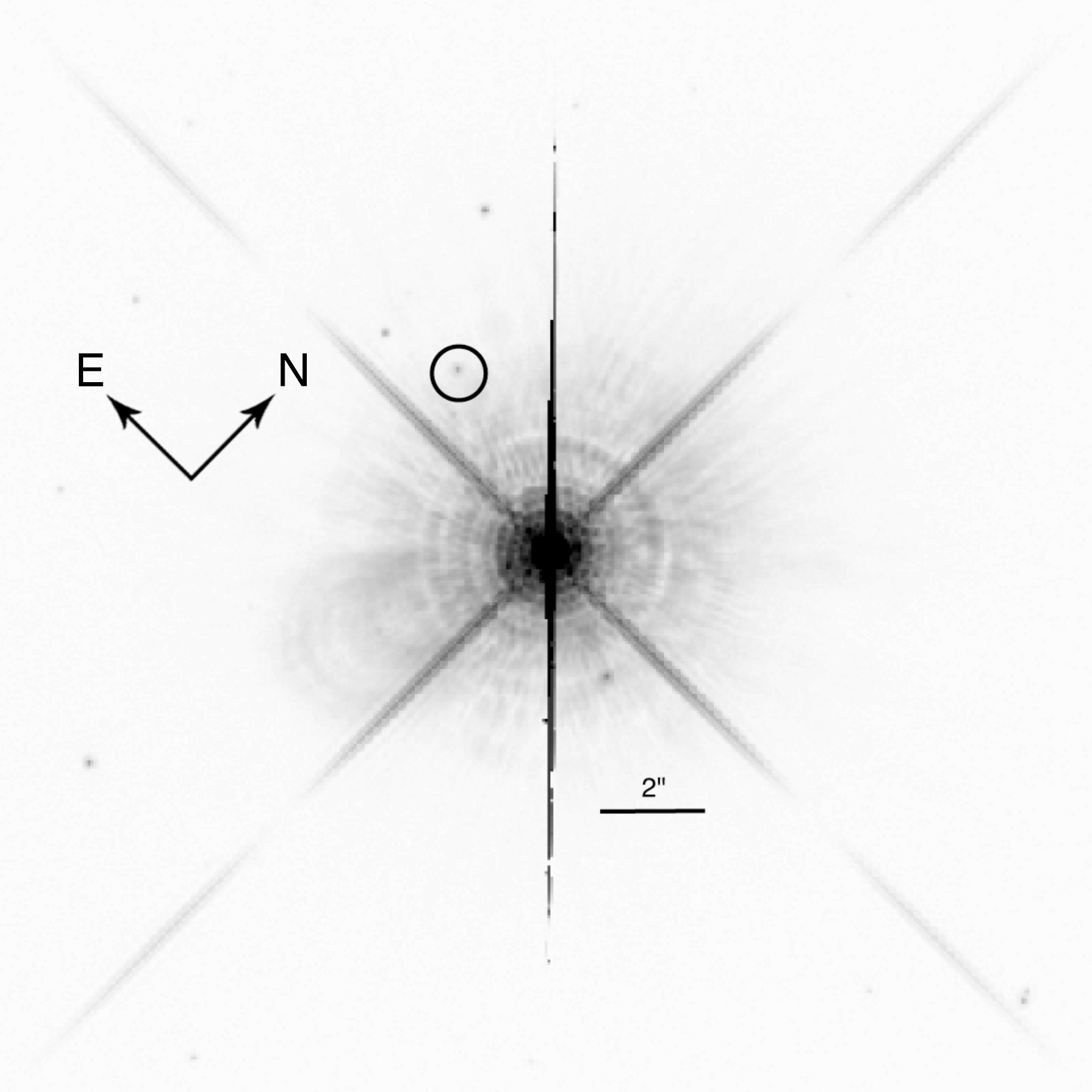}{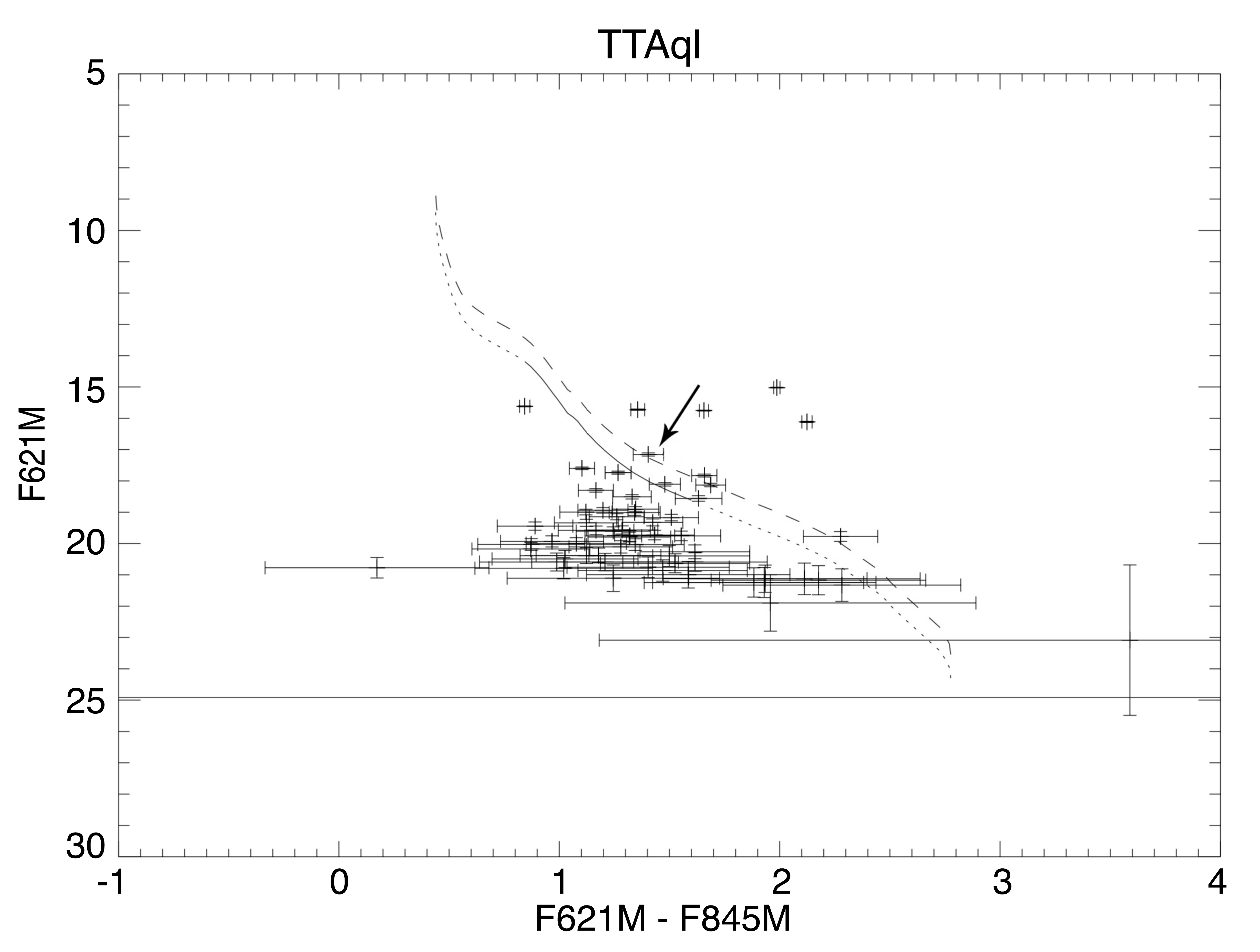}
\caption{(left) The inner portion of the F845M 
WFC3 image  of TT Aql.  The possible companion is
circled.
A log scale is used, and  2\arcsec\/ is indicated by the bar. 
(right)  The CMD from the  F845M and F621M WFC3 images.  The lower line is
the ZAMS at the distance and with the reddening of the Cepheid.  The 
solid portion is the spectral region of F2 to K7; dotted parts are
the extension to other spectral types.  The dashed line is 0.75 mag 
brighter to include binaries.  The arrow perpendicular to the ZAMS
indicates the possible companion.  Other stars within this band are 
more widely separated from the Cepheid which  were discussed in Paper II.  
\label{ttaql}}
\end{figure}

\begin{figure}[hb!]
\plottwo{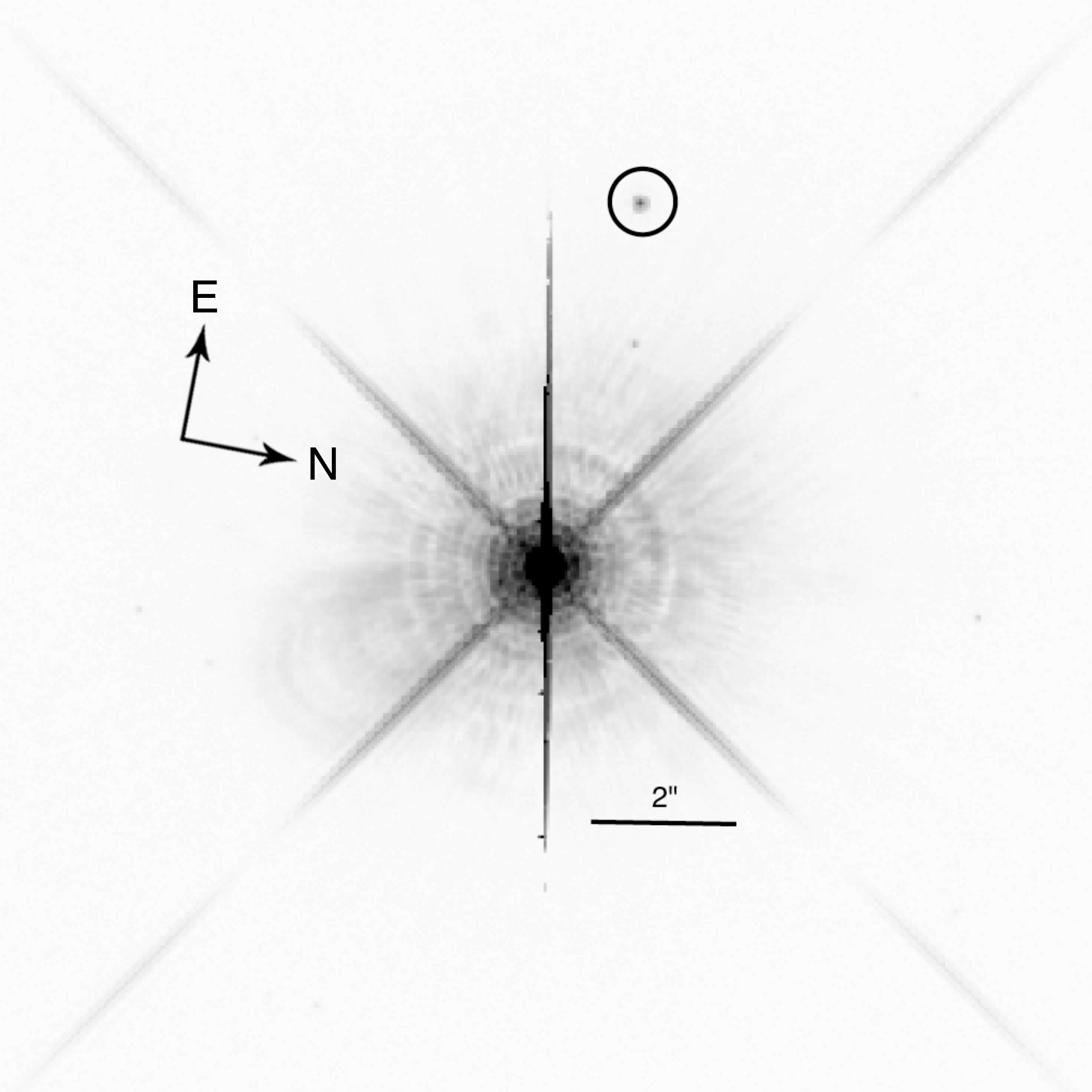}{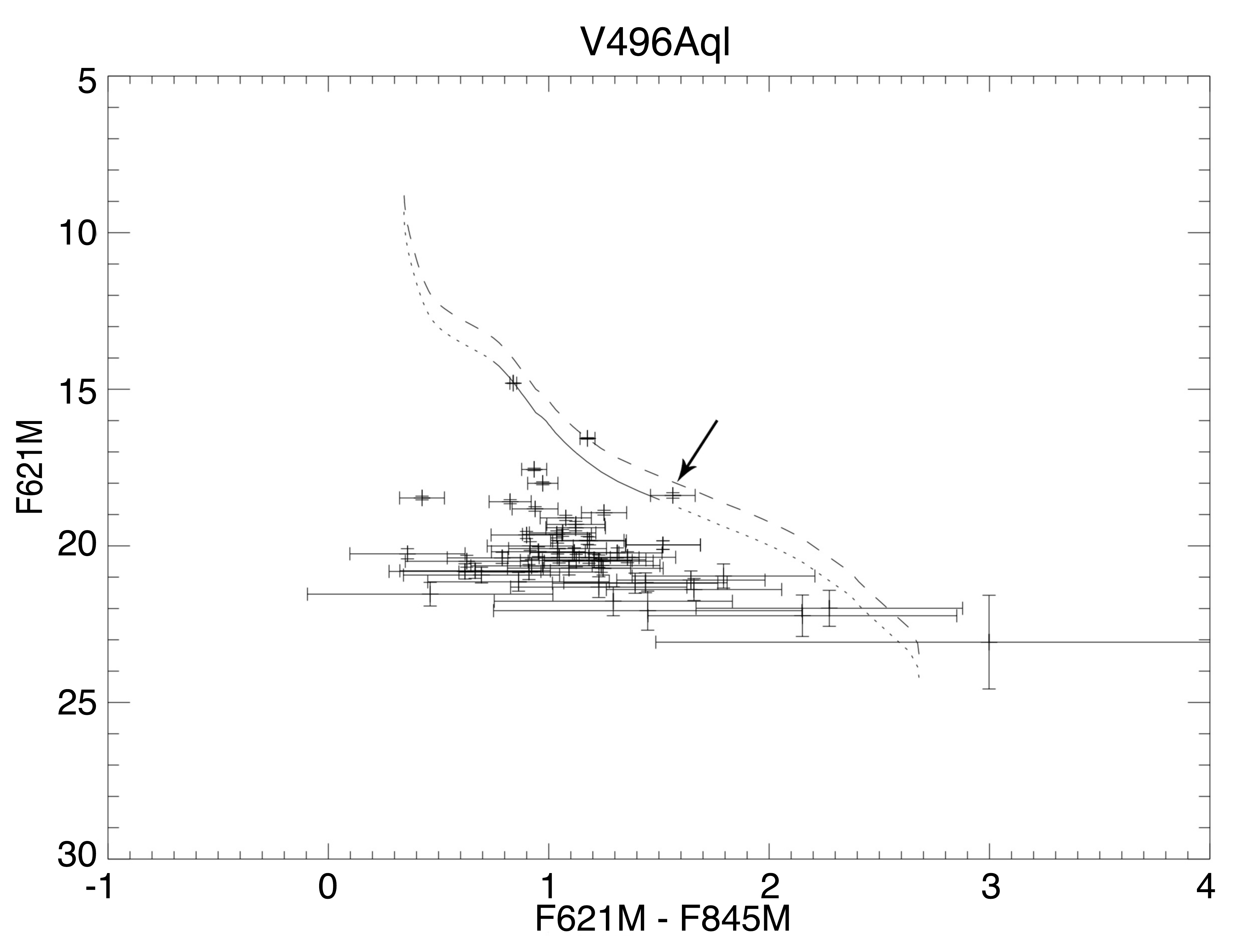}
\caption{(left) The inner portion of the F845M 
WFC3 image  of V496 Aql.  The possible companion is
circled.
A log scale is used, and  2\arcsec\/ is indicated by the bar. 
(right)  The CMD from the  F845M and F621M WFC3 images for V496 Aql.  Symbols
are the same as for Figure~\ref{ttaql}.
\label{v496aql}}
\end{figure}

\begin{figure}
\plottwo{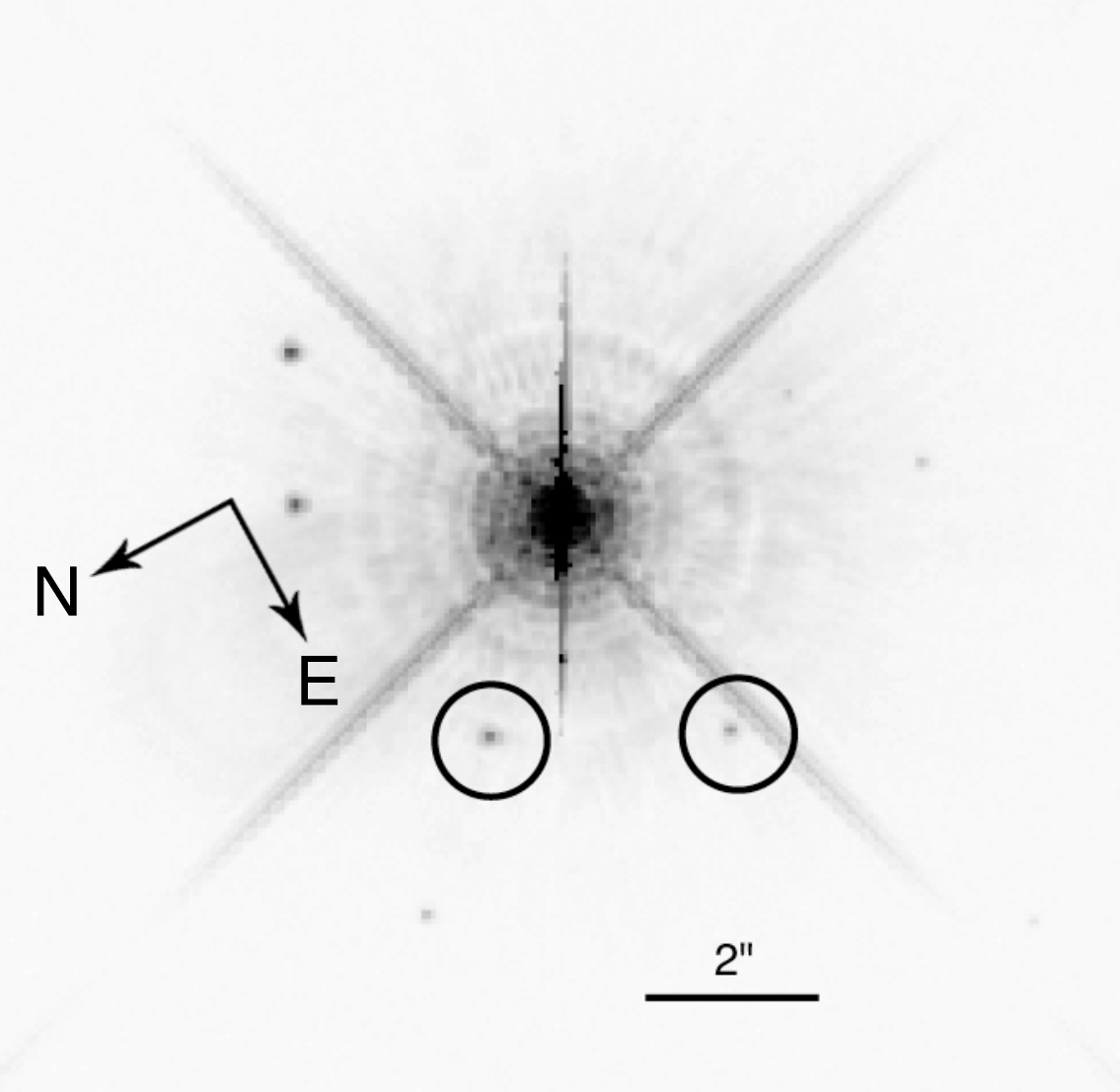}{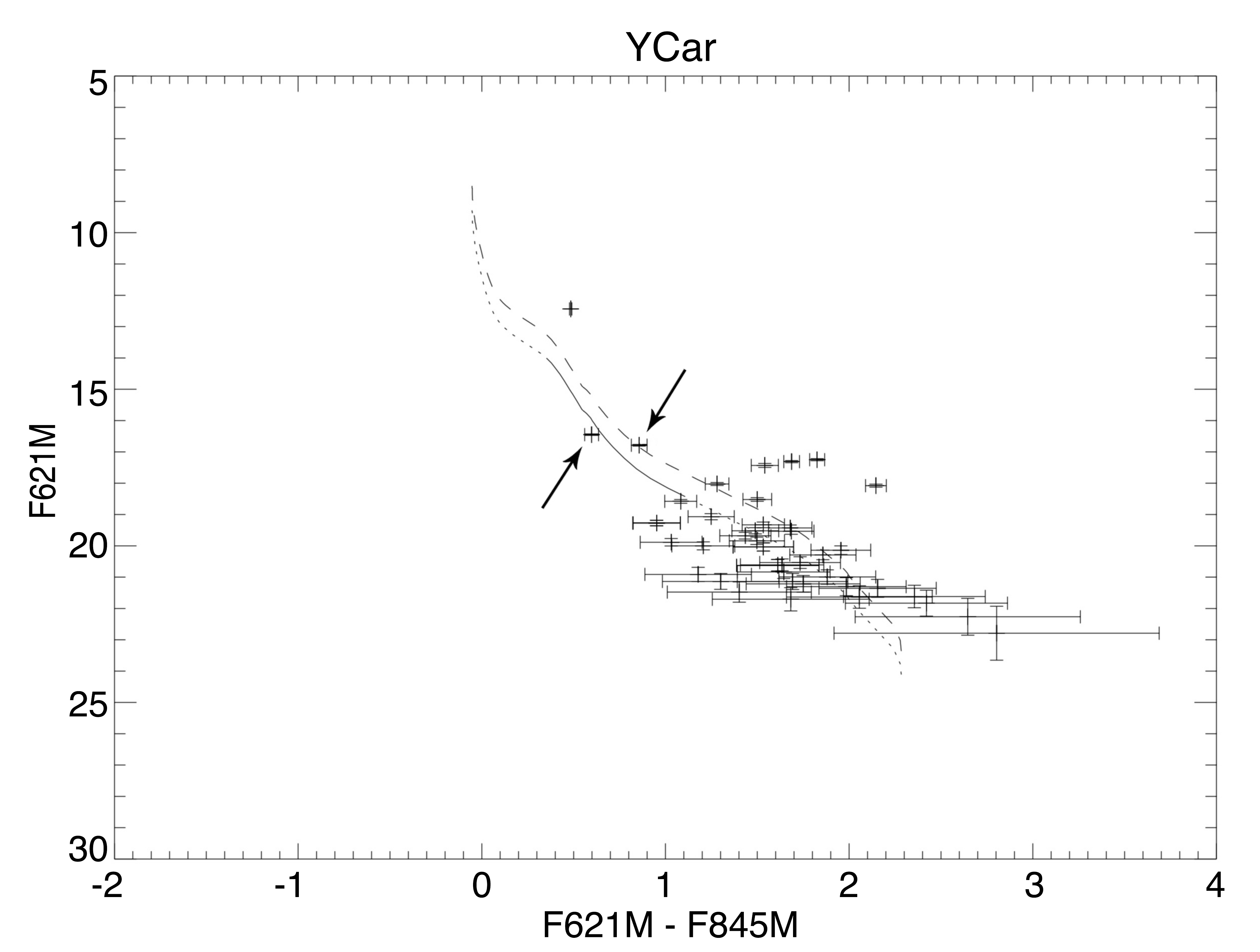}
\caption{(left) The inner portion of the F845M 
WFC3 image  of Y Car.  The possible companions are
circled.
A log scale is used, and  2\arcsec\/ is indicated by the bar. 
(right)  The CMD from the  F845M and F621M WFC3 images for Y Car.  Symbols
are the same as for Figure~\ref{ttaql}.
\label{ycar}}
\end{figure}

\begin{figure}
\plottwo{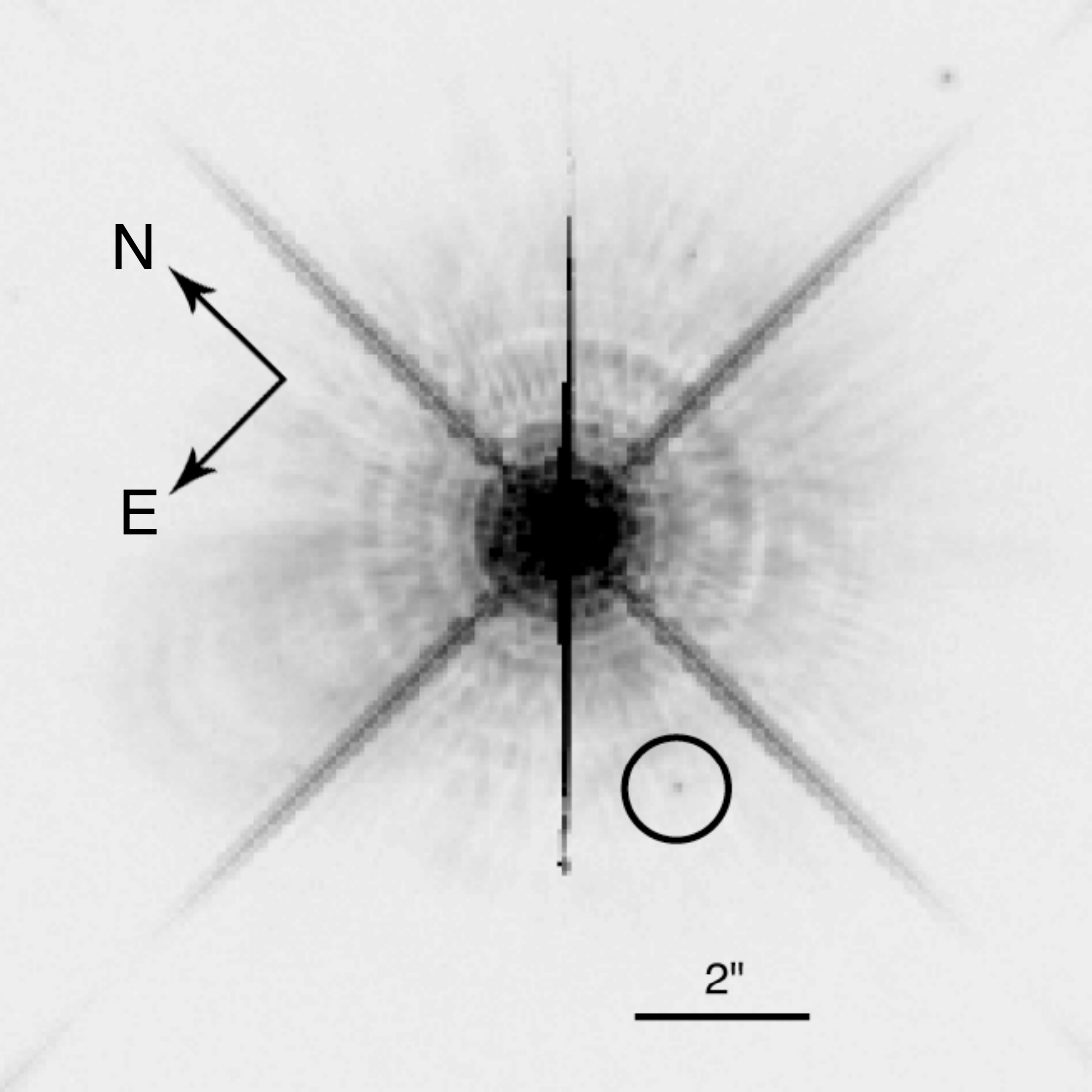}{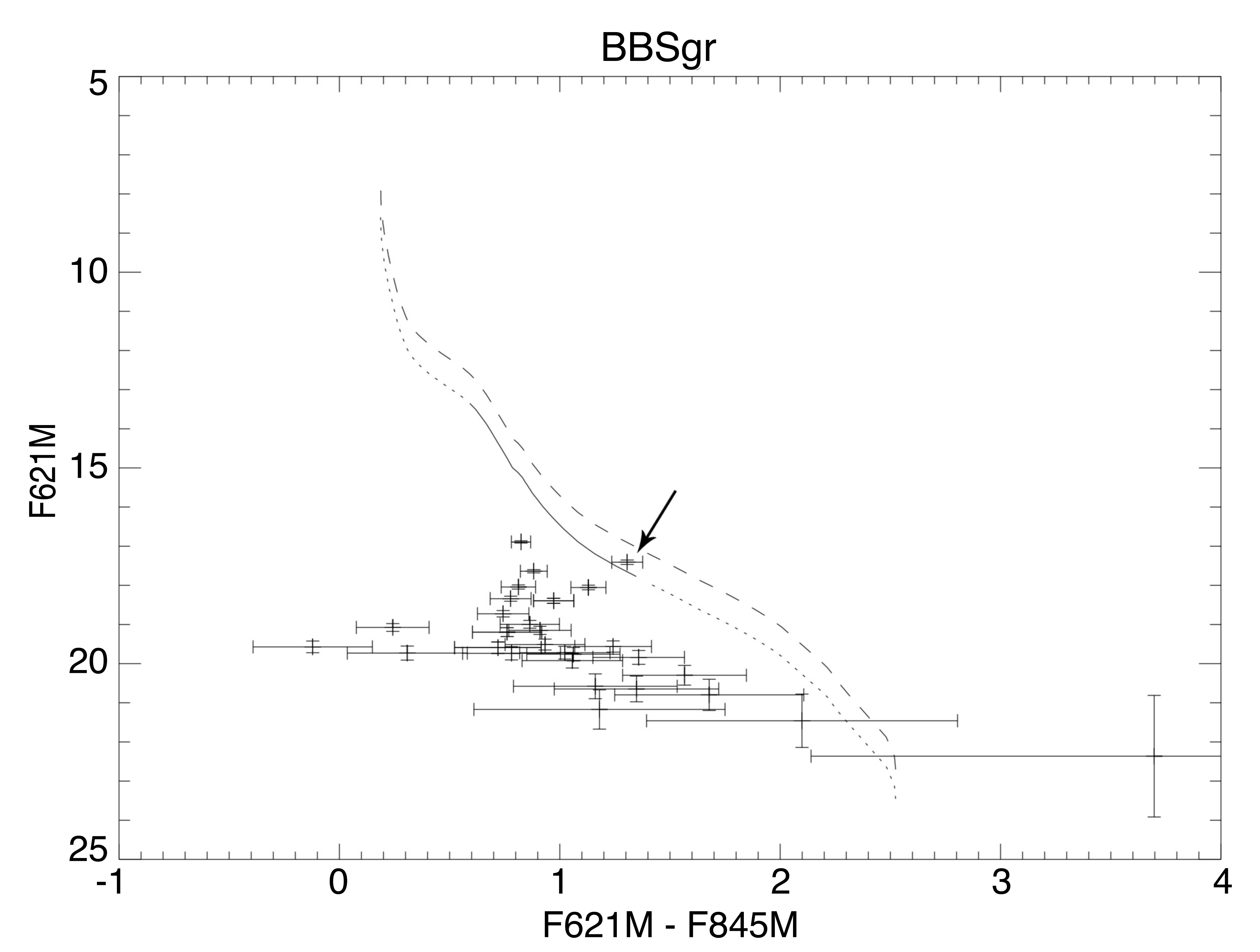}
\caption{(left) The inner portion of the F845M 
WFC3 image  of BB Sgr.  The possible companion is
circled.
A log scale is used, and  2\arcsec\/ is indicated by the bar. 
(right)  The CMD from the  F845M and F621M WFC3 images for BB Sgr.  Symbols
are the same as for Figure~\ref{ttaql}.
\label{bbsgr}}
\end{figure}

\begin{figure}
\plottwo{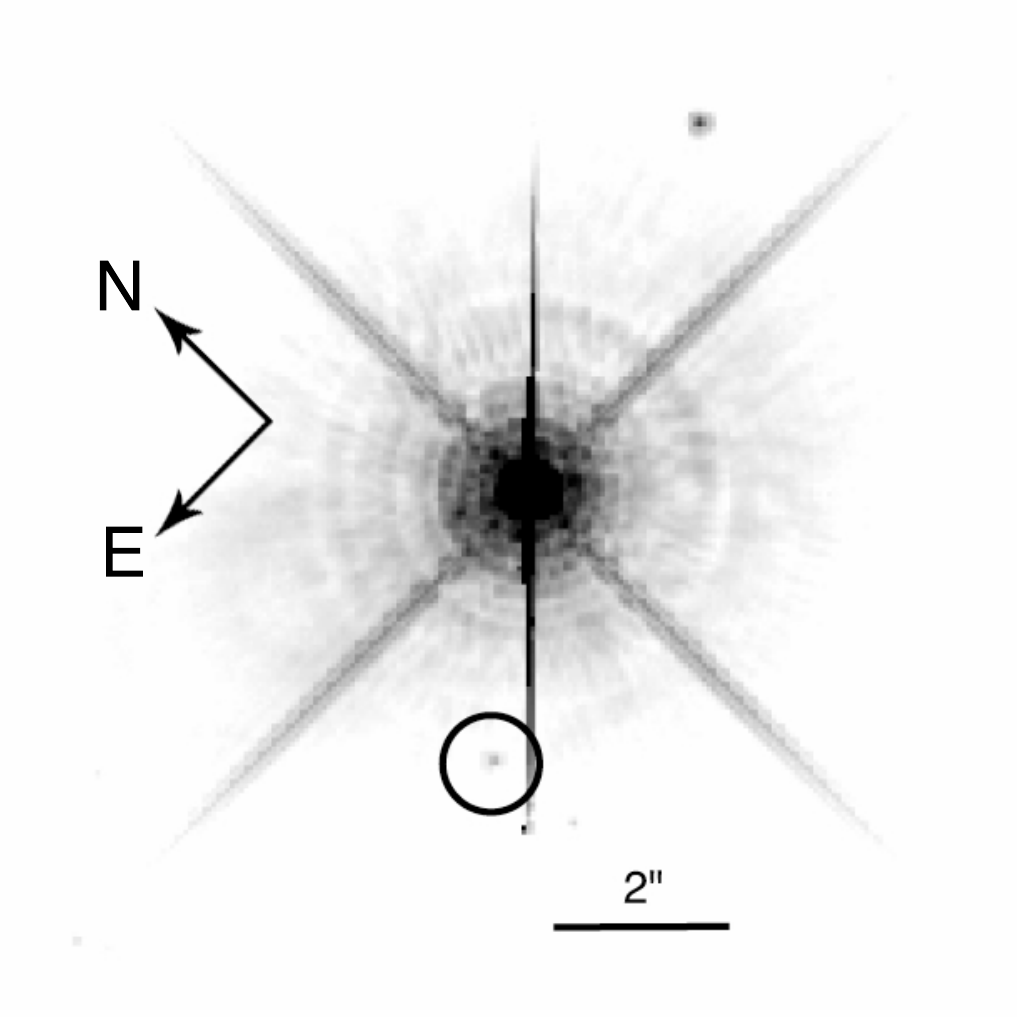}{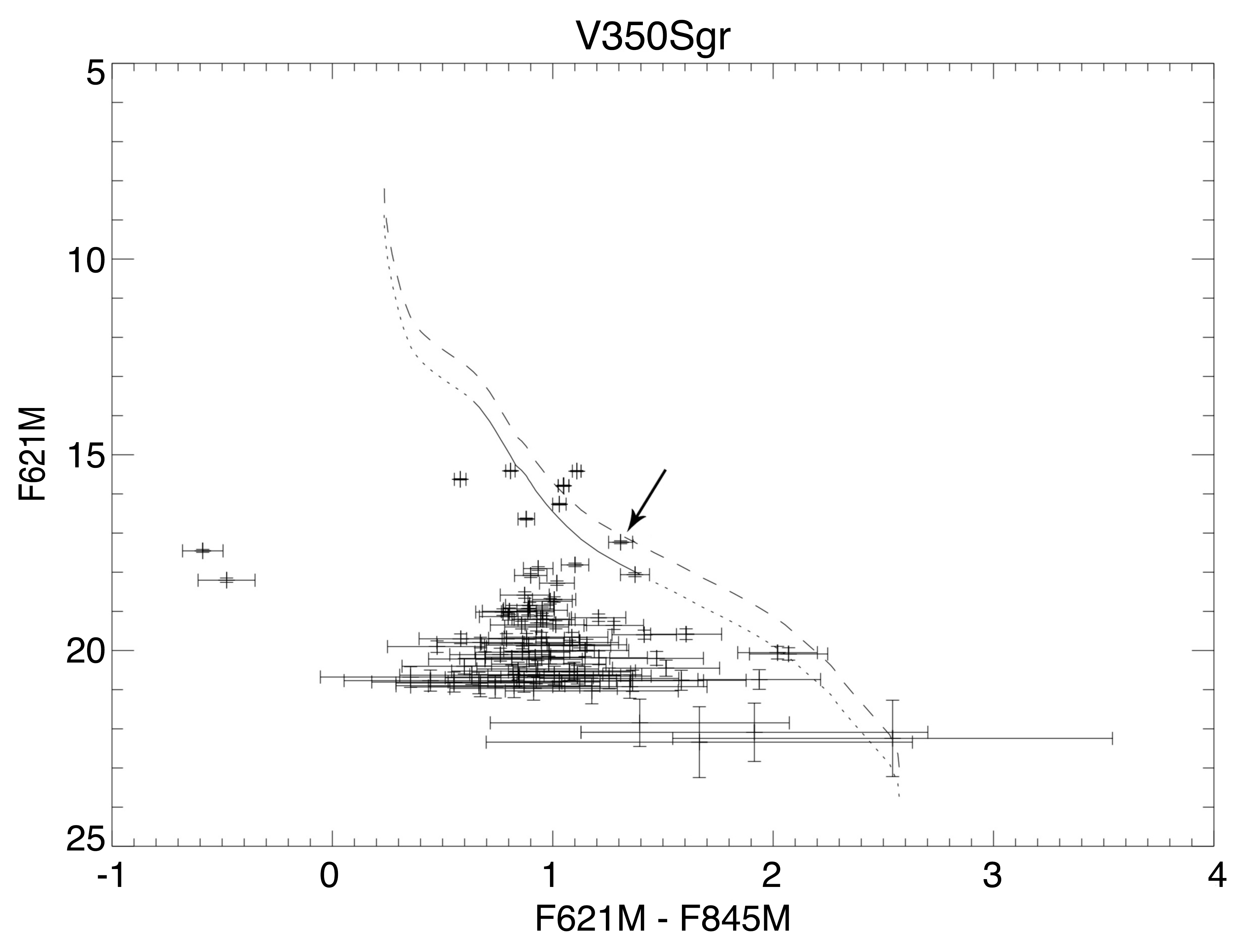}
\caption{(left) The inner portion of the F845M 
WFC3 image  of V350 Sgr.  The possible companion is
circled.
A log scale is used, and  2\arcsec\/ is indicated by the bar. 
(right)  The CMD from the  F845M and F621M WFC3 images for V350 Sgr.  Symbols
are the same as for Figure~\ref{ttaql}.
\label{v350sgr}}
\end{figure}

\begin{figure}
\plottwo{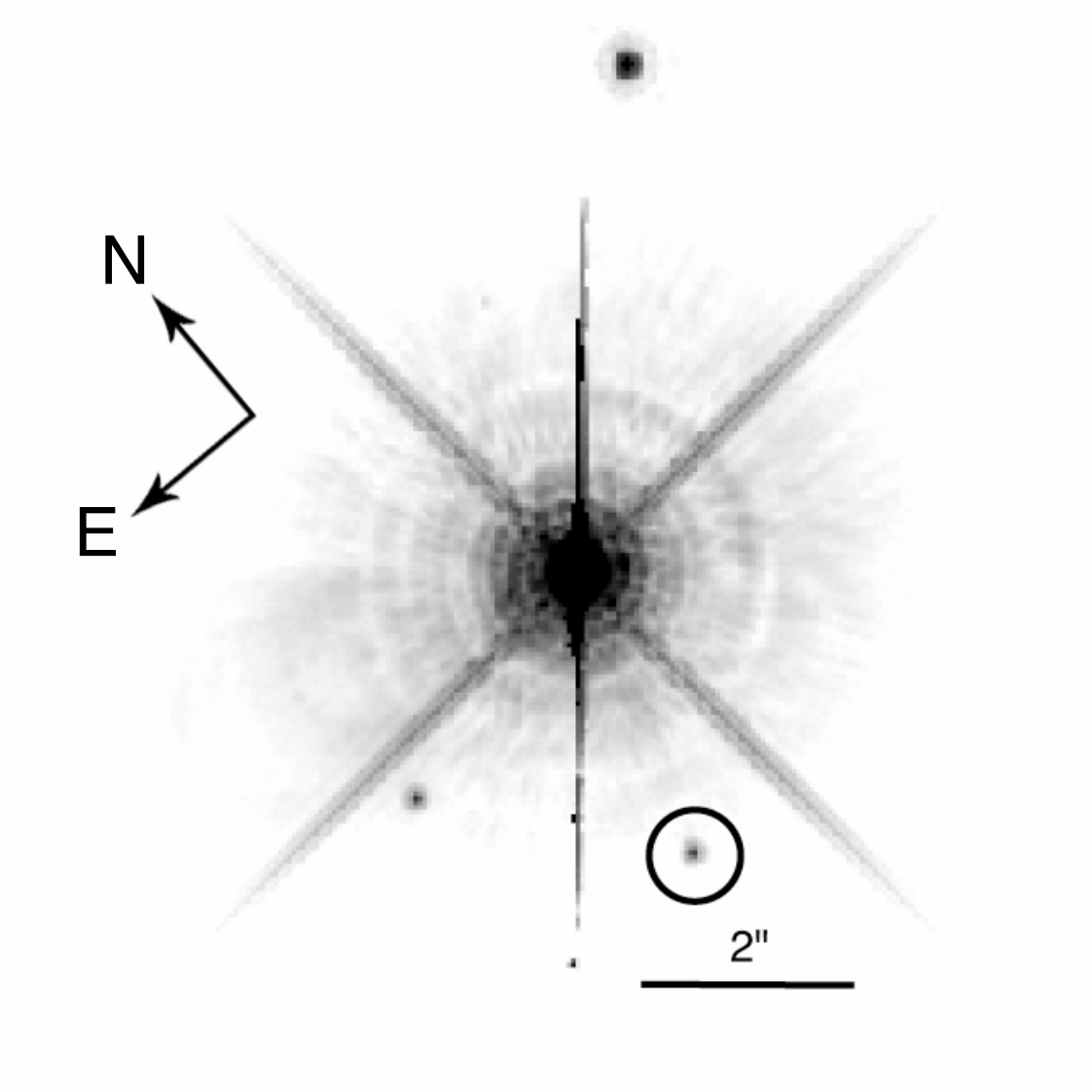}{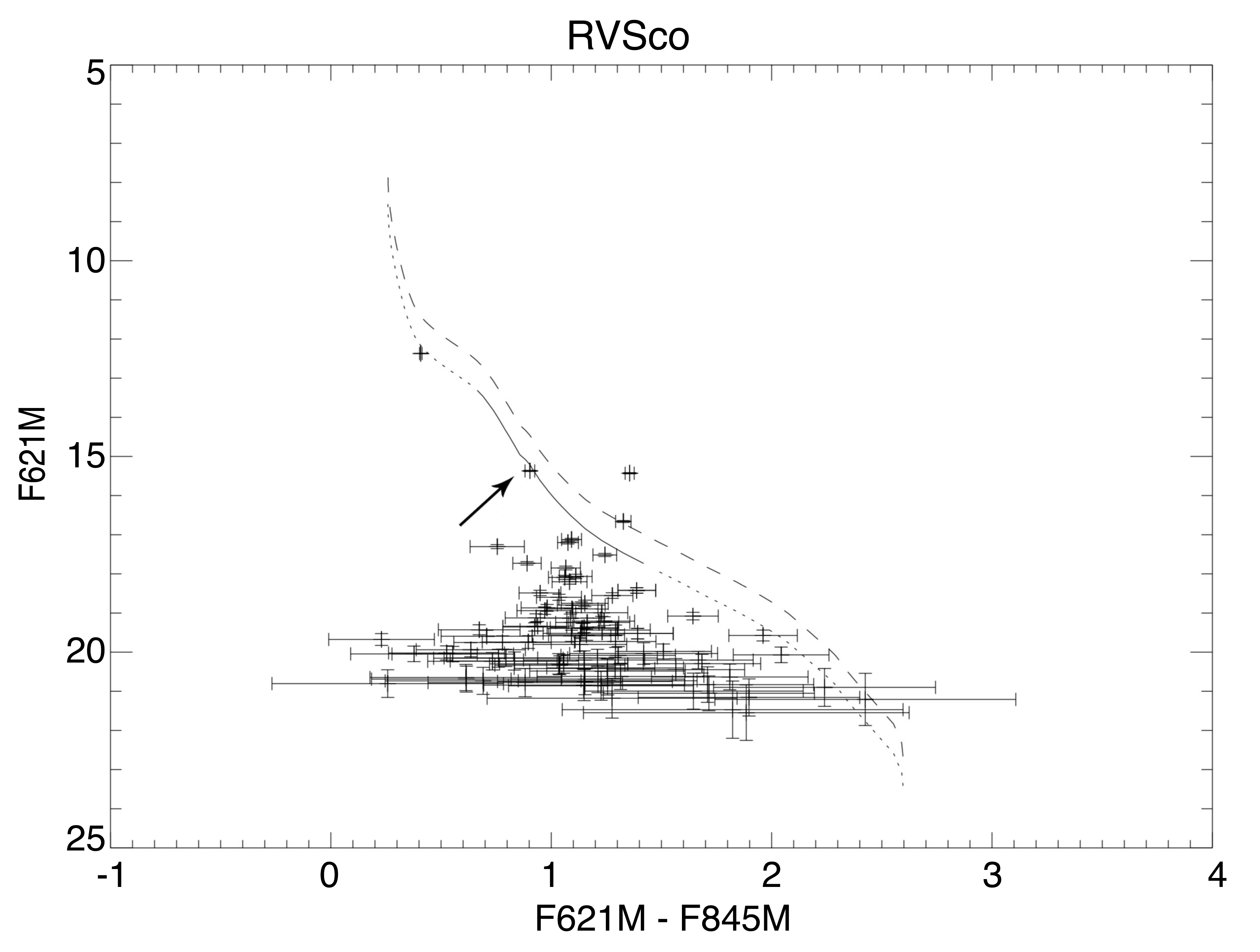}
\caption{(left) The inner portion of the F845M 
WFC3 image  of RV Sco.  The possible companion is
circled.
A log scale is used, and  2\arcsec\/ is indicated by the bar. 
(right)  The CMD from the  F845M and F621M WFC3 images for RV Sco.  Symbols
are the same as for Figure~\ref{ttaql}.
\label{rvsco}}
\end{figure}

\clearpage


\section{Candidate Companions within $2''$: Resolved Close Companions}
\label{appcom2}

The \HST\/ images  for companion candidates within $2''$, discussed in 
Section~\ref{detect.2}, are presented in this Appendix both before and after PSF correction.


\begin{figure}[hb!]
\begin{center}
\includegraphics[width=1.0\textwidth]{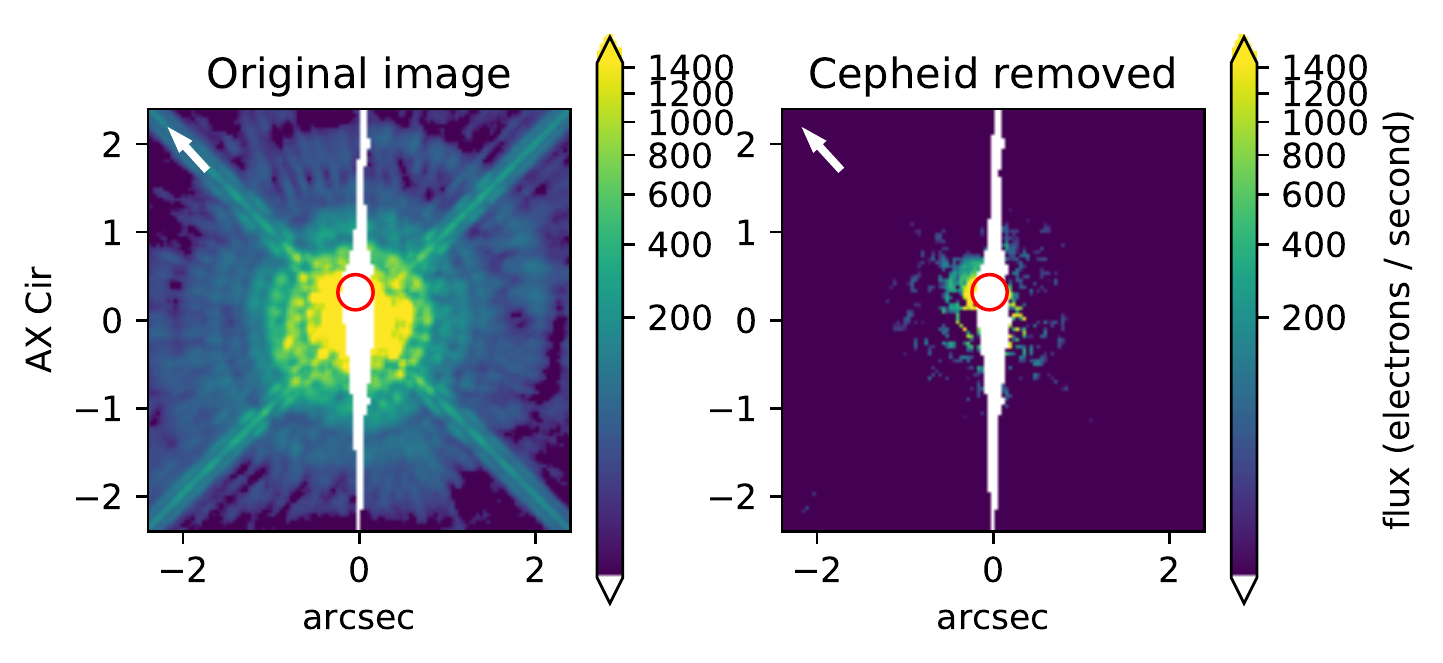}
\caption{ Image of AX~Cir in the F845M band. The image is displayed 
in the same way as for Figure~\ref{fig:EtaAql845} ($\eta$ Aql).  Left: Original image. Right:
after LOCI subtraction.
\label{fig:AXCir845}}
\end{center}
\end{figure}

\begin{figure}
\begin{center}
\includegraphics[width=1.0\textwidth]{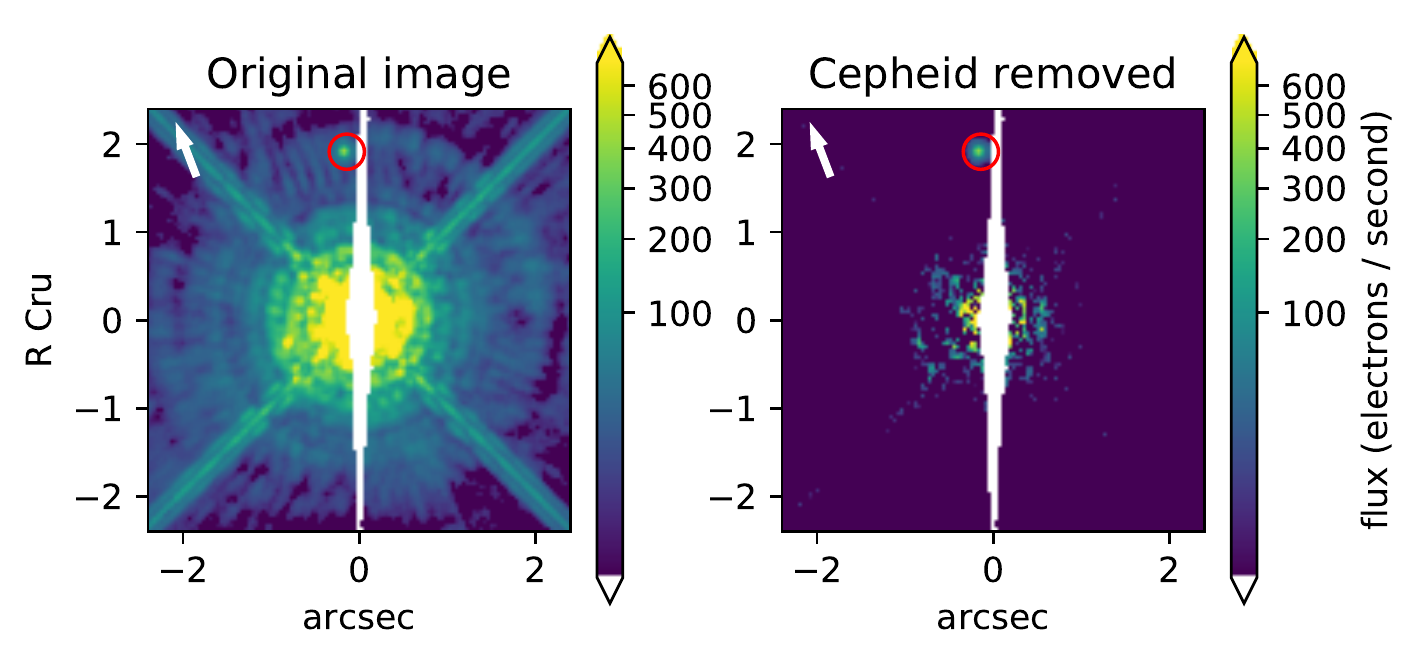}
\caption{ Image of R Cru in the F845M band. The image is displayed 
in the same way as for Figure~\ref{fig:EtaAql845} ($\eta$ Aql).  Left: Original image. Right:
after LOCI subtraction.
\label{fig:RCru845}}
\end{center}
\end{figure}

\begin{figure}
\begin{center}
\includegraphics[width=1.0\textwidth]{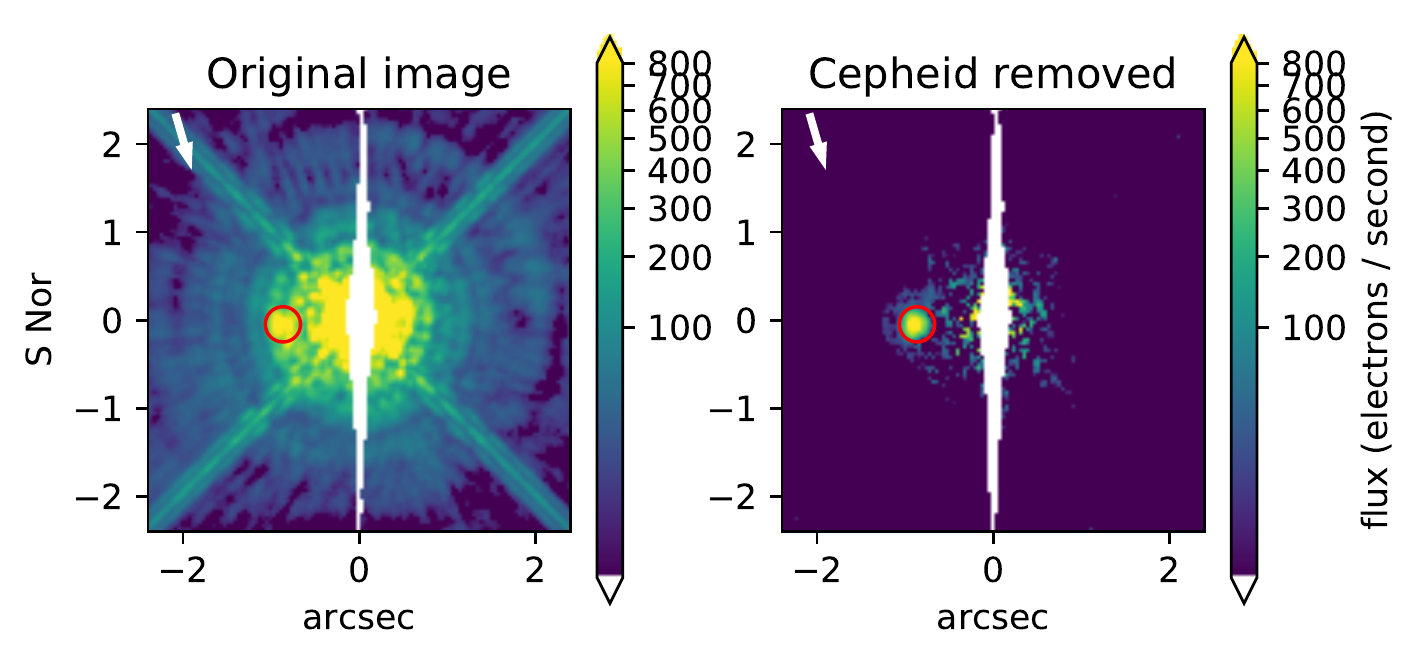}
\caption{ Image of S~Nor in the F845M band. 
The image is displayed 
in the same way as for Figure~\ref{fig:EtaAql845} ($\eta$ Aql).  Left: Original image. Right:
after LOCI subtraction.
\label{fig:SNor845}}
\end{center}
\end{figure}

\begin{figure}
\begin{center}
\includegraphics[width=1.0\textwidth]{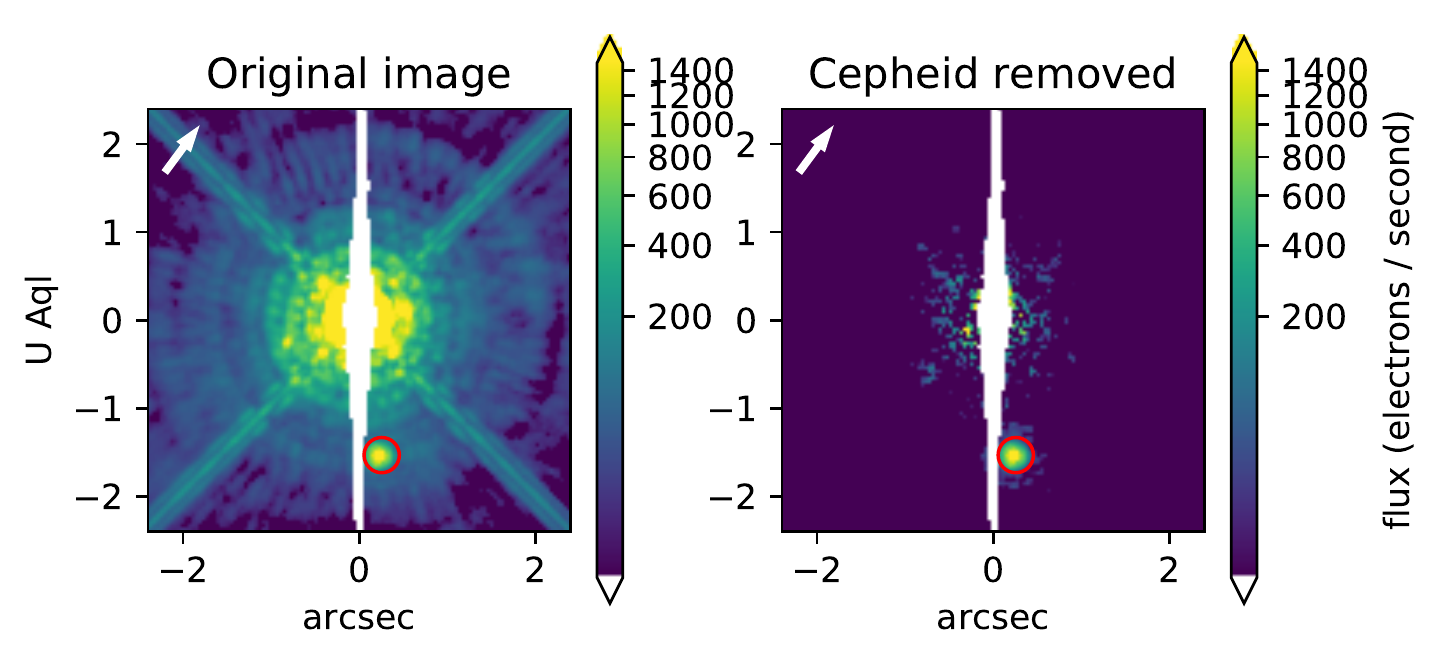}
\caption{ Image of U~Aql in the F845M band. 
The image is displayed 
in the same way as for Figure~\ref{fig:EtaAql845} ($\eta$ Aql).  Left: Original image. Right:
after LOCI subtraction.
\label{fig:uaql845}}
\end{center}
\end{figure}

\begin{figure}
\begin{center}
\includegraphics[width=1.0\textwidth]{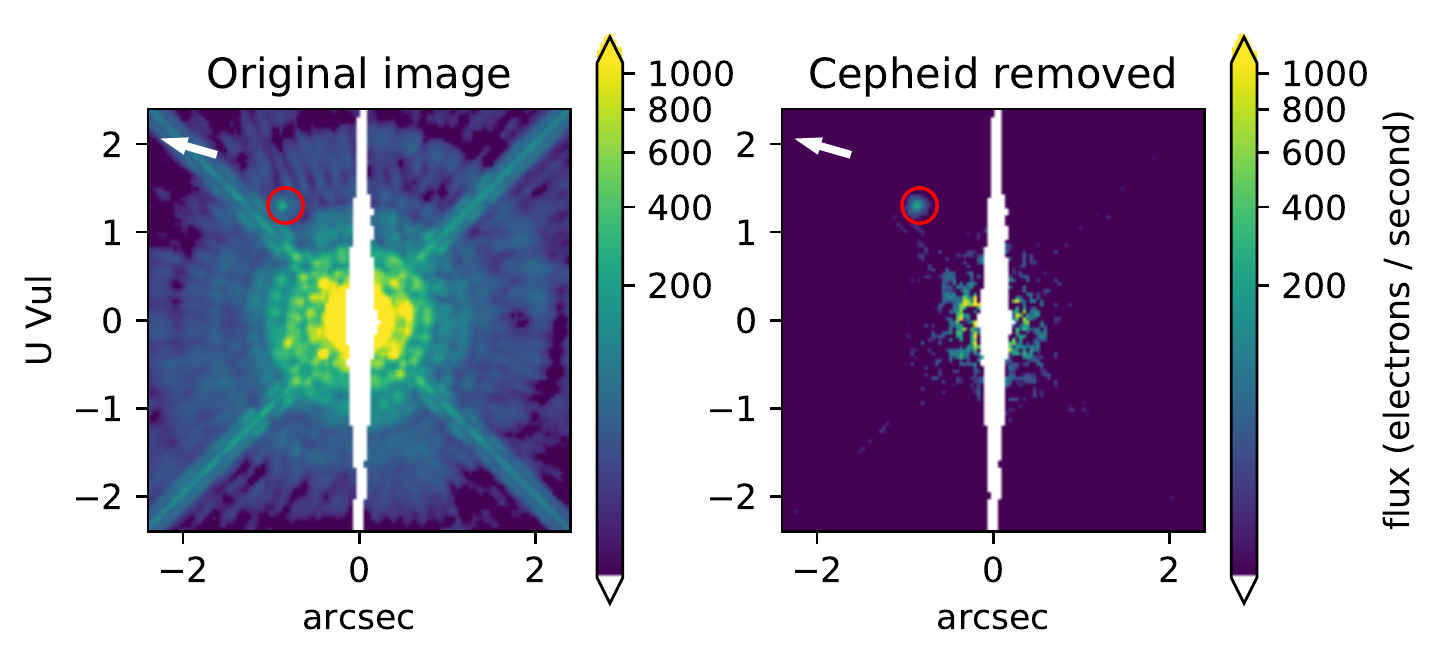}
\caption{ Image of U~Vul in the F845M band. 
The image is displayed 
in the same way as for Figure~\ref{fig:EtaAql845} ($\eta$ Aql).  Left: Original image. Right:
after LOCI subtraction.
\label{fig:uvul845}}
\end{center}
\end{figure}

\begin{figure}
\begin{center}
\includegraphics[width=1.0\textwidth]{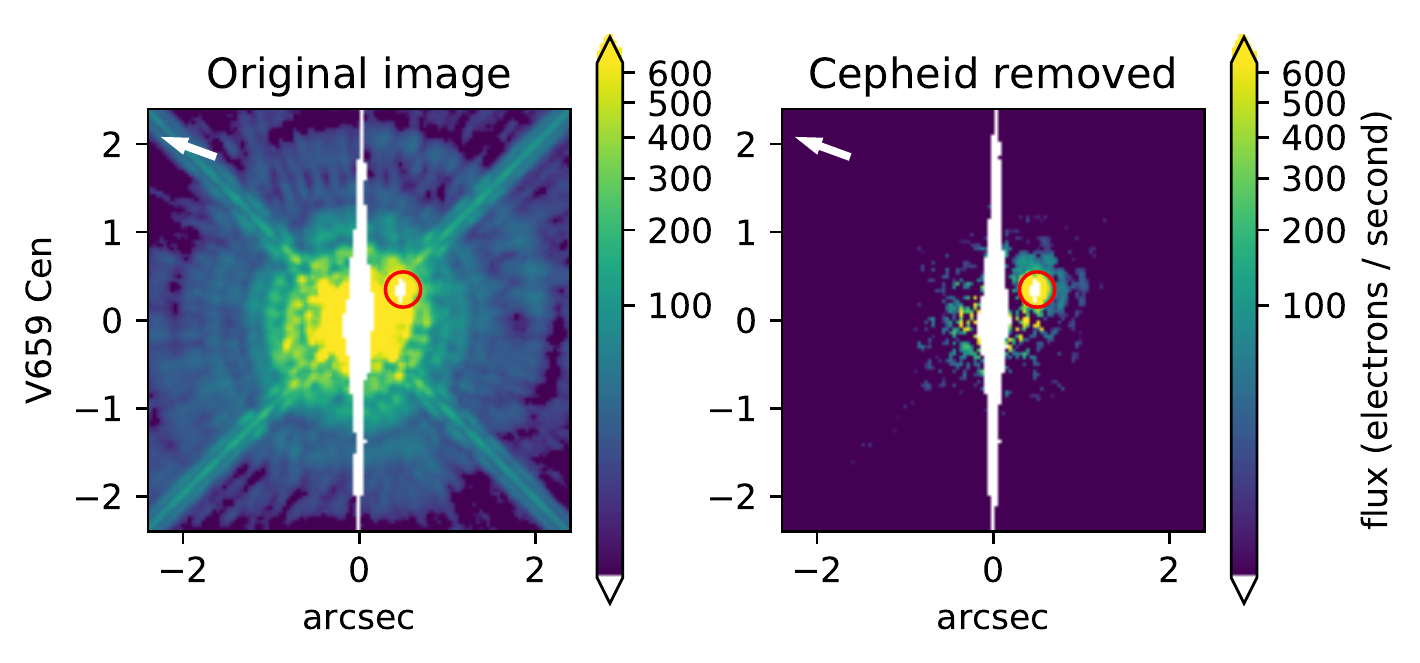}
\caption{ Image of V659~Cen in the F845M band. 
The image is displayed 
in the same way as for Figure~\ref{fig:EtaAql845} ($\eta$ Aql).  Left: Original image. Right:
after LOCI subtraction.
\label{v659cen845}}
\end{center}
\end{figure}

\begin{figure}
\begin{center}
\includegraphics[width=1.0\textwidth]{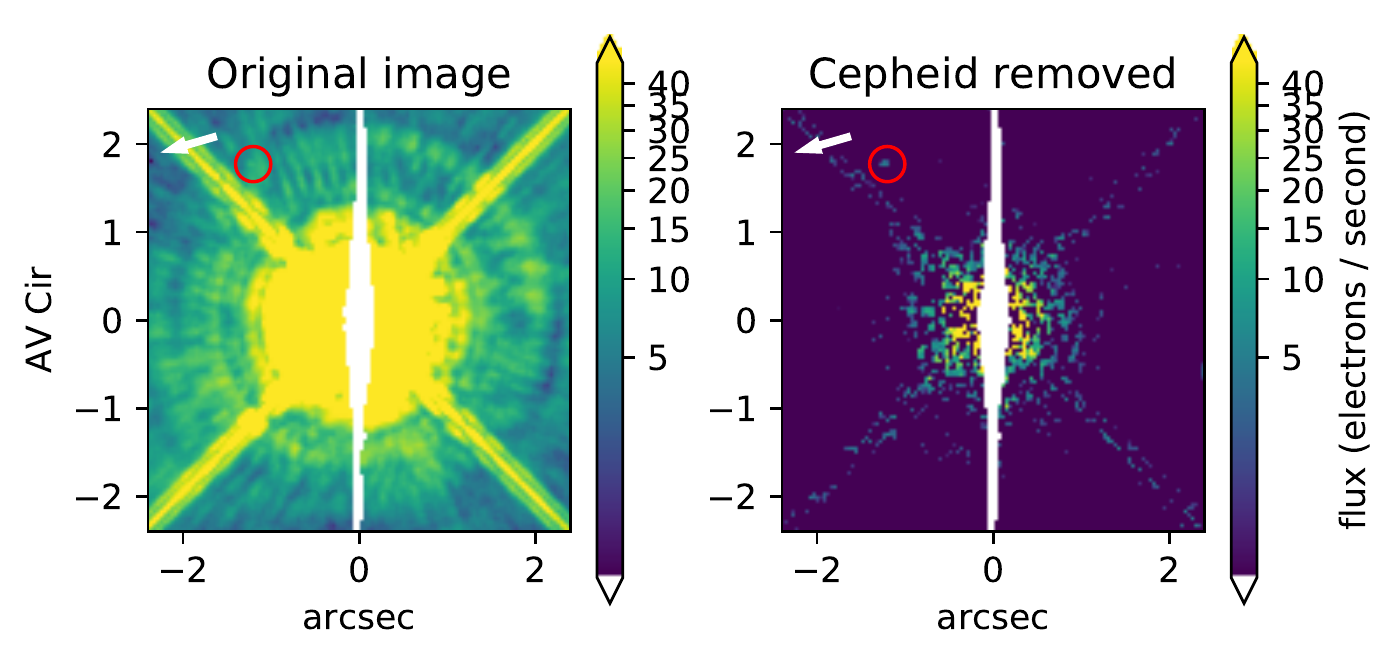}
\caption{ Image of AV~Cir in the F845M band. 
The image is displayed 
in the same way as for Figure~\ref{fig:EtaAql845} ($\eta$ Aql).  Left: Original image. Right:
after LOCI subtraction.
\label{fig:avcir845}}
\end{center}
\end{figure}

\begin{figure}
\begin{center}
\includegraphics[width=1.0\textwidth]{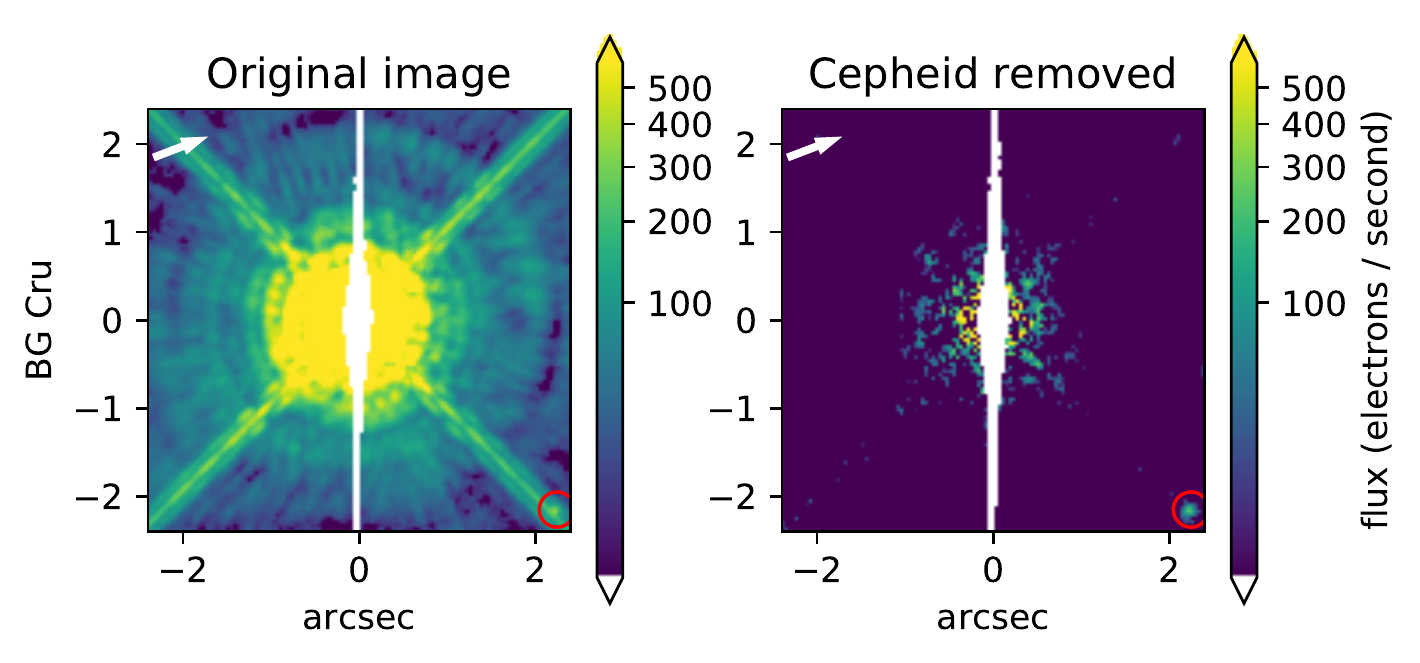}
\caption{ Image of BG~Cru in the F845M band. 
The image is displayed 
in the same way as for Figure~\ref{fig:EtaAql845} ($\eta$ Aql).  Left: Original image. Right:
after LOCI subtraction.
\label{fig:bgcru845}}
\end{center}
\end{figure}

\begin{figure}
\begin{center}
\includegraphics[width=1.0\textwidth]{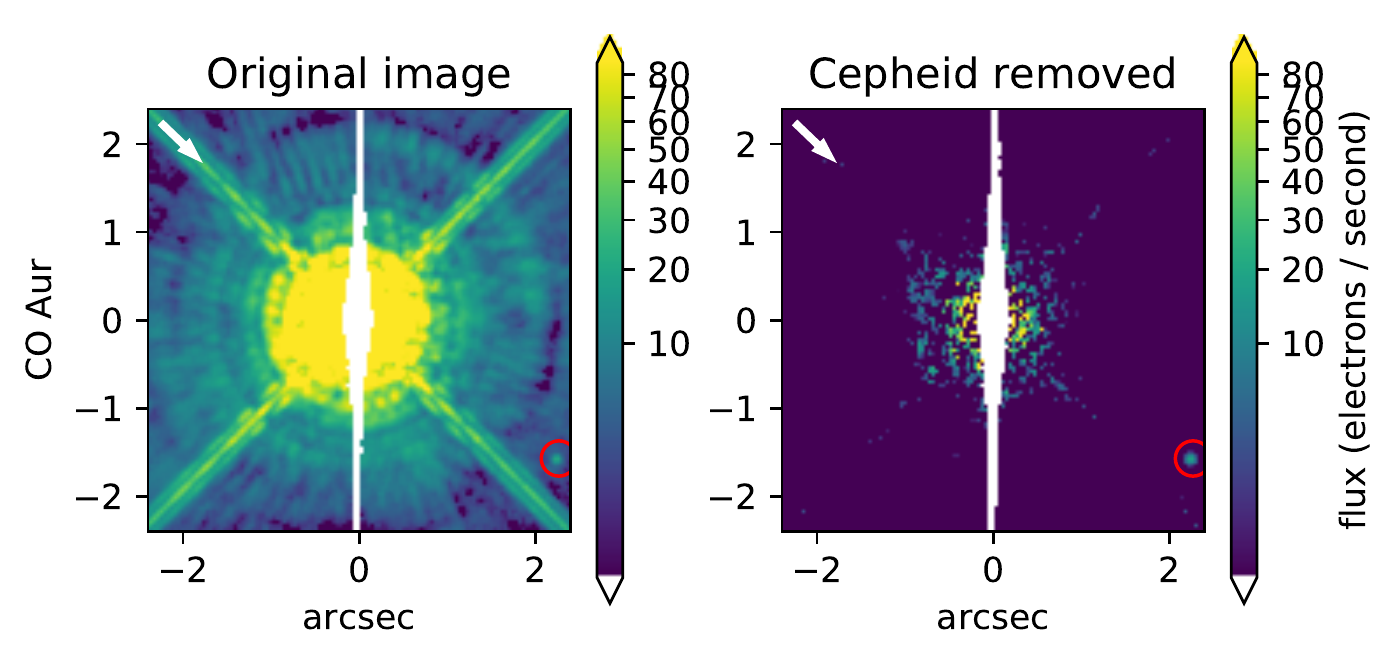}
\caption{ Image of CO~Aur in the F845M band. 
The image is displayed 
in the same way as for Figure~\ref{fig:EtaAql845} ($\eta$ Aql).  Left: Original image. Right:
after LOCI subtraction.
\label{fig:coaur845}}
\end{center}
\end{figure}

\clearpage

\section{{\it Chandra} Observations}\label{chandra}

The goal of this paper is to identify resolved companions in the Cepheid {\it HST\/} snapshot survey and to examine the properties of members of the systems they belong to. 
As discussed in Paper~IV, X-ray observations are an excellent way to distinguish 
between an X-ray-active star young enough to be a physical Cepheid companion, and 
an old field star without appreciable X-ray activity.  
Since Paper~IV, we have added {\it Chandra\/} observations for two 
stars which add to the understanding of system parameters.  The two systems were  
found to be X-ray sources in Paper~IV: R~Cru 
 (Table~\ref{tab:comp.mag})  and S~Mus.  
Among the {\it XMM\/} targets, they had X-ray detections 
at approximately the position of the Cepheid, but with a possible
companion which was not resolved from the Cepheid in the {\it XMM\/} images.  
We followed up these detections with {\it Chandra\/} observations to identify the 
source of the X-rays, using the higher resolution of {\it Chandra}.  
S Mus and R Cru were observed with the
Advanced CCD Imaging Spectrometer (ACIS) using the four ``I'' 
CCD chips.  Details are provided in 
Table~\ref{chan}




\begin{deluxetable}{lccc}[hb!]
\footnotesize
\tablecaption{{\it Chandra\/} Observations \label{chan}}
\tablewidth{0pt}
\tablehead{
\colhead{Star}  & \colhead{OBSID} & \colhead{JD}  
  & \colhead{Exp.} \\
\colhead{} &  \colhead{} & \colhead{$-$2,400,000}  
  & \colhead{[ks]} 
}
\startdata
S Mus  & 17740 &  57,388.068    &   5.0    \\
R Cru & 17741 &  57,578.170   & 14.9  \\
\enddata

\end{deluxetable}

\subsection{R Cru}

The {\it HST\/} and {\it XMM\/} images are shown in Figure~\ref{rcrustart}, with the locations
of the Cepheid and the possible optical companions circled.  Either of the possible 
companions could be the source of the X-rays in the XMM image (although the 
closer one is more likely).

\begin{figure}[hb!]
\centering
\includegraphics[width=5in]{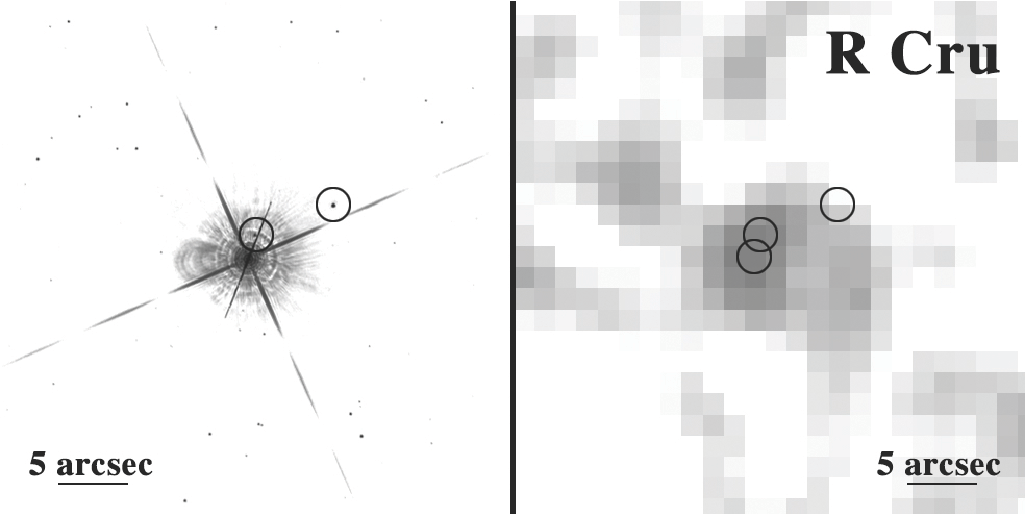}
\caption{Possible companions of R Cru.  Left: {\it HST\/} 
WFC3 $I$-band image of R Cru.  Possible companions are circled.
  The stretch is logarithmic to emphasize faint 
features.  Right: {\it XMM\/} image, again with circles indicating 
locations of the Cepheid and its two possible companions (including
a small correction derived from the X-ray positions of 2MASS sources.) The orientation of both panels is the same: north up and east on the left.
 \label{rcrustart}}
\end{figure}

The {\it Chandra\/} image of R~Cru is shown in Figure~\ref{rcruchan},
which compares the location of the X-ray source to that 
of the Cepheid and the possible close companion (including a small shift to align the 
Cepheid in the {\it HST\/} image with its coordinates from SIMBAD).   
The shape of the X-ray source based on a relatively weak detection has some uncertainty, 
but it is closer to the Cepheid than to the possible close companion. 

\begin{figure}[hb!]
\centering
\includegraphics[width=5in]{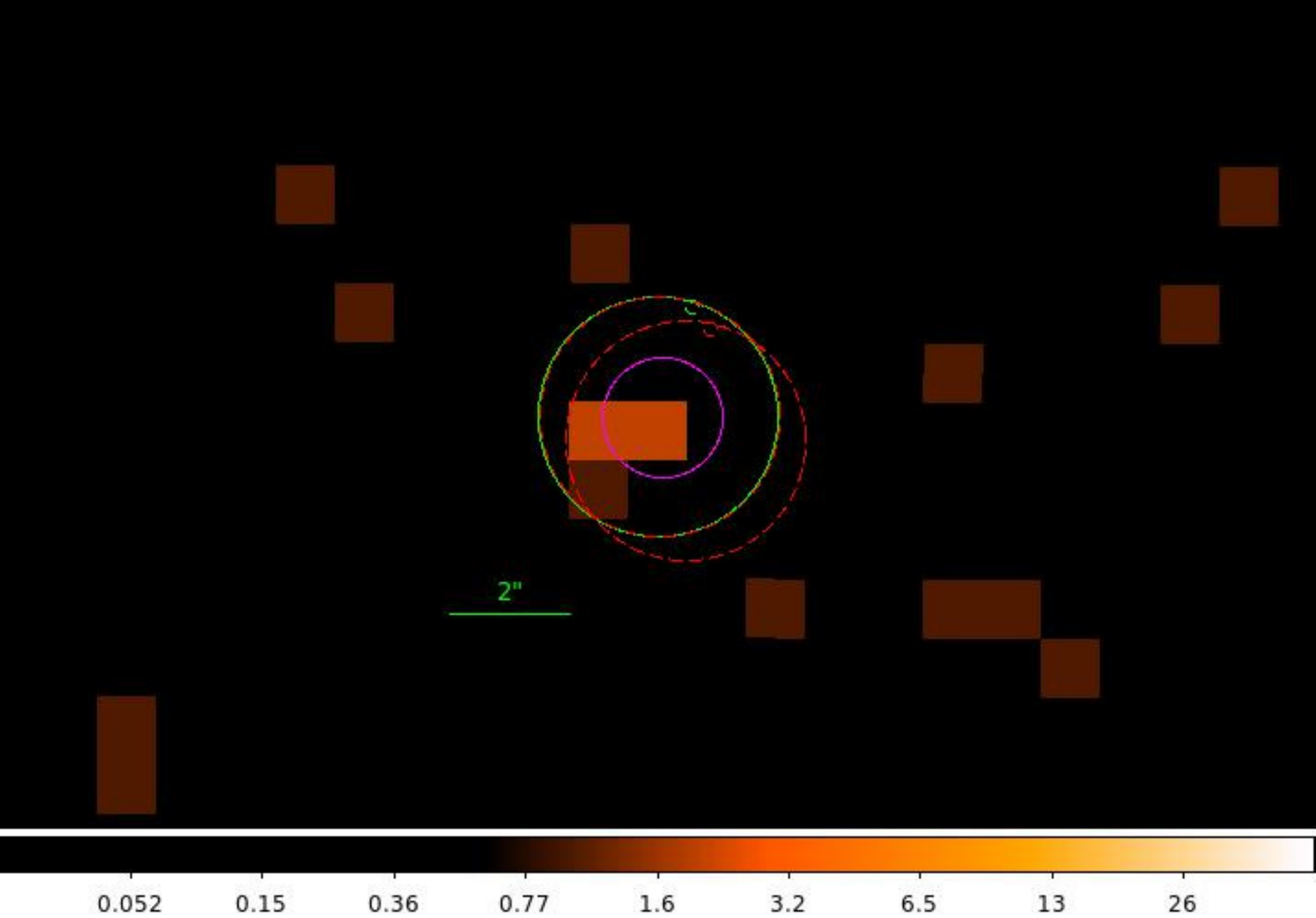}
\caption{The center of the {\it Chandra\/} image of R Cru, showing the Cepheid and the closer companion.  The orientation is 
the same as in Figure~\ref{rcrustart}. The bright red rectangle is the location of 
the {\it Chandra\/} source. 
The circles are as follows: green: $2''$ radius
centered on the position of the Cepheid from SIMBAD (including proper motion); red: the 
same radius centered on the Cepheid in the {\it HST\/} image; 
magenta:  $1''$ radius within
the green position of the Cepheid.  The position of the companion 
from the {\it HST\/} image is shown by 
 the very small red and green circles 
(about 1 o'clock) corresponding to 
the positions of the Cepheid in those colors.  
The image is a log stretch. 
 \label{rcruchan}}
\end{figure}


Are the X-rays from the Cepheid itself?  We now have a pattern to X-ray activity 
in Cepheids (Engle et al.\ 2017), that it is relatively constant at about $\log L_X =
28.8\,\rm ergs\, s^{-1}$ for most of the pulsation cycle, but has a burst of X-rays just
after maximum radius (approximately phase 0.5).  The X-ray flux 
from {\it XMM\/} for R Cru  
 ($\log L_X = 29.8\,\rm ergs\, s^{-1}$; Paper~IV)
is larger than the quiescent level. The 
phase of the {\it Chandra\/} observation is 0.94 (using the period 5.825701$^d$
and the epoch of maximum light of JD 2455172.5100 from Usenko et al.\ [2014]), 
not the phase of X-ray maximum.  Thus the X-ray flux is larger than expected 
for the Cepheid, particularly at a phase other than maximum radius.  

There is, however, another possibility.  R Cru was observed with the CORAVEL 
radial-velocity spectrograph for 4 seasons, approximately 1996 July, and 1997 March,
July, and December (Bersier 2002).  The results are shown in 
Figure~\ref{rcruvr}, with different symbols for each season.  The velocities
for the third season (and to some extent the fourth) fall below the first 
two, particularly considering that a typical error for the velocities is 
$0.33\,\rm km\, s^{-1}$.  This indicates a systemic velocity change within a year, 
presumably due to orbital motion. This implies that there is 
a second companion candidate
in a close  orbit, which 
 would be unresolved from the Cepheid in the {\it Chandra\/} 
observation.  An orbit with a period near a year corresponds to a separation 
near one AU.

There is one further piece of information which gives insight into the 
R Cru system.  It was observed with {\it IUE\/} 
 (Evans 1992a), with the result  that any companion must have a later 
spectral type than A2, corresponding to a mass less than $2.5\, M_\odot$.

To summarize companion possibilities for R Cru, there are two reasonably 
close companions, one at a separation of 1580 AU (Table~\ref{com.sum}), and the 
spectroscopic binary companion (Figure~\ref{rcruvr})
at approximately 1~AU\null.  Note that these two 
would be dynamically compatible.  The {\it Chandra\/} observation in 
Figure~\ref{rcruchan} provides an indication favoring the closer companion, but this
is tentative.  However {\it Gaia\/} data (Kervella et al.\ 2019b) 
provide additional evidence that the candidate at 1580 AU is a physical 
companion, as discussed in Appendix~\ref{sys.2}.  
Therefore we continue to include it in Table~\ref{com.sum}. 
The {\it IUE\/} observation 
shows that any companion is only about half as massive as the Cepheid (or
less).

\begin{figure}[ht!]
\centering
\includegraphics[width=5in]{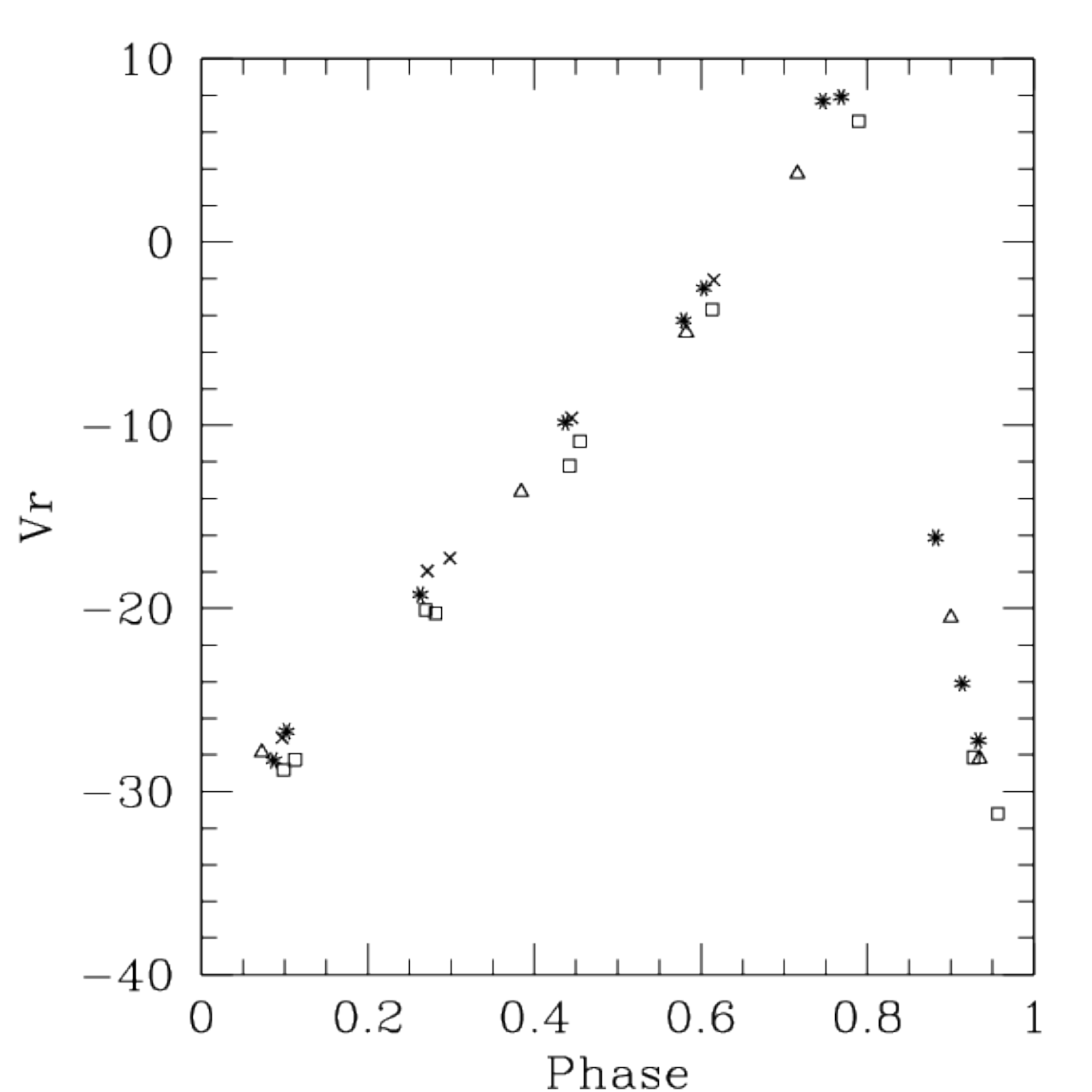}
\caption{CORAVEL radial velocities of R Cru.  The sequence of the four 
seasons of observations from first to last is: x's, asterisks, squares, 
and triangles (see text).
The displacement of the squares from x's and asterisks is particularly apparent.  Velocities are 
in km s$^{-1}$.
 \label{rcruvr}}
\end{figure}

\subsection{S Mus}

The {\it HST\/} and {\it XMM\/} images are shown in Figure~\ref{smusstart}, with the locations
of the Cepheid and the possible companion circled.  Figure~\ref{smuschan} shows
the {\it Chandra\/} image with the same two locations overlaid.  It has 
been mildly smoothed with the {\it CIAO\/} tool smooth.  

\begin{figure}[hb!]
\centering
\includegraphics[width=5in]{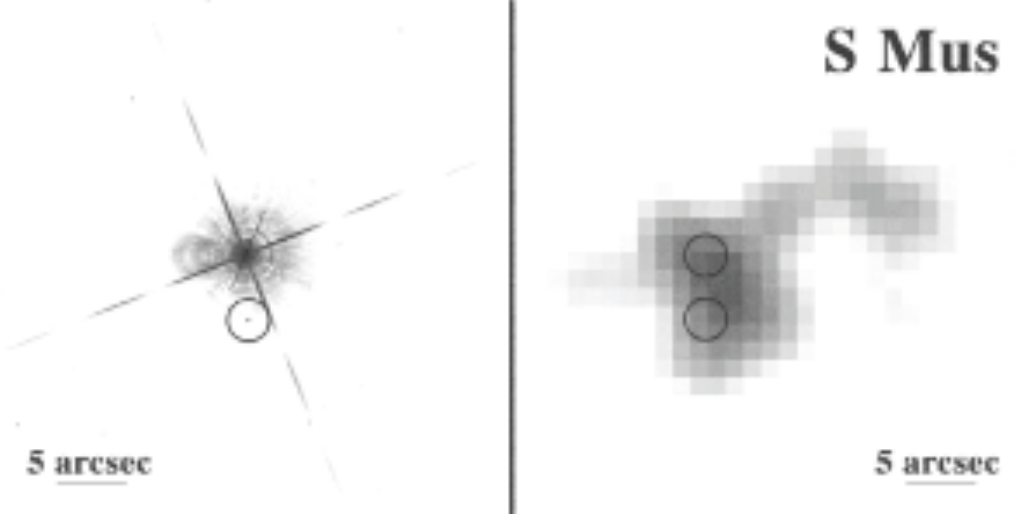}
\caption{Possible companion of S Mus.  Left: The {\it HST\/} 
WFC3 $I$-band image of S Mus.  The possible companion is circled.
 The scale is logarithmic to emphasize faint 
features.  Right: The {\it XMM\/} image, again with circles indicating 
the locations of the Cepheid and the possible companion (including
a small correction derived from the X-ray positions of 2MASS sources.)
The orientation of both figures is the same with N up and E 
on the left. 
 \label{smusstart}}
\end{figure}

\begin{figure}[hb!]
\centering
\includegraphics[width=5in]{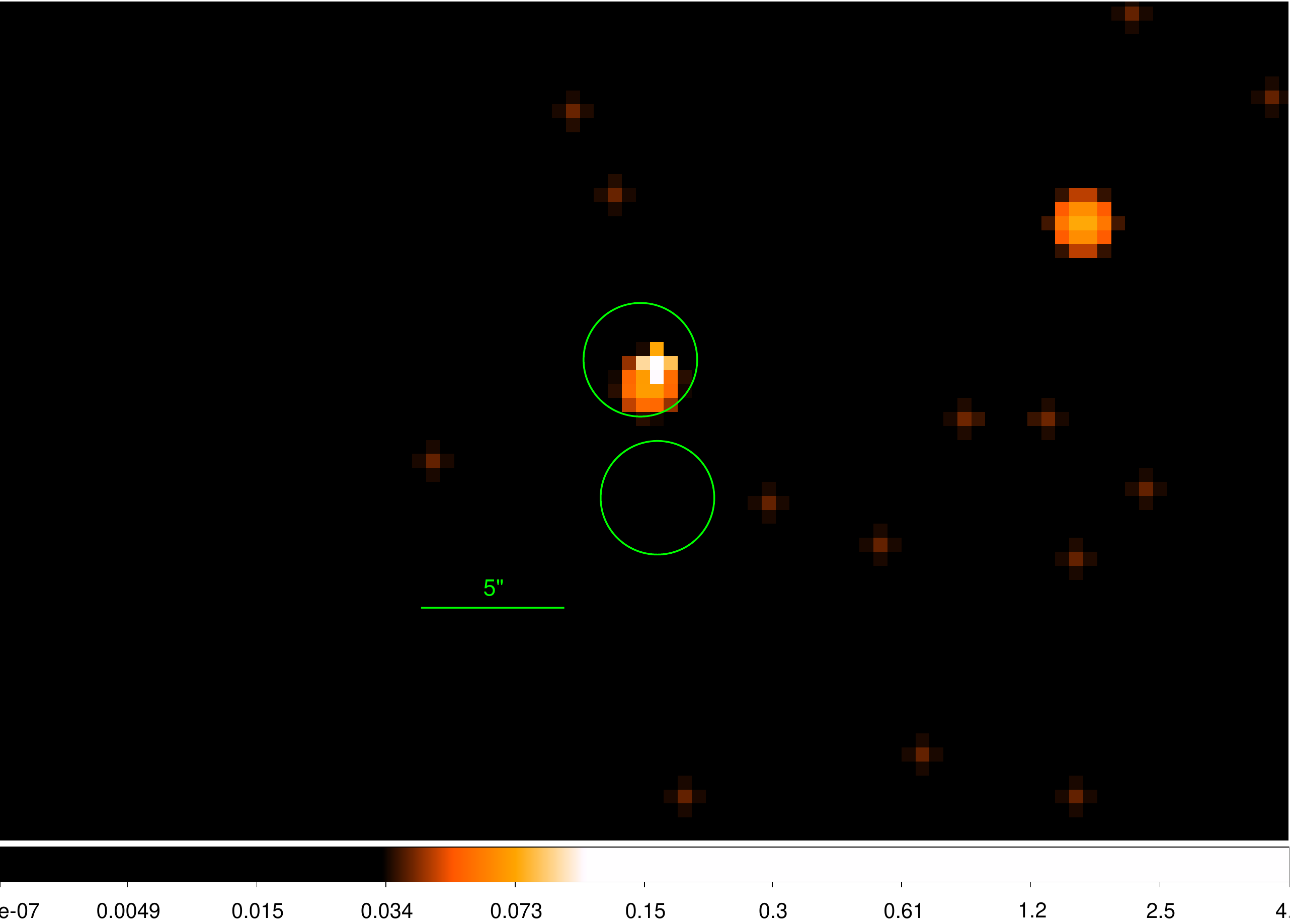}
\caption{The center of the {\it Chandra} image of S Mus.  The orientation is the same
as Fig~\ref{smusstart}.  The two circles (2$"$ radius) show the positions of the Cepheid
and the possible companion in  Fig~\ref{smusstart}.  The image counts are on a log scale 
and have been mildly adaptively smoothed.
 \label{smuschan}}
\end{figure}

 The {\it Chandra\/} image rules out the $5\farcs2$ (4540~AU) companion 
as  young, and hence it is not a  
physical companion of the Cepheid S Mus.  The Cepheid is a member of a 
spectroscopic  binary system with a period of 505~days ($a \sin i = 2.8$~AU; Paper~I).    
The companion is the hottest known Cepheid companion: B3~V (Evans et al.\ 2006).
Stars this hot produce X~rays through wind shocks
(Harnden et al.\ 1979), typically with 
$L_x/L_{\rm bol} = 10^{-7}$. Using $M_{\rm bol}$ for a B3 star of $-4.0$ 
(Drilling \& Landolt 2000), the luminosity is 
$1.2\times10^{37}$ ergs~s$^{-1}$.  Thus the X-ray 
luminosity of $\log L_X = 30.46$ (Paper~IV) is close to the expected ratio 
for a hot star.  
 The phase of the {\it Chandra\/} observation is 0.06, using the pulsation  
period and epoch of Petterson et al.\ (2004).  
Hence, S Mus is much brighter 
than the quiescent phases.  Thus the flux must be at least heavily dominated by 
the hot companion.

\clearpage

\section{Systems with a Resolved Companion Closer than $2''$}\label{sys.2}

In this Appendix we discuss the components in systems with a resolved close companion. The results are summarized in Table~\ref{mcomp}. 

 {\bf S Nor}:  Although the companion is slightly closer than 1$\arcsec$, 
the photometry in Table~\ref{com.sum}  from the WFC3 images is 
used.  From an {\it IUE\/} spectrum, Evans (1992b) found a B9.5 V companion, which has 
essentially the same $M_V$ as in Table~\ref{tab:prop.comp}, implying that it is 
the resolved companion.

Are there any other components
in the system besides the Cepheid and the resolved component?  
An inner binary in the S Nor system is likely but not proven. 
 Mermilliod, Mayor,  \& Burki
(1994) and Bersier et al.\ (2002) find little evidence of orbital motion in three 
well-covered seasons of CORAVEL data.  Groenewegen (2008)
found a low-amplitude orbit (2.53 km s$^{-1}$) with a period of 3584 days.  
However, the three seasons of CORAVEL data span 2700 days, so  that 
orbital period seems
unlikely (though not impossible).  Gallenne et al.\ (2019) added radial-velocity data spanning 1700 days, but were unable to find an orbital 
signature. 
Kervella et al.\ (2019a)  found differential proper motions between 
{\it Hipparcos} and {\it Gaia} marginally consistent with orbital motion.
Combined  with the orbit from  Groenewegen, they found a companion mass
of $1.5 \,M_\odot$, approximately an F0~V star. The orbit has 
a semimajor axis of 8.87~AU\null.  Gallenne et al.\ (2019)
observed the system with interferometry from the VLTI.   If the hot star
were the spectroscopic companion, it might have been marginally 
detectable but they found no detection at a level greater than 2.3 $\sigma$.
An F0 V star would not be expected to be detected.  
  (The resolved 
companion in Table~\ref{tab:comp.mag}  is outside the field.)
In summary, 
this confirms  that the resolved companion is probably the one dominating
the {\it IUE} spectrum. Estimated properties of the inner binary are in 
Table.~\ref{mcomp}.  Thus, there is evidence of a short-period binary; 
however an orbit is still uncertain, which  may be due to a low-mass
companion, the orbital inclination, or high eccentricity.  

  S Nor is a member of a cluster 
(NGC 6087) which may  affect either the formation scenario or its dynamical history.
 One of the possible companions 
in our WFC3 survey is an X-ray source 
(Paper IV; Table~\ref{mcomp}),  with a separation from 
the Cepheid of 15$\arcsec$  or 13,300 AU\null.  The spectral type is derived from the 
$(V-I)_0$ color, using the calibration of Drilling \& Landolt (2000).
However, since it has an 
unusually large separation from the Cepheid (Figure~\ref{comp.mult}), the conlusion in Paper IV was
that it is likely to be a chance alignment with a cluster star, similar to the case
of $\delta$ Cep.  Kervella et al.\ (2019b) find additional  companions
at 41,100 and 38,800 AU (Table~\ref{gcomp}), and note that many stars have similar 
proper-motion vectors.
 They do not confirm the companion discussed by Evans \& Udalski (1994)
at a separation of 32,000 AU\null.  The most likely companion properties 
for companions within 15$\arcsec$
 are summarized in Table~\ref{mcomp},
although  associated cluster members clearly complicate identification. 


 {\bf $\eta$ Aql}:
 Two companions to the Cepheid are known in this system. 
The companion in Table~\ref{com.mix} is illustrated in Paper I
and Figure~\ref{fig:EtaAql845}.
Gallenne et al.\ (2014a) used  VLT/NACO to measure the properties of 
the companion at 
$0\farcs65$, which they identify as an F1~V to F5~V star, corresponding
to $(V-I)_0= 0.37$ to 0.53 (Bessell \& Brett 1988).  
They measure  $H_0$ = 9.34.
Using the  calibration of Bessell and Brett, this corresponds to
$V_0 = 10.1$ to 10.5.  This is  good agreement with a $V_0 = 10.0$
(Table~\ref{com.mix}), even at a separation of only $0\farcs6$ from the Cepheid.  
At  a distance of 
273 pc, this corresponds to $M_V= +3.0$ to +3.3. 
In Tables~\ref{com.mix}   and 
\ref{mcomp} and Figures~\ref{cmd} and  \ref{sep}  
we substitute the color $(V-I)_0=0.4$ and $M_V=+3.2$
for the contaminated color and magnitude in Table~\ref{tab:comp.mag}

 There is a second, hotter companion detected in an {\it IUE\/} spectrum, 
with a spectral type of B9.8~V (Evans 1991). In Table~\ref{mcomp} the 
$V_0$ and $M_V$ are listed from that source.
 This second companion must be
closer to the Cepheid and 
unresolved by either Gallenne et al.\ or in this study.  However, neither it nor
the Cepheid itself affect the measurement of
 companion in Table~\ref{com.sum} using the 
resolved photometry of Gallenne et al.  
Unfortunately, {\it Gaia\/} data are poor  for a star as bright as $\eta$ Aql.
Benedict et 
al. 2018 reported a preliminary orbit for the inner companion based on high precision radial 
velocities of $\eta$ Aql.
Additional radial velocities (Barnes et al. 2021 in preparation) have not confirmed that orbit as 
yet.
We include in    
Table~\ref{mcomp} a very preliminary value of the semimajor axis
estimated from the masses, and a period of 4 years
which can at least be used  in the schematic distribution of 
separations.  While orbital information is sketchy, there is no 
doubt that there is a third star in the system, and that it 
is closer to the Cepheid than the resolved companion in 
Table 4.  Preliminary indications are that the orbit is nearly face-on and 
possibly eccentric.





{\bf V659 Cen:} As can be seen from the discovery image (Paper I,
 Figure 1 and Figure~\ref{v659cen845}),
 the companion in that image lies on a diffraction 
spike (in addition to being very close to the Cepheid), 
so the photometry is unreliable.  
However, there is information about the hottest companion from 
an {\it IUE\/} spectrum (Evans 1992b), in which the companion is found to be 
a B6.0 V star.  This results in a corrected $E(B-V)$ for the Cepheid 
of 0.21 mag, and an absolute magnitude for the companion of $-0.32$ mag. 
A mild extrapolation from Table 15.11 of Drilling \& Landolt (2000)
provides $V-I = -0.16$. 

The system was also resolved by {\it Hipparcos\/} (HIP 65970), with very similar 
parameters:  PA  $234^\circ$, separation  $0\farcs574$.

To search for a possible inner companion,
radial velocities were investigated.
Possible orbital motion for V659 Cen  
must be small, and Lloyd Evans (1968) rated it as  questionable at best. 
Lloyd Evans (1982) added new velocities to previous data, and concluded that the 
velocity differences between three seasons were probably not significant. 
However, R. I. Anderson 
(in preparation 2020) has identified orbital motion in 
a more recent series of observations.  Kervella et al.\ (2019a) investigated 
{\it Gaia} DR2 and {\it Hipparcos} proper-motion anomalies as indications 
of orbital motion.  They found marginal evidence in the  {\it Hipparcos}
data.  In the {\it Gaia} data there is no significant anomaly, but the error
in the proper-motion vector is unusually large, which could be due 
to orbital motion, particularly of a close companion. Based on this we 
have added a tentative entry in Table~\ref{mcomp} and Figure~\ref{comp.mult} 
of 3 AU, reasonable for an orbit of a few years.  
Pending confirmation of orbital motion, V659 Cen would have 
both a resolved companion and a closer spectroscopic companion.
Just prior to submission, we obtained an {\it HST} STIS G140L  spectrum oriented to separate the resolved companion from the Cepheid.  This spectrum shows that the resolved companion is the hottest star in the system, with the spectroscopic binary companion being much cooler. 
Entries in Table~\ref{com.mix} are based on 
the {\it IUE\/} spectral type.  Kervella et al.\ (2019b) find 
an additional very wide (62$\arcsec$)  gravitationally bound low-mass (M3~V) companion.

{\bf AX Cir}:  This is  the closest companion 
to a Cepheid identified here, and has 
particular problems
for that reason.  It is clear from the F845M image, however,  that there is a star partly
obscured by the bleed column and a diffraction spike.  For this reason, the F621M
filter  is not usable, hence a color is not available.  In the discussion below, we 
use the results from an {\it IUE} spectrum for further information
about the companion.  

The AX Cir system has been resolved 3 times:  2018 (Tokovinin et al.\ 2019),  1991.25 
(HIP 72773), and 1913 by Innes.

AX Cir is also a member of a binary system (Petterson et al.\ 2004).  
An \IUE\/ spectrum 
shows that the hottest star in the system is a B6 V star (Evans 1994). The 
system was resolved with  VLTI/PIONIER interferometry (Gallenne et al.\ 2014b)
with a projected separation of 29.2 $\pm$ 0.2 mas, which corresponds 
to a separation of 15.4 AU at a distance of 527 pc. The separation 
of the wider companion ($0\farcs3$ in Table~\ref{com.sum}) corresponds
to 158 AU. An {\it HST} STIS spectrum 
(Gallenne, Evans, Kervella et al.\ 2020 in preparation)
 showed that the B6 V star is the wider companion.  From the STIS 
spectrum the spectral  type of the close spectroscopic companion was found to be B9 V. 
A tentative velocity measure of the close companion from the 
STIS spectrum indicates that the most likely interpretation is that the inner 
companion is itself a short-period binary, making a total of four components 
in the system.  
Kervella et al.\ 2019a find a strong indication from {\it Gaia\/} 
and {\it Hipparcos\/} proper motions of an anomaly indicating a short-period orbit of the close companion.  Combining this with the Petterson
orbit and an estimated mass for the Cepheid, they find a companion 
mass larger than the Cepheid mass, confirming that the companion 
is itself a binary.  Kervella et al.\ (2019b) also find a wider ($81\farcs5$ 
= 43000 AU) likely gravitationally bound  system member, which is
a low-mass M3.5~V star (Table~\ref{gcomp}).   
As with V659 Cen, we include the $M_V$ and $(V-I)_0$ inferred from the {\it IUE\/}
spectrum in Table~\ref{mcomp}

{\bf R Cru}:  There are three companion candidates to this system.

1. The $M_V$ and $(V-I)_0$ in Table~\ref{com.mix} correspond to an early K star. 
(The WFC3 photometry for a companion 
at this separation is much less likely to have
 contamination from the Cepheid than closer companions.) 

2. To check for additional system members, radial velocities were investigated as 
 discussed above.  The orbital motion in Figure~\ref{rcruvr}  
  partly guided by the position of the X-ray
source (Appendix~\ref{chandra}) provides evidence for a companion inside the one identified
in this study (Table~\ref{tab:prop.comp}).  
Kervella et al.\ (2019b) also find 
find marginal evidence in 
{\it Gaia} and {\it Hipparcos} proper motions
of short-period orbital motion,  
 which would be consistent with the radial-velocity orbit.  
As a preliminary estimate of the separation for an inner
companion, orbital motion within a year suggests a period of a couple of years. 
We have used this approximation to add this companion to Figure~\ref{comp.mult}
and Table~\ref{mcomp}.
 We know from {\it IUE\/} observations (Evans 1992a) that the companion in the
system is cooler than A2,  corresponding to a mass
of $2.7 \,M_\odot$.  

3.  We have also reinstated the  $7\farcs6$ companion
from Paper II
for the following reasons. Kervella et al.\ (2019b) find that it is at the 
same distance as the Cepheid using  {\it Gaia} data and is a 
gravitationally bound companion (spectral type G8 V).  Possible orbital motion
is detected. 
We note that there are X-ray counts
from {\it XMM} in the 
position of that companion in Paper IV, although it is not the most prominent
source in the field.  The {\it Chandra} image (Figure~\ref{rcruchan}) is a 
shallower image, hence this companion is not detected.

 {\bf U Aql}:
  In Appendix~\ref{appuaql} there is extensive new 
  material about U Aql, for which we provide a detailed
  discussion. 
  There are three companions
to the Cepheid in this system.  
 $M_V$ for  the companions is from Table~\ref{ucomp}.
 The spectral type of the $1\farcs6$
companion is  estimated to be A5 V from the STIS spectrum.
 We use the calibration of Drilling \& Landolt (2000) of 
$(V-I)_0  = 0.16$.  This is in agreement with the value in 
Table~\ref{res.sum} for a companion with a separation of $1\farcs6$.
Component B (close) has an estimated projected separation of 66 AU 
and spectral type of A3-4~V (Appendix~\ref{appuaql}).

 The $1\farcs6$ pair of U Aql has been measured twelve times 
 since the first resolution by Kuiper in 1934. 
  It has always been at about the same
 relative position.

The U Aql spectroscopic binary was resolved at three epochs with 
the VLTI PIONIER combination (Gallenne et al.\ 2019).  Using a 
distance of 592 $\pm$ 19 pc, they determined preliminary masses for 
the Cepheid and the spectroscopic companion of $6.2\pm 0.8$ and 
$2.2 \pm 0.2\, M_\odot$, respectively.  This is somewhat smaller 
than the mass inferred from the B9.8~V spectral type using the 
Drilling \& Landolt calibration. However it agrees with the 
Harmanec (1988) calibration within the errors (decreased 
by 0.02 in log M since the companions of Cepheids will be younger 
than average main-sequence stars of their spectral type 
[Evans et al.\ 2018]). A strong anomaly 
was detected in the {\it Gaia} and {\it Hipparcos} proper motions
 (Kervella et al.\ 2019a).  
The inferred separation and companion 
mass (assuming a Cepheid mass) are in agreement with those of 
Gallenne et al.\ (2019).
  The derived separation  in Table~\ref{mcomp} is the combined 
interferometric separation from Gallenne et al.\ and the 
distance from Paper II.
No wider resolved companions were 
detected in the  {\it Gaia} data (Kervella et al.\ 2019b).

 {\bf U Vul:}
The $M_V$ of the U Vul companion in Table~\ref{com.mix} corresponds to a K0 star
(Drilling \& Landolt 2000).  
Another system member is indicated by 
an orbit for U Vul which was  published by Imbert (1996),
which has  a high eccentricity and low amplitude.
Kervella et al.
(2019a) found a significant proper-motion anomaly 
 in the {\it Gaia} and {\it Hipparcos} data.  Assuming 
a Cepheid mass, they found a mass for  
the spectroscopic companion of $2.4 \pm 0.4 \,M_\odot$
and a semimajor axis of 7.1 AU\null.
The upper limit of spectral
type from an {\it IUE\/} spectrum is A1 V corresponding to 
a mass of $2.5 \,M_\odot$, which is in agreement with (and very 
close to)  the Kervella
mass within the errors.  Note that this 
upper limit pertains to any companion in the system, both
the spectroscopic companion and the resolved companion 
 (Table~\ref{com.sum}). 
 No wider resolved companions 
were found in the {\it Gaia} DR2 data (Kervella et al.\ 2018b)


\section{U Aql}\label{appuaql}


The quest for masses and luminosities for Cepheid
variables has benefited greatly from the 
availability of the ultraviolet spectrum using 
satellites ({\it IUE} and {\it HST}). Through these studies, 
the picture of multiplicity has become 
increasingly complex.     
 U Aql = HD 183344   is a good case in point.  Its substantial 
orbital motion was only recognized in 1979 (Slovak et al.\ 1979).  
Welch et al.\ (1987) 
provide an orbit, as well as a summary of previous 
velocity information. The spectrum of the hottest
star in the system  dominates in the ultraviolet
below about 2000~\AA, and  {\it IUE} observations have  
been discussed by B\"ohm-Vitense
\& Proffitt (1985) and Evans (1992).
These provided a temperature of 9300 $\pm$ 100 K
and a spectral type of B9.8 V respectively.  

\subsection{Companions}

New insight into the system came with {\it HST} Faint Object 
Camera (FOC) and  Space Telescope Imaging Spectrograph (STIS)
images (PI: D. Massa). The STIS image was made with the G230L
grating and the STIS NUV-MAMA detector.  A series of spectra 
were taken at varying roll angles and times in the orbit.
Similar data are fully discussed for AW Per (Massa and Evans 2008)
and W Sgr (Evans, Massa, \& Proffitt 2009).
One of the images is presented as an example in  Figure~\ref{uaql.img}. 
 Unexpected complexity in the system is immediately 
evident.  The spectroscopic binary is the bottom spectrum,
with the Cepheid Aa as the brightest star in the long-wavelength
region and the companion Ab as the brightest star in the 
short wavelength region.  At the top is the resolved 
companion C found in the {\it HST} WFC3 imaging (Table 2). Between 
them, but very close to the spectroscopic binary is a 
previously unsuspected companion B.     The components are 
summarized in Table~\ref{mcomp}.  In addition to the separation of C
from Table 2, the projected separation of B has been measured
from the STIS image.
The separation in the spectroscopic binary Aa and Ab has been taken 
from the interferometry of Gallenne et al. (2019),         10.06$\pm$0.16 mas.
Using the distance from Paper II (613 pc), this becomes 6.2 AU.

  Companion B may be the interferometric
  pair resolved three times by Ismailov (1992).
 However, it was not resolved by
speckle at SOAR in 2008 and 2016 (Tokovinin et al.\ 2010 and 
2018c respectively).


Spectra were extracted from the STIS image as described in 
Evans et al.\ (2009).  
 The U Aql observations were done with the G230L and the unfiltered 25MAMA
aperture.   For the PSF comparison star we used 
 the data set o6hr01040, an observation
of the hot  subdwarf GRW+70D5824. This exposure 
has relatively high S/N and was taken
in February 2002, only a few months before the first of the U Aql observations.
The image of U Aql which was analyzed was o6f101020, selected because
it has the best projected separation between components.
The projected separation between the spectroscopic binary with 
the Cepheid and the closest companion is 0.107$\arcsec$ using a plate
scale of 0.0247$\arcsec$/px, which corresponds to 66 AU using the distance
from Paper II.
The results  are shown in  Figure~\ref{stis.spec}.  The Cepheid 
dominates in the long wavelength section, but the companions
dominate at the short wavelength end.   Components B and C are
resolved and the Cepheid makes no contribution to the spectrum. 
For the spectroscopic binary Aa and Ab, however, in the region
near 1800\AA\/
there is still a contribution from the Cepheid, although the 
companion Ab dominates.  In this section we estimate the 
spectral types of all 3 companions, and we correct for 
the contribution of the Cepheid to the spectrum of Ab as 
follows.  We use the {\it IUE} spectra of standard stars to create
a spectrum B9.8 V (midway between B9.5 V and A0 V) 
 as in Evans (1992b). (The B9.8 V spectrum has been
reddened to match the U Aql spectrum and then scaled for 
comparison.)  Figure~\ref{iue.spec} shows that the B9.8 V spectrum 
represents the U Aql companion Aa well at wavelengths shorter 
than 1500 \AA, however the additional contribution from the 
Cepheid can be seen at longer wavelengths.  Therefore we  will 
use the flux from the B9.8 V spectrum to estimate the magnitude
difference between the three companions in the U Aql system.

For all three companions in the U Aql system (Figure~\ref{stis.spec.lg}) 
the mean flux from 1650 to 1950  was created 
(Table~\ref{f18}) and from these the magnitude difference between the 
companions and the U Aql spectrum was created.  At the bottom of 
Table~\ref{f18} the fluxes from the {\it IUE} spectra are given.  
From the difference between the B9.8 V spectrum and the 
composite U Aql spectrum (Aa + Ab at 1800\AA), a correction of 
0.15 mag was derived to account for the contribution to 
the U Aql spectrum from the Cepheid.  The far right column 
for the Far (C) and Close (B) companions from the STIS spectra 
includes this correction.  Note that the {\it IUE} spectrum 
of the U Aql system  includes all three companions 
although the hottest star in the system (Ab) is only affected 
in a minor way at the shortest wavelengths.

The spectral type of companion Aa was derived in Evans (1992b).  
To estimate the spectral type of 
the magnitude differences (Table~\ref{f18}),  Table~\ref{m18}
shows the magnitude differences for main sequence spectral
types from the {\it V}--ultraviolet colors (column 2) from Wesselius et al
(1980) and the absolute magnitude calibration (column 3) from Drilling
\& Landolt (2000) for the ZAMS, which is essentially the same as 
used in Evans (1991, 1992b).  The Far component C is approximately A5 V, and 
the Close component B is A3-4 V, corresponding to $M_V = +2.1$ and +1.8, 
respectively.  This magnitude for C is very similar to that from the 
WFC3 value (Table~\ref{com.mix}).   


\begin{figure}
\centering
 \includegraphics[width=3in]{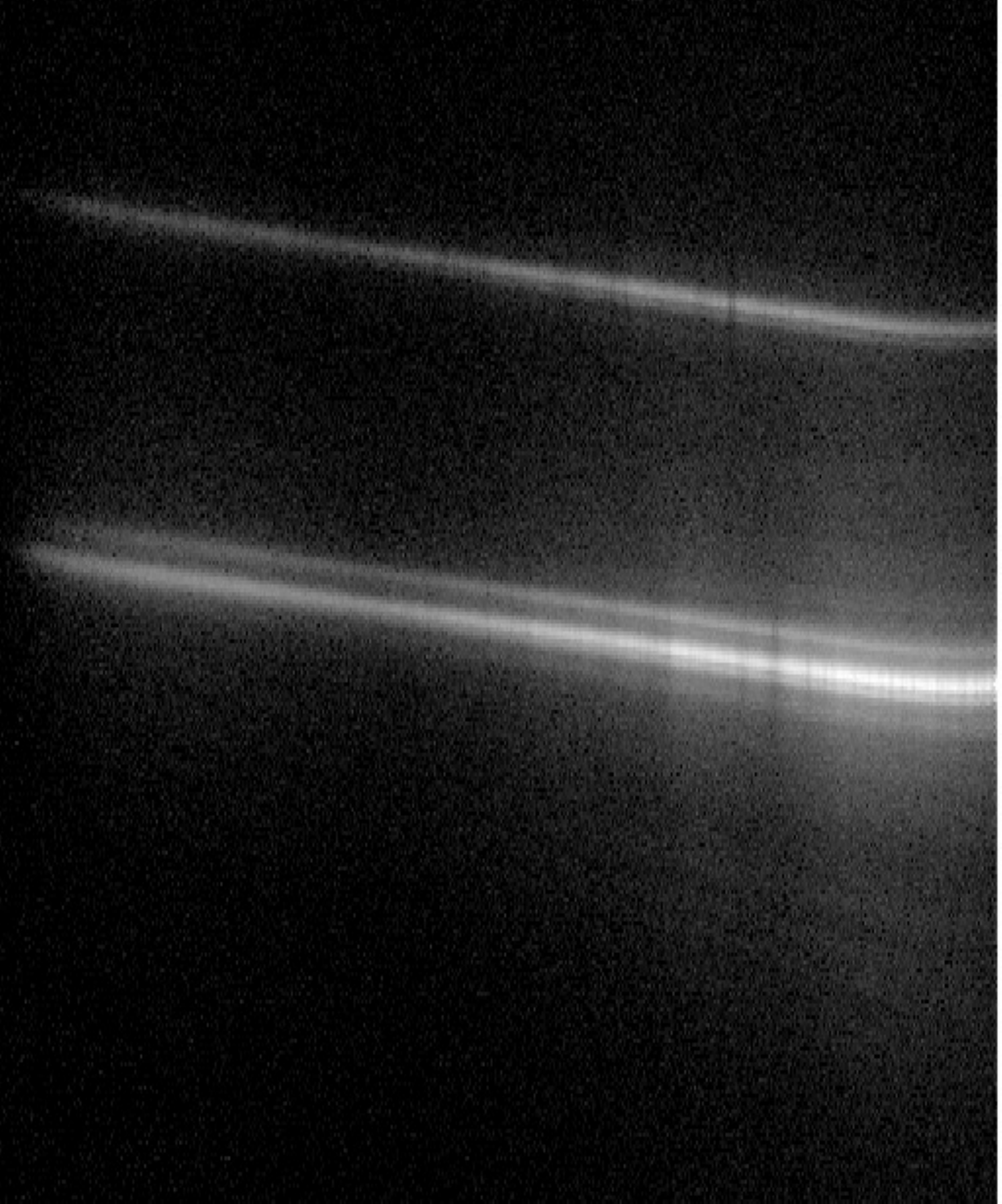}
\caption{A flat-fielded image o6f101020 of the U Aql system.
The wavelength increases to the right from about 1796 to 3382 \AA.
The lower component has the Cepheid (Aa; brightest at long wavelengths)
and also the hottest star in the system (Ab; brightest at short
wavelengths).  The component at the top (C) is the one 
identified in the {\it HST} imaging (Table~\ref{tab:comp.mag}).  A third component
(B) is between them, close to the spectroscopic binary containing
the Cepheid.  The spectrum is on a log scale.  The strongest 
feature in the Cepheid spectrum is 2800 \AA. 
 \label{uaql.img}}
\end{figure}

\begin{figure}
\centering
\includegraphics[width=3in]{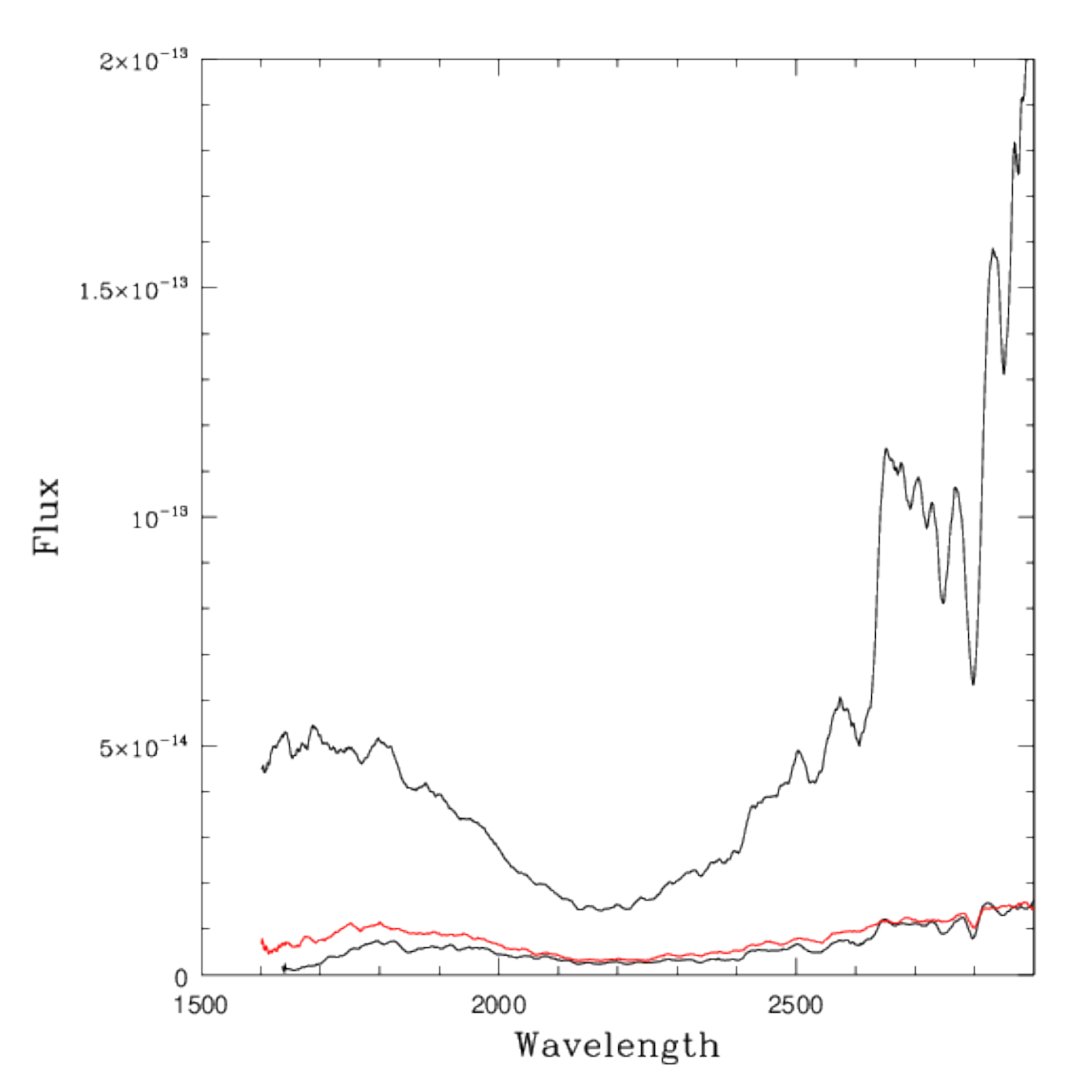}  
\caption{The extracted spectra for the U Aql system.  The 
brightest spectrum is the spectroscopic binary containing 
the Cepheid and the hottest companion.  The faintest 
spectrum is the wide companion.  The spectrum in red is the close
resolved companion.
 \label{stis.spec}}
\end{figure}

\begin{figure}
\centering
 \includegraphics[width=3in]{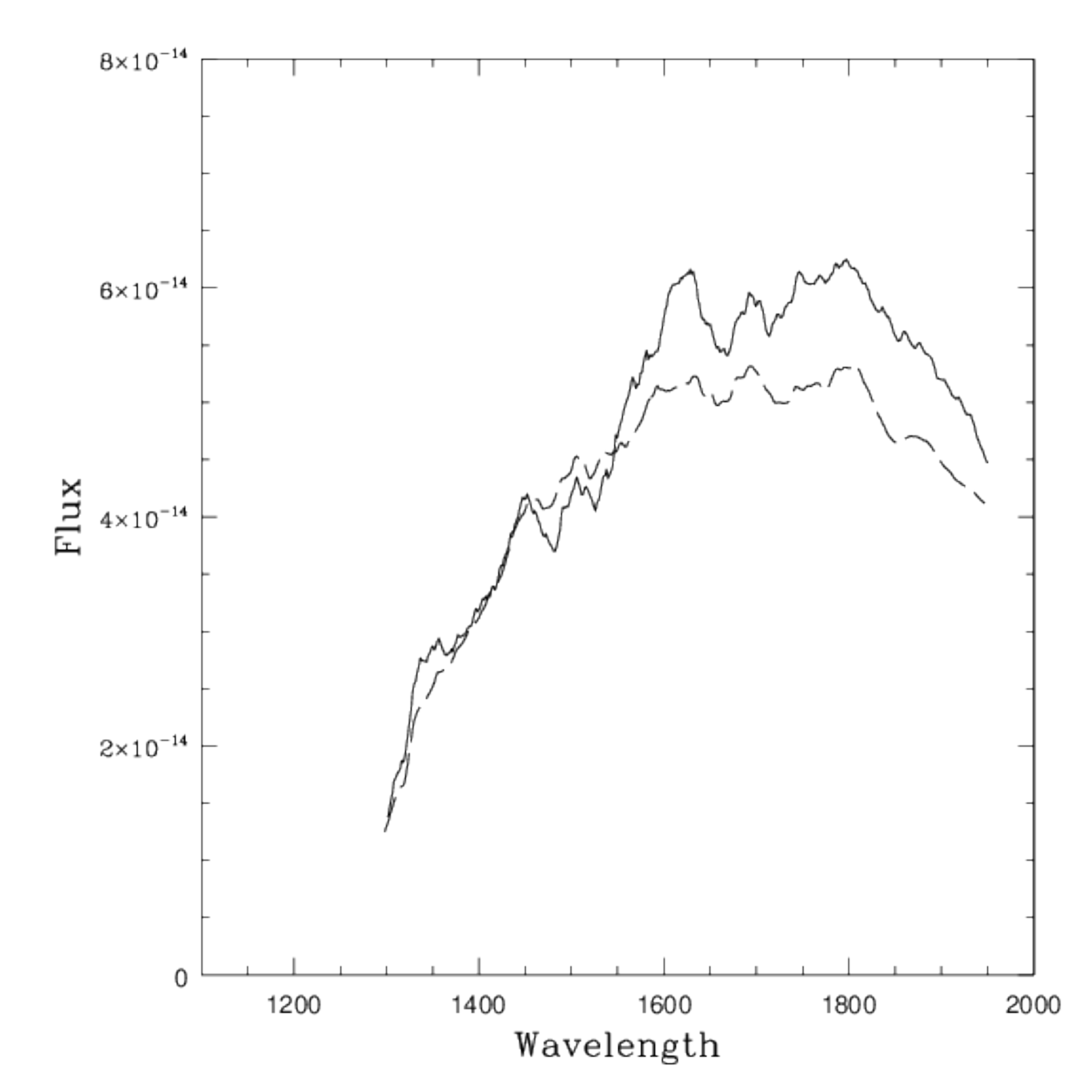} 
\caption{ Comparison of the {\it IUE} spectra of U Aql and B9.8 V. 
The solid line is the U Aql spectrum; the dashed line is the
B9.8 V spectrum.
  \label{iue.spec}}
\end{figure}

\begin{figure}
\centering
 \includegraphics[width=3in]{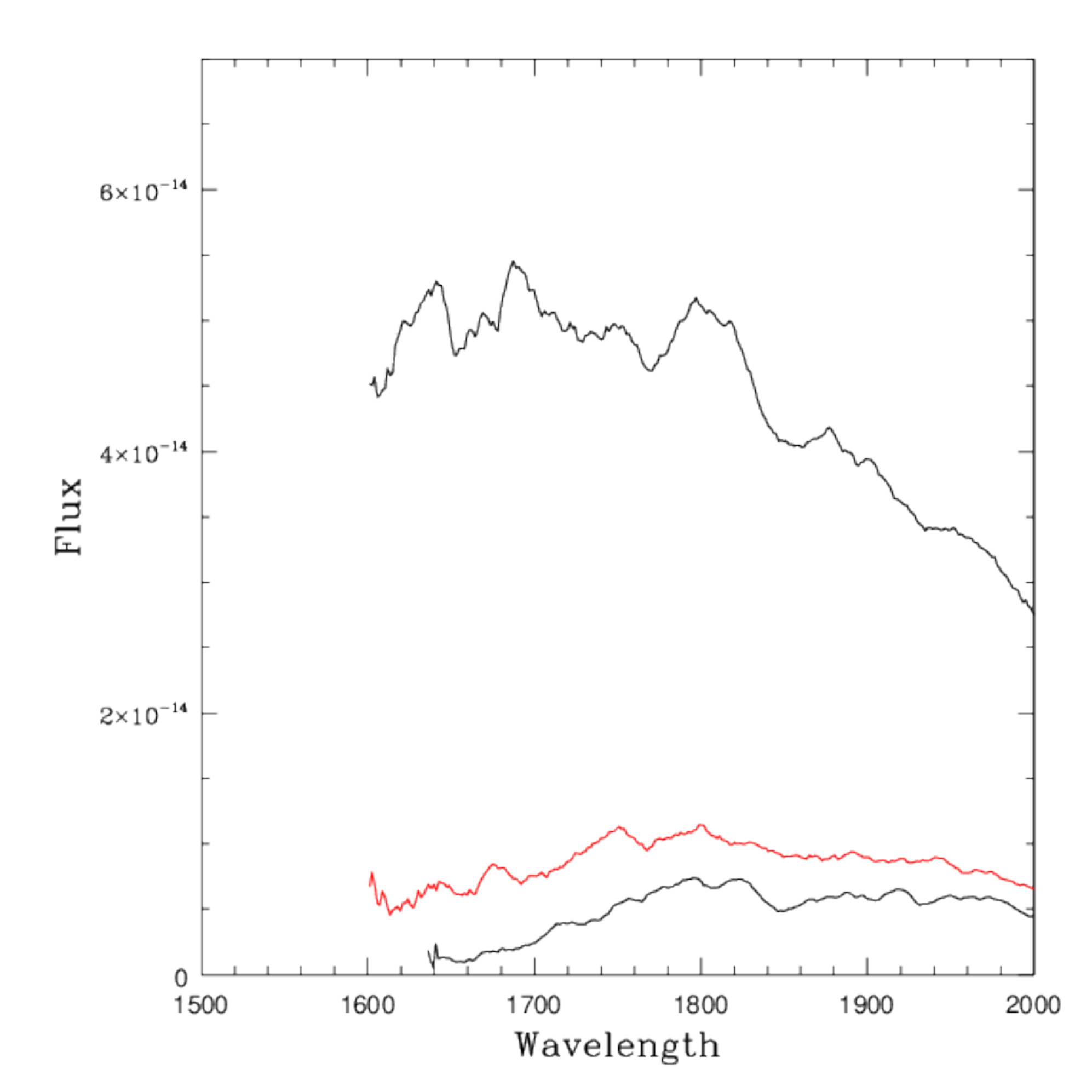} 
\caption{The three companions in the U Aql system
in the short wavelength region.  As in 
Figure~\ref{stis.spec}, the spectrum in red is from the close
resolved companion (B).  The spectroscopic binary is represented
by the B9.8 V spectrum.
  \label{stis.spec.lg}}
\end{figure}

\begin{deluxetable}{lrrrr}
\footnotesize
\tablecaption{Components of the U Aql System  \qquad\qquad\qquad\qquad \label{ucomp}}
\tablewidth{0pt}
\tablehead{
\colhead{ID}
  & \colhead{Sep }  & \colhead {Sep }    & $M_V$  & SpTy   \\
\colhead{}
  & \colhead{ $\arcsec$}  & \colhead { AU}    & mag  & 
}
\startdata
  Ab  &  0.0101     &   6.2        &    1.2  & B9.8 V     \\
  B    &  0.107      & 66 &      1.8    &   A3-4 V     \\
 C  &     1.6   &    981      & 2.1   & A5 V    \\
  \enddata

\end{deluxetable}




\begin{deluxetable}{lccc}
\footnotesize
\tablecaption{Flux at 1800 \AA  \label{f18}}
\tablewidth{0pt}
\tablehead{
\colhead{Star}  & \colhead{ f(1800)}  & \colhead {$\Delta$ M(1800)} &
\colhead{Corr}   \\
 \colhead {}  
 & \colhead{ ergs cm$^{-2}$ s$^{-1}$} $\AA^{-1}$  
 & \colhead {wrt U Aql} & 
}
\startdata
\noalign{\medskip}
\multispan4{\hfil \HST/STIS \hfil} \\
Far: C  &  $5.01\times 10^{-15}$   &    2.39   &        2.24 \\           
Close: B &    $9.14\times 10^{-15}$   &   1.74   &        1.59  \\
U Aql &   $4.52 \times10^{-14}$   &         &        \\         
\noalign{\medskip}
\multispan4{\hfil \IUE\/ \hfil} \\
U Aql &    $5.64 \times 10^{-14}$  &   &  \\
B9.8 V  &     $4.89 \times 10^{-14}$  & &  \\
\enddata
\end{deluxetable}

\begin{deluxetable}{lcccc}
\footnotesize
\tablecaption{Energy Distributions of Companions \label{m18}}
\tablewidth{0pt}
\tablehead{
\colhead{
Spec.\ Type}  & \colhead{ M(1800 -{\it V})}  & \colhead {$M_V$} &
\colhead{M(1800)}  & \colhead{M - M(B9.8)(1800)}   
}
\startdata
B9 V   &  $-0.90$ &   +0.9  &  +0.0 & \\
B9.5 V & $-0.70$ &   +1.1 &  +0.4 & \\
B9.8 V & $-0.60$  &  +1.2 &  +0.6 &  \\
A0 V  & $-0.51$  &  +1.3  &  +0.8 & 0.2    \\
A1 V &  $-0.21$  &  +1.4  &  +1.2 &  0.6   \\
A2 V &  0.04  &  +1.5  &  +1.5  & 0.9      \\
A3 V &   0.27  &  +1.7  & +2.0  & 1.4   B \\
A4 V &   0.46  &  +1.9 &  +2.4 &  1.8   B \\
A5 V &   0.68  &  +2.1  & +2.8 &  2.2 C \\
  \enddata

\end{deluxetable}


\subsection{Velocities}

Previously Evans et al.\ (1998) obtained spectra with the 
Goddard High Resolution Spectrograph (GHRS) to measure 
the orbital velocity amplitude of the spectroscopic binary
companion with the medium resolution grating G200M.  
The velocity difference between the two is -29.1 km s$^{-1}$
with an estimated error of 3.8 km s$^{-1}$.
Combining this with the orbital velocity 
amplitude of the Cepheid (Welch et al.\ 1987) and the 
mass of the main sequence companion they derived a mass 
for the Cepheid of $5.1 \pm 0.7 \,M_\odot$.

More recently, a  spectrum was obtained with STIS\null.
The spectrum used the E230H grating, and was taken over 2 days in 2003:
August 1 (exposure time 12109 sec) and 14 hours later on August 2 
(exposure time 7375 sec). Using the orbit of Welch et al.\ (1987) the
orbital phase is 0.44. 
The data used are shown 
in   Figure~\ref{d1.d2}.  A portion of the spectrum was selected
where the blaze correction is satisfactory and there are spectral
features.  
It was used in two ways.  First the spectra for the two days 
were cross correlated to look for a velocity difference which 
could be caused by motion in a short period binary orbit.   
In addition the STIS spectrum was cross correlated with each 
of the GHRS spectra to confirm the orbital velocity 
amplitude in the 5 year orbit

Although 
the spectrum on each day was weak, cross correlation of these 
two spectra provided the opportunity to check for possible 
orbital motion from a short period binary.
The spectral lines 
are reasonably broad, which is not surprising for a late B 
main sequence star, resulting in a broad velocity peak.
However Figure~\ref{d1.d2.gauss}
shows no evidence of a change in velocity which would be an
indication of short term binary motion.  


This is in agreement with the mass of the secondary in Kervella et al.\ (2019a)
incorporating {\it Gaia} and {\it Hipparcos} proper motions with the orbit.
The mass is appropriate for a B9.8 V star, and would not accommodate 
an additional star.

To check the result from the two GHRS spectra (Evans et al.\ 1998), 
the STIS spectrum was cross correlated with each of the GHRS 
spectra.  The result (Figure~\ref{v1.v2.gauss}) showed a similar 
shift between the two orbital phases to that found previously.
On the other hand, it is possible that the STIS spectrum 
is as much as 30 km s$^{-1}$ smaller than both the GHRS 
spectra.  In particular, the STIS spectrum should have a 
very similar orbital velocity to GHRS Visit 1.  

In summary, the STIS spectra taken a day apart do not 
show any velocity signal that the companion is 
 a very short-period binary.  Comparison of the STIS 
and GHRS velocities (Figure~\ref{v1.v2.gauss}) leaves open
the possibility of a longer period binary with smaller 
orbital velocity.  However, because of the broad lines, 
and hence the broad correlation, this is only a possibility.  

\begin{figure}
\centering
 \includegraphics[width=3in]{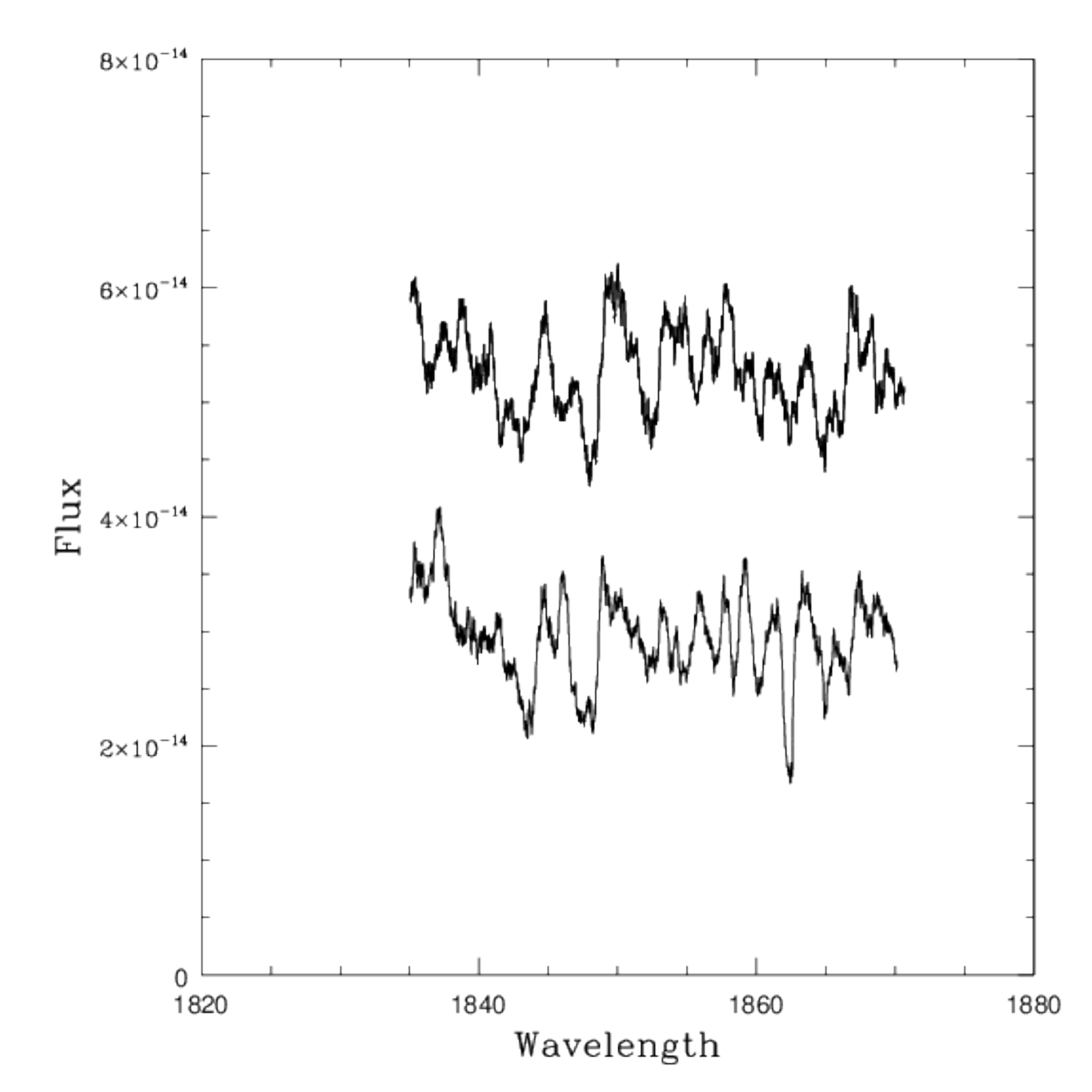}  
\caption{The spectral region used in this analysis.  The bottom is 
the spectrum from day 1 (see text); the top is the spectrum from day 2 
shifted by 0.2 10$^{-14}$ in flux for clarity.  Both spectra have been 
smoothed by 100 points.
 \label{d1.d2}}
\end{figure}





\begin{figure}
\centering
 \includegraphics[width=3in]{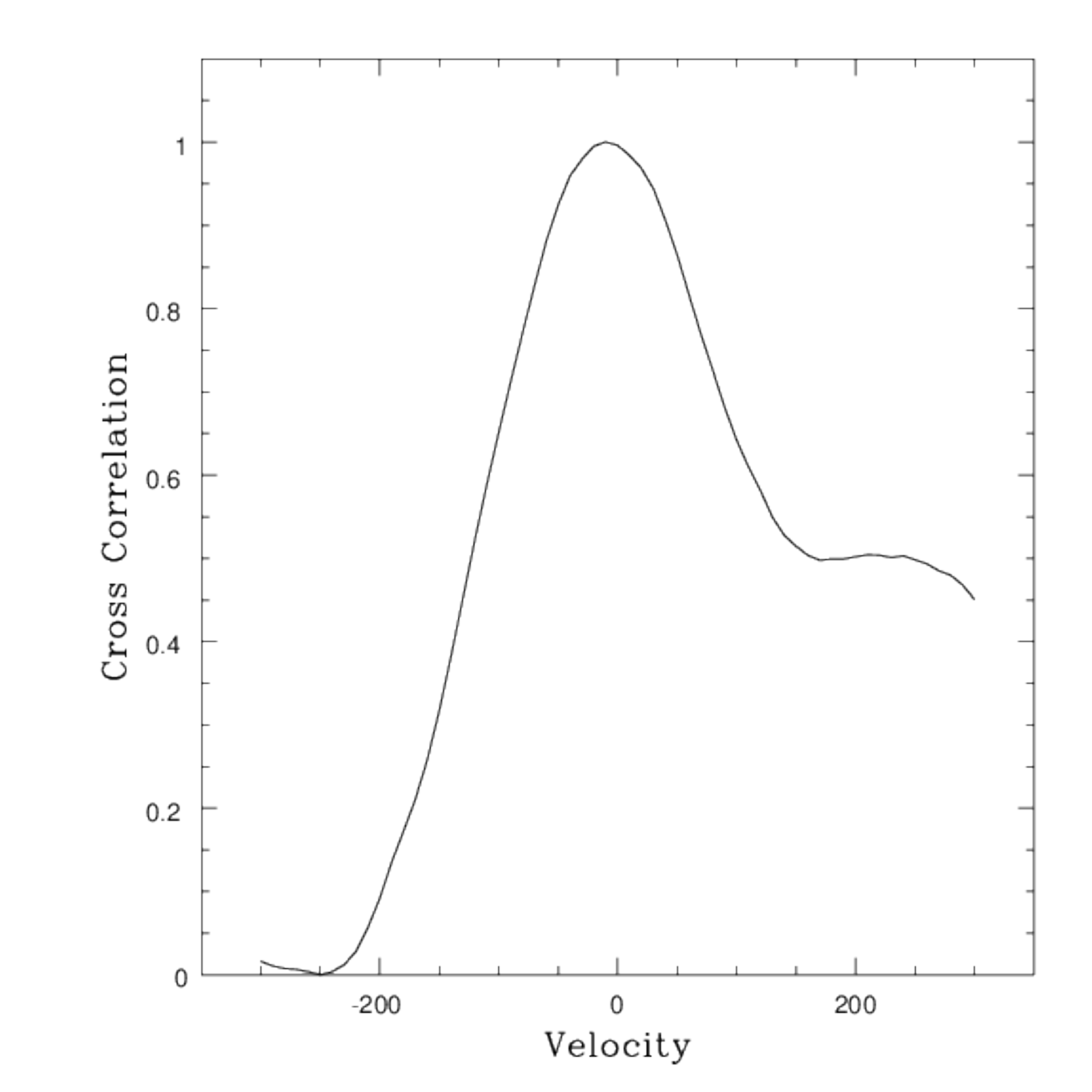} 
\caption{The cross correlation between the exposures on the 
day 1 and day 2.  
 \label{d1.d2.gauss}}
\end{figure}

\begin{figure}
\centering
 \includegraphics[width=3in]{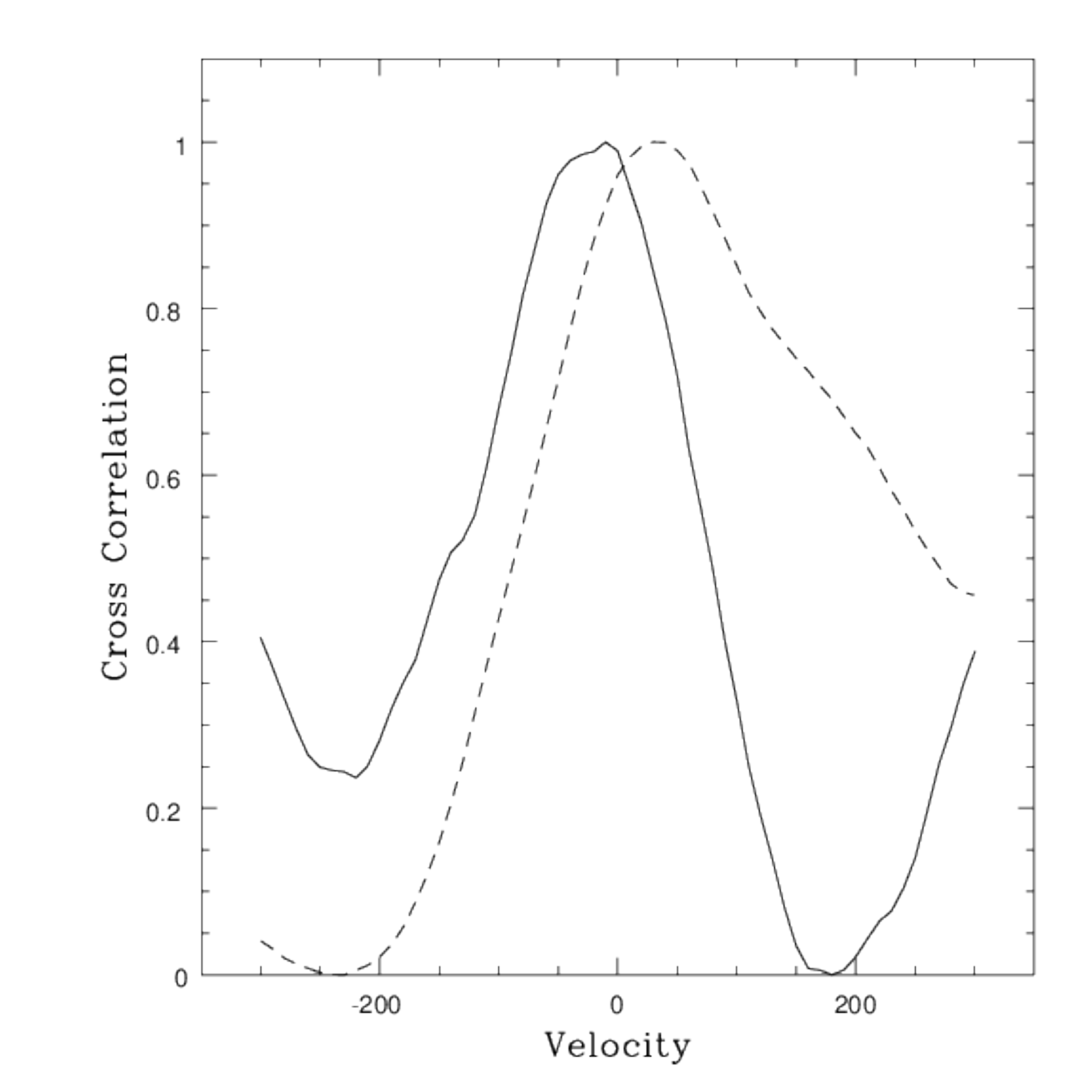}  
\caption{The cross correlation of the STIS spectrum with 
the two GHRS spectra.  Solid line: the velocity of the GHRS
Visit 1 spectrum with respect to the STIS spectrum; dashed 
line: GHRS Visit 2 with respect to STIS.  The program 
returns velocities that are 
in the sense that the velocity of the companion has 
increased between Visit 1 and Visit 2.
 \label{v1.v2.gauss}}
\end{figure}

\subsection{Discussion} The STIS image (Figure~\ref{uaql.img})
provides an explanation for a surprisingly large discrepancy between 
an absolute magnitude for the Cepheid from the {\it IUE}
spectrum of the companion and the Leavitt Law 
(Period-Luminosity relation) of Feast and Walker (1987).  As 
discussed in Evans (1992b) typical differences between the 
{\it IUE} absolute magnitudes and those of Feast and Walker
are 0.2 mag for well exposed companion spectra with well
known reddening.   For U Aql the difference was more 
than twice that, even though both conditions were met.  
However, the addition of the flux from the two additional 
companions which were unknown at the time (but would have been 
in the {\it IUE} aperture)
accounts for the discrepancy. 

\clearpage


\section{Companions in Systems with a Resolved Wide Companion 
(Wider than $2''$)}\label{sys.5}

In this Appendix we discuss the components in systems with a resolved wide companion. The 
results are summarized in Table~\ref{mcompw}.


{\bf FF Aql}:  The interpretation of the companion in Table~\ref{com.sum} from the
ground has been ``challenging''.  Udalski and Evans (1993) concluded 
that it is not a physical companion.  However the WFC3 measures (Paper II)
 found the photometry consistent with a physical companion.  The 
Washington Double Star Catalog (WDS) finds no relative motion between 
the companion and Cepheid, indicating that they are moving together.  In the 
recent discussion of {\it Gaia} DR2 data Kervella et al.\ (2019b) find that 
the parallax of the companion is consistent with that of the Cepheid, 
but that the proper motion has a projected relative velocity between 
the two of $\simeq$ 30 km s$^{-1}$.  However, they describe the position and 
parallax of the two stars as ``quite remarkable from a statistical point 
of view'', and suggest that binary motion in the companion is a possible
explanation.   FF Aql is a well known spectroscopic binary.
Astrometric motion has been detected by Benedict, et al. (2007).
It has been tentatively resolved by CHARA MIRC (Gallenne et al.\ 2019), with a separation of 8.9 mas. It also has 
strong indication of orbital motion in  
{\it Hipparcos}--{\it Gaia} proper motions (Kervella et al.\ 2019a).  
Combining proper motion information with the orbit, the separation
(8.2 mas corresponding to 4.5 AU)
and mass of the companion ($0.8 \pm 0.1 \,M_\odot$) were derived (Kervella 
et al.\ 2019; in Table~\ref{mcompw}).  This corresponds to 
a spectral type approximately K0 V, which is later than the 
range estimated from {\it IUE\/} spectra (Evans et al.\ 1990).  
At present the inferred masses from Benedict, et al., 
Kervella, et al., and Evans et al. are not consistent, however improvement from the next {\it Gaia} release should resolve this. 
Kervella et al.\ (2019b) do not find likely wider companions.

{\bf RV Sco}: {\it Hipparcos}--{\it Gaia} proper motions  clearly 
show orbital motion (Kervella et al.\ 2019a), 
as was suspected by Szabados (1989), identifying
a spectroscopic binary.  As a very preliminary estimate of the 
separation (for plotting purposes), we use estimated orbital 
period of 8000$^d$, corresponding to a semimajor axis of 14 AU for
reasonable masses.  
 WDS measures for RV Sco indicate that the 
brighter resolved candidate at a separation of 6.0$\arcsec$ has no relative 
motion with the Cepheid, and hence is likely bound, making three 
components in the system.  However, it is 
dynamically unlikely that both possible companions in   Table~\ref{com.sum}
are bound, hence the  6.0$\arcsec$ is the most likely physical companion.  
An {\it IUE} observation of the system showed that all 
stars in the system are A3~V or later, corresponding to a mass of $1.9 \,M_\odot$
or less.  

 
{\bf AP Sgr}: {\it Hipparcos}--{\it Gaia} proper motions show 
weak indication of orbital motion (Kervella et al.\ (2019a).  
Radial velocity observations 
indicate  possible  low amplitude orbital variation 
(Gieren 1982; Lloyd Evans 1982; 
Szabados 1989), with a period estimate of 7000 days.  As with RV Sco, this
has been used to derive a very preliminary estimate of the semimajor axis
in Table~\ref{mcompw}.
Additional velocity observations are needed to confirm this.  
This would continue the pattern of an interior orbit found in 
systems with resolved companions.  Alternately, the separation of 
the resolved companion is the second largest in Table~\ref{com.sum}, and 
the color of the companion is the reddest.  This would be 
consistent with the companion being a line of sight coincidence
with a field star. An {\it IUE} observation shows that any companion
in the system is  A5 V or later, corresponding to a mass of
$\leq\!1.8 \,M_\odot$. Two wider possibly comoving candidates (49,000
and 28,200 AU) have been identified by Kervella et al.\ (2019b). However, even 
the brightest (49,000 AU) is cooler than even a late M main sequence star.
Both are fainter and cooler than the limit for our
survey (hotter than M0 stars), so they are not included in Table~\ref{gcomp2}.


{\bf BB Sgr}:  
As with AP Sgr, there is some indication of 
low level
orbital velocity (Gieren 1982; Barnes et al.\ 1988; Szabados 1989),
but it is not conclusive. Velocities are available from 
Gorynya et al.\ (1996, 1998) from 1994 to 1997.  As discussed 
by Evans et al.\ (2015), they have a $\gamma$ velocity of 
6.6 km s$^{-1}$, which is very similar to velocities assembled
in Szabados (1989).  Furthermore, there is 
 no orbital motion to within less than 
1 km s$^{-1}$ in the Gorynya et al.\ data.  Anderson (2020, in prep.) also sees no orbital motion.
 {\it Hipparcos}--{\it Gaia\/} 
proper motions show no 
anomaly consistent with orbital motion (Kervella et al.\ 2019a). 
An {\it IUE} observation shows that any companion
in the system is  A0 V or later, corresponding to a mass of 
$\leq\!2.1 \,M_\odot$.  BB Sgr is a member of the open cluster 
Collinder 394 (Anderson et al.\ 2013) 
As with S Nor this makes possible a line-of-sight coincidence with a 
cluster member, rather than a bound companion.

{\bf Y Car}:  Table~\ref{com.sum} has two possible companions for Y Car, with 
similar separations from the Cepheid, but  it is dynamically unlikely that
both are bound companions.  At larger separations, Kervella et al.\ 
(2019b) identify several stars with similar proper motions from {\it Gaia},
indicating that there may be a comoving group, making a chance
line of sight alignment rather than a gravitationally bound companion 
likely for at least one of the candidates  Table~\ref{mcompw}. 
Y Car is a well-known spectroscopic binary (Petterson et al.\ 2004).
A velocity from {\it HST\/} STIS spectrum of the companion indicated that
the companion is itself a binary (Evans et al.\ 2005).  The spectral
type and mass of the hottest star in the system was determined
from an {\it IUE\/} spectrum to be  B9.0 V and $2.4 \,M_\odot$.
Preliminary estimates of the inclination and mass of the companion 
using {\it Hipparcos}--{\it Gaia\/} proper motions (Kervella et al.\ 
2019a) provide a total mass of the two companion stars of 
$2.8 \pm 0.4  \,M_\odot$.  (This is an estimate because the orbital 
period is relatively short, although there is a highly significant
detection of orbital motion.)  Since this is only slightly larger 
than the mass of the hotter star of the pair from the {\it IUE}
spectrum, this implies that the second star in the companion pair
has a low mass. The spectroscopic system with the Cepheid has 
been tentatively resolved (Gallenne et al.\ 2019) with 
the VLTI PIONIER, with a separation of 2.5 mas, corresponding 
to 3.6 AU.  Table~\ref{mcompw} summarizes the component properties 
considered most probable.

{\bf V350 Sgr}: There is one possible resolved companion to V350 Sgr
in Table~\ref{com.sum}.  V350 Sgr is a well known binary system, 
where the mass of the Cepheid has been measured (Evans et al.\ 2018).
The mass of $5.2 \pm 0.3  \,M_\odot$ comes from combining the 
ground-based orbit of the Cepheid with the orbital velocity 
amplitude of the companion from {\it HST\/} STIS spectra, 
and the mass of the companion inferred from  {\it IUE\/} spectra
of $2.5 \,M_\odot$.  The mass of the companion determined from 
the spectroscopic orbit and the {\it Hipparcos}--{\it Gaia\/}
proper motions is approximately 2 $\sigma$ from this:
$3.4 \pm 0.5 \,M_\odot$
(Kervella et al.\ 2019b using the orbit derived by 
Gallenne et al.\ 2019).
The proper motions have a strong orbital signature.
A tentative resolution of the system was made with VLTI PIONIER
(Gallenne et al.\ 2019) with a separation of 3.0 mas or 2.7 AU.
Outside the {\it HST} survey here, which 
includes $20''$ around the Cepheid,
two candidate bound companions are found from proper motion 
agreement (Kervella et al.\ 2019b), 
making it possible that the Cepheid is 
 part of a 4 or 5 member system. 

\end{document}